\pdfoutput=1

\documentclass[
  reprint,
  superscriptaddress,
 amsmath,amssymb,
 aps,prb
]{revtex4-1}

\usepackage{graphicx}
\usepackage{dcolumn}
\usepackage{bm}
\usepackage{color}

\bmdefine{\bdi}{i}
\bmdefine{\bdj}{j}
\bmdefine{\bdx}{x}
\bmdefine{\bdy}{y}
\bmdefine{\bdr}{r}
\bmdefine{\bdR}{R}
\bmdefine{\bdS}{S}
\bmdefine{\bdL}{L}
\bmdefine{\bdJ}{J}
\bmdefine{\bdA}{A}
\bmdefine{\bdE}{E}
\bmdefine{\bdD}{D}
\bmdefine{\bdQ}{Q}
\bmdefine{\bdq}{q}
\bmdefine{\bdK}{K}
\bmdefine{\bda}{a}
\bmdefine{\bdb}{b}
\bmdefine{\bdzero}{0}
\bmdefine{\bddelta}{\delta}

\begin{document}

\title{
  Polarization-dependent magnetic properties of periodically driven $\alpha$-RuCl$_{3}$
}

\author{Naoya Arakawa}
\email{arakawa@phys.chuo-u.ac.jp}
\affiliation{The Institute of Science and Engineering,
  Chuo University, Bunkyo, Tokyo 112-8551, Japan}
\author{Kenji Yonemitsu}
\affiliation{The Institute of Science and Engineering,
  Chuo University, Bunkyo, Tokyo 112-8551, Japan}
\affiliation{Department of Physics,
  Chuo University, Bunkyo, Tokyo 112-8551, Japan}

\date{\today}

\begin{abstract}

  We study magnetic properties of
  a periodically driven Mott insulator with strong spin-orbit coupling
  and show some properties characteristic of linearly polarized light. 
  We consider a $t_{2g}$-orbital Hubbard model 
  driven by circularly or linearly polarized light
  with strong spin-orbit coupling 
  and derive its effective Hamiltonian
  in the strong-interaction limit for a high-frequency case.
  We show that linearly polarized light can change
  not only the magnitudes and signs of the exchange interactions,
  but also their bond anisotropy
  even without the bond-anisotropic hopping integrals. 
  Because of this property, 
  the honeycomb-network spin system 
  could be transformed into weakly coupled zigzag or step spin chains
  for the light field polarized along the $b$- or $a$-axis, respectively.
  Then,
  analyzing how the light fields affect
  several magnetic states in a mean-field approximation, 
  we show that
  linearly polarized light can change
  the relative stability of the competing magnetic states,
  whereas such a change is absent 
  for circularly polarized light.
  We also analyze
  the effects of both
  the bond anisotropy of nearest-neighbor hopping integrals
  and a third-neighbor hopping integral on the magnetic states
  and show that
  the results obtained in a simple model,
  in which the bond-averaged nearest-neighbor hopping integrals are considered, 
  remain qualitatively unchanged
  except for the stability of zigzag states in the non-driven case
  and the degeneracy lifting of the zigzag or stripy states.
  
\end{abstract}
\maketitle

\section{Introduction}

In periodically driven systems,
magnetic properties 
can be controlled via a time-periodic field.
In the presence of a time-periodic field
the solution to the Schr\"{o}dinger equation satisfies
the Floquet theorem
and
the Floquet Hamiltonian, a time-independent Hamiltonian,
can describe the time evolution
in steps of the driving period $T$~\cite{Floq1,Floq2,review1}.
Such a description may be appropriate
if the effects of heating
due to the driving field are negligible; 
such a situation could be realized
before the system approaches
an infinite-temperature state~\cite{Heat-Floq1,Heat-Floq2}.
Since the Floquet Hamiltonian usually depends on
parameters of the driving field,
it is possible to control magnetic properties
of a periodically driven system. 
For example,
in a single-orbital Mott insulator
driven by $E(t)=E_{0}\cos\omega t$,
we can change the magnitude and sign of 
the antiferromagnetic Heisenberg interaction
by tuning $E_{0}$ and $\omega$~\cite{Floquet-1orbMott}.
This or an extended method could be used 
to control magnetic properties of Mott insulators.

If a periodically driven Mott insulator has strong spin-orbit coupling (SOC),
it may offer possibility for controlling various exchange interactions
and magnetic states.
The low-energy excitations of a Mott insulator with strong SOC 
can be described
by the spin and orbital
entangled degrees of freedom~\cite{StrongLS-Mott,StrongLS-Mott-review,Takagi-review}
and its effective Hamiltonian has
not only the Heisenberg interaction,
but also the anisotropic exchange
interactions~\cite{StrongLS-Ex1,StrongLS-Ex2,Rau-PRL,Valenti-PRB,NA-Jeff}.
Then,
various magnetic states appear,
depending on the values of the Heisenberg interaction
and
the anisotropic exchange interactions~\cite{Rau-PRL,Valenti-PRB,PD1,PD2,PD3,PD4,PD5}.
In our previous paper~\cite{Floquet-NA}
we showed that
by applying a circularly polarized light field
$\bdE(t)={}^{t}(E_{0}\cos\omega t\ -E_{0}\sin\omega t)$
to a multiorbital Mott insulator with strong SOC
and tuning $\omega$ and $E_{0}$, 
the magnitudes and signs of three exchange interactions
can be changed simultaneously;
these interactions are the Heisenberg interaction $J$,
the Kitaev interaction $K$,
and the off-diagonal symmetric exchange interaction $\Gamma$.

The aim of this paper is to clarify 
the polarization dependences of magnetic properties
for a periodically driven Mott insulator with strong SOC.
First,
it is essential to understand
the similarities and differences between 
the effects of circularly and linearly polarized light.
In general, magnetic properties of solids
depend on the polarization of light~\cite{review-opt}.
It is also necessary to clarify
how the changes in exchange interactions due to a light field
affect
energies of several magnetic states.
These topics are not discussed in our previous paper~\cite{Floquet-NA}. 

In this paper
we study the exchange interactions and the energies of several magnetic states
for a periodically driven Mott insulator with strong SOC. 
We consider  
a $t_{2g}$-orbital Hubbard model on the honeycomb lattice
with strong SOC and a field of linearly or circularly polarized light
as a model of periodically driven $\alpha$-RuCl$_{3}$.
To analyze the effects of both the bond anisotropy of nearest-neighbor hopping integrals
and a third-neighbor hopping integral, 
we consider five cases of our model,
including a simple case considered in our previous paper~\cite{Floquet-NA};
the differences among them are about the hopping integrals. 
Treating the effects of one of the light fields as Peierls phase factors
and using
the Floquet theory~\cite{Floquet-1orbMott,Floquet-MultiMott,Floquet-NA},
we derive the effective Hamiltonian of periodically driven $\alpha$-RuCl$_{3}$
in the strong-interaction limit for a high-frequency case.
Evaluating the exchange interactions numerically in the first two cases of our model, 
we show that
linearly polarized light
can be used to change
not only the magnitudes and signs of $J$, $K$, and $\Gamma$,
but also their bond anisotropy
even without the bond anisotropy of the hopping integrals.
This property,
which is distinct from the effects of circularly polarized light~\cite{Floquet-NA}, 
could be used to
transform the honeycomb-network spin system
into weakly coupled zigzag or step spin chains. 
We also show that
the effects of the bond anisotropy of the nearest-neighbor hopping integrals
are just quantitative
in the sense that it only
induces weak bond anisotropy of the exchange interactions. 
Then,
by using a mean-field approximation (MFA),
we evaluate the expectation value of our effective Hamiltonian. 
Analyzing the effects of the light fields on
the energies of several magnetic states
in the five cases of our model, 
we show that
linearly polarized light
can change the relative stability of the competing magnetic states,
whereas circularly polarized light does not.
Furthermore,
we show that
the results obtained in the simple case remain qualitatively unchanged 
except that
the stability of the zigzag states or the ferromagnetic state
depends on the strength of the bond anisotropy of the hopping integrals
and the value of the third-neighbor hopping integral
and that the degeneracy of the zigzag or stripy states
is lifted not only by the fields of linearly polarized light,
but also by the bond anisotropy of the hopping integrals. 

The rest of the paper is organized as follows.
In Sec. II we introduce the Hamiltonian of our model. 
It consists of the hopping integrals of the $t_{2g}$-orbital electrons
on the honeycomb lattice with a light field $\bdE(t)$,
their $LS$ coupling,
and the $t_{2g}$-orbital Hubbard interactions.
There are three $\bdE(t)$'s considered: 
$\bdE_{\textrm{circ}}(t)$, $\bdE_{\textrm{linear-}b}(t)$,
and $\bdE_{\textrm{linear-}a}(t)$ [Eqs. (\ref{eq:E_circ}){--}(\ref{eq:E_lin-a})]. 
We consider five cases of our model,
one of which is used in our previous paper~\cite{Floquet-NA};
the others are used
to study the effects of 
the bond-anisotropic nearest-neighbor hopping integrals
and the third-neighbor hopping integral.
In Sec. III A we express the effective Hamiltonian
of our periodically driven Mott insulator
in terms of the isotropic and the anisotropic exchange interactions. 
In Sec. III B we present
the $|u_{ij}|(=eE_{0}/\omega)$ dependences of 
the exchange interactions estimated numerically
in the first two cases of our model.
We compare the results for the three $\bdE(t)$'s and
discuss the properties characteristic of the linearly polarized light. 
In Sec. IV A,
by applying the MFA to our effective Hamiltonian,
we derive an expression of the energy of a magnetic state characterized
by the ordering vector.
We also explain the characteristics of the magnetic states considered.
In Sec. IV B
we present the $|u_{ij}|$ dependences
of the numerically evaluated energies of several magnetic states
in the five cases of our model.
Comparing the results for the three $\bdE(t)$'s,
we discuss the similarities and differences between the effects
of the fields of circularly and linearly polarized light.
We also analyze how the bond anisotropy of the nearest-neighbor hopping integrals
and the third-neighbor hopping integral affect the magnetic states.
In Sec. V
we discuss the validity of our model,
the effects of heating,
and an experimental observation of our results.
Furthermore,
we remark on 
a property induced by a field of circularly polarized light
and several directions for further relevant research.
In Sec. VI
we summarize the main results and their implications. 

Throughout this paper,
we set $\hbar=1$ and $a_{\textrm{NN}}=1$,
where $a_{\textrm{NN}}$ denotes
the distance between nearest-neighbor sites
on the honeycomb lattice (Fig. \ref{fig1}).
This choice of $a_{\textrm{NN}}$ leads to
$a_{\textrm{2nd}}=\sqrt{3}$ and $a_{\textrm{3rd}}=2$,
where $a_{\textrm{2nd}}$ and $a_{\textrm{3rd}}$ denote
the distances
between second-neighbor and third-neighbor, respectively, sites.
For simplicity,
we neglect a small difference between the lengths of a $Z$ bond and
of an $X$ or $Y$ bond. 

\section{Model}

\begin{figure}
  \includegraphics[width=50mm]{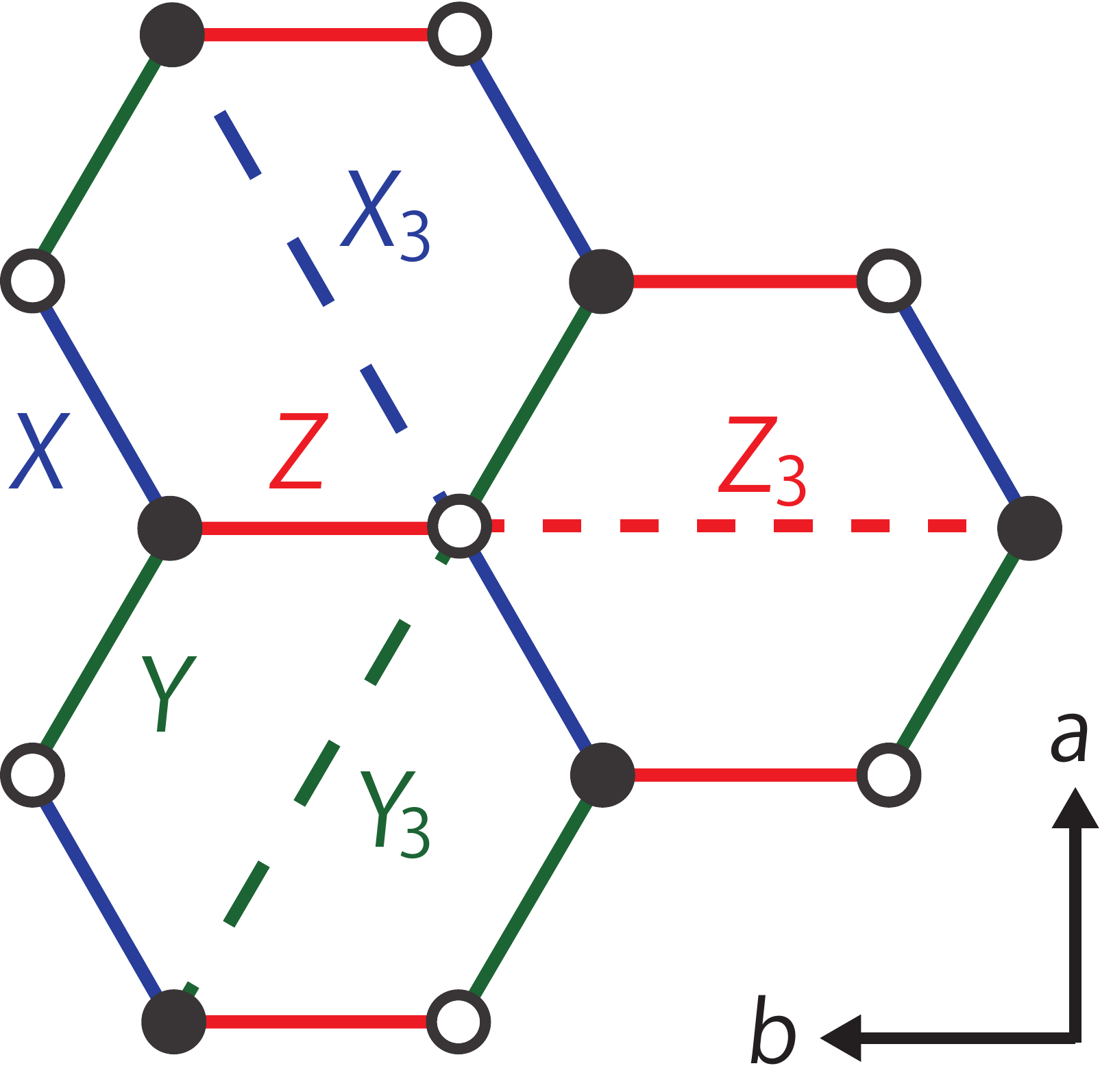}
  \caption{\label{fig1}
    Structure of the honeycomb lattice.
    Blue, green, and red lines
    represent 
    three nearest-neighbor bonds,
    i.e., $X$, $Y$, and $Z$ bonds, respectively;
    blue, green, and red dashed lines
    represent
    three third-neighbor bonds, i.e., 
    $X_{3}$, $Y_{3}$, and $Z_{3}$ bonds, respectively. 
    Black and white circles denote
    the $A$ and $B$ sublattices, respectively.
    This sublattice structure is necessary because 
    the honeycomb lattice,
    which is not a Bravais lattice,
    can be represented as a triangular Bravais lattice
    with a two-point basis~\cite{Ashcro-Merm}. 
    The $a$ and $b$ axes of the crystal are also shown. 
  }
\end{figure}

Our model Hamiltonian consists of three parts:
\begin{align}
  H=H_{\textrm{KE}}+H_{\textrm{SOC}}+H_{\textrm{int}},
\end{align}
where $H_{\textrm{KE}}$ represents the kinetic energy,
$H_{\textrm{SOC}}$ represents the atomic SOC~\cite{StrongLS-Mott-review},
and $H_{\textrm{int}}$ represents the Coulomb interactions~\cite{Kanamori}.

\begin{table*}
  \caption{\label{tab}Five cases of our model.
    Bond anisotropy of the nearest-neighbor hopping integrals
    is absent in the first and third cases and present in the other three cases. 
    The third nearest-neighbor hopping integral
    is neglected in the first two cases and considered in the last three cases. 
  }
  \begin{ruledtabular}
    \begin{tabular}{cccccc}
      Parameter & $1$st & $2$nd & $3$rd & $4$th & $5$th\\ \hline
      Bond anisotropy & Absent & Present & Absent & Present & Present \\
      $t_{\textrm{3rd}}$ (meV) & $0$ & $0$ & $-40$ & $-40$ & $-60$
    \end{tabular}
  \end{ruledtabular}
\end{table*}

The kinetic energy is given by the hopping integrals of the $t_{2g}$-orbital electrons
on the honeycomb lattice (Fig. \ref{fig1})
in the presence of a light field $\bdE(t)={}^{t}(E_{\bar{b}}(t)\ E_{a}(t))$.
Here
$E_{\bar{b}}(t)$ is the component antiparallel to the $b$ axis
and $E_{a}(t)$ is that parallel to the $a$ axis. 
The $\bdE(t)$ for circularly polarized light is given by
\begin{align}
  \bdE_{\textrm{circ}}(t)={}^{t}(E_{0}\cos\omega t\ -E_{0}\sin\omega t),\label{eq:E_circ}
\end{align}
and that for linearly polarized light is given by
\begin{align}
  \bdE_{\textrm{linear-}b}(t)={}^{t}(E_{0}\sin\omega t\ 0),\label{eq:E_lin-b}
\end{align}
or
\begin{align}
  \bdE_{\textrm{linear-}a}(t)={}^{t}(0\ E_{0}\sin\omega t).\label{eq:E_lin-a}
\end{align}
We do not consider the helicity of circularly polarized light 
because the magnetic properties shown in this paper for $\bdE(t)=\bdE_{\textrm{circ}}(t)$
remain unchanged even for $\bdE(t)={}^{t}(E_{0}\cos\omega t\ E_{0}\sin\omega t)$.
Then, we treat the effects of $\bdE(t)$ as Peierls phase factors: 
\begin{align}
  H_{\textrm{KE}}=
  \sum_{i,j}\sum_{a,b}\sum_{\sigma}
  t_{iajb}e^{-ie(\bdR_{i}-\bdR_{j})\cdot\bdA(t)}c_{ia\sigma}^{\dagger}c_{jb\sigma},\label{eq:H_KE}
\end{align}
where $\bdA(t)={}^{t}(A_{\bar{b}}(t)\ A_{a}(t))$
for $\bdE(t)=\bdE_{\textrm{circ}}(t)$, $\bdE_{\textrm{linear-}b}(t)$,
or $\bdE_{\textrm{linear-}a}(t)$
is given by
\begin{align}
  &\bdA_{\textrm{circ}}(t)=
      {}^{t}\Bigl(-\frac{E_{0}}{\omega}\sin\omega t\ -\frac{E_{0}}{\omega}\cos\omega t\Bigr),\label{eq:A_c}\\
  &\bdA_{\textrm{linear-}b}(t)=
      {}^{t}\Bigl(\frac{E_{0}}{\omega}\cos\omega t\ 0\Bigr),\label{eq:A_lb}
\end{align}
or 
\begin{align}
  \hspace{-20pt}
  &\bdA_{\textrm{linear-}a}(t)=
      {}^{t}\Bigl(0\ \frac{E_{0}}{\omega}\cos\omega t\Bigr),\label{eq:A_la}
\end{align}
respectively.
As for $t_{iajb}$'s,
we consider five cases (TABLE \ref{tab}). 
In the first case 
$t_{iajb}$'s are parametrized by
three nearest-neighbor hopping integrals~\cite{Rau-PRL,Floquet-NA}:
the finite $t_{iajb}$'s for the $Z$ bonds (Fig. \ref{fig1}) are
\begin{align}
  &t_{id_{yz}jd_{yz}}=t_{id_{zx}jd_{zx}}=t_{1},\label{eq:t1}\\
  &t_{id_{yz}jd_{zx}}=t_{id_{zx}jd_{yz}}=t_{2},\label{eq:t2}\\
  &t_{id_{xy}jd_{xy}}=t_{3},\label{eq:t3}
\end{align}
and 
those for the $X$ and the $Y$ bonds are obtained
by replacing $(x,y,z)$ in Eqs. (\ref{eq:t1}){--}(\ref{eq:t3}) 
by $(y,z,x)$ and $(z,x,y)$, respectively.
This case, which is used also in our previous paper~\cite{Floquet-NA},
corresponds to a minimal model of $\alpha$-RuCl$_{3}$. 
In the second case
the bond anisotropy of $t_{1}$, $t_{2}$, and $t_{3}$ is considered:
the finite $t_{iajb}$'s for the $Z$ bonds are
the same as Eqs. (\ref{eq:t1}){--}(\ref{eq:t3}),
whereas those for the $X$ or the $Y$ bonds are
\begin{align}
  &t_{id_{zx}jd_{zx}}=t_{id_{xy}jd_{xy}}=t_{1}^{\prime},\label{eq:t1-X}\\
  &t_{id_{zx}jd_{xy}}=t_{id_{xy}jd_{zx}}=t_{2}^{\prime},\label{eq:t2-X}\\
  &t_{id_{yz}jd_{yz}}=t_{3}^{\prime},\label{eq:t3-X}
\end{align}
or
\begin{align}
  &t_{id_{xy}jd_{xy}}=t_{id_{yz}jd_{yz}}=t_{1}^{\prime},\label{eq:t1-Y}\\  
  &t_{id_{xy}jd_{yz}}=t_{id_{yz}jd_{xy}}=t_{2}^{\prime},\label{eq:t2-Y}\\
  &t_{id_{zx}jd_{zx}}=t_{3}^{\prime},\label{eq:t3-Y}
\end{align}
respectively.
This case is used to study 
the effects of the weak bond anisotropy~\cite{Valenti-PRB} of $\alpha$-RuCl$_{3}$
on the magnetic properties. 
In the third or the fourth case 
we consider not only the nearest-neighbor hopping integrals considered
in the first or the second case, respectively,  
but also a third-neighbor hopping integral $t_{\textrm{3rd}}$.
The $t_{\textrm{3rd}}$ for the $Z_{3}$, $X_{3}$, or $Y_{3}$ bonds (Fig. \ref{fig1}) 
is the intraorbital hopping integral
of the $d_{xy}$, $d_{yz}$, or $d_{zx}$ orbital, respectively. 
(We consider only this among the third neighbor hopping integrals
because it is the largest~\cite{Valenti-PRB}.)
The fifth case is almost the same as the fourth case
except for the value of $t_{\textrm{3rd}}$ (see the first paragraph of Sec. IV B).  
The last three cases are used 
to analyze the effects of $J_{\textrm{3rd}}$,
the Heisenberg interaction between third neighbors,
on several magnetic states.
In Sec. V
we will compare our choices of $t_{iajb}$'s with
the result obtained in first-principles calculations~\cite{Valenti-PRB}. 

The atomic SOC is given by
the $LS$ coupling of the $t_{2g}$-orbital electrons~\cite{StrongLS-Mott-review,Takagi-review}.
Because of its nonperturbative effect,
the $t_{2g}$-orbital states with spin degrees of freedom
are converted into
the $j_{\textrm{eff}}=1/2$ doublet and
the $j_{\textrm{eff}}=3/2$ quartet.
(In this argument
we have omitted the components of the $LS$ coupling
between the $t_{2g}$ and the $e_{g}$ orbitals
because the crystal field energy between them 
is supposed to be sufficiently large;
this is 
the reason why the total angular momentum is not $j$, but $j_{\textrm{eff}}$.)
For $\alpha$-RuCl$_{3}$
the low-energy properties can be described by 
the $j_{\textrm{eff}}=1/2$ doublet~\cite{Jeff-RuCl3-1,Jeff-RuCl3-2},
which is occupied by an electron (or a hole) per site, 
\begin{align}
  &|+\rangle_{i}
  =\frac{1}{\sqrt{3}}
  (c_{id_{yz}\downarrow}^{\dagger}+ic_{id_{zx}\downarrow}^{\dagger}+c_{id_{xy}\uparrow}^{\dagger})
  |0\rangle,\label{eq:jeff-p}\\
  &|-\rangle_{i}
  =\frac{1}{\sqrt{3}}
  (c_{id_{yz}\uparrow}^{\dagger}-ic_{id_{zx}\uparrow}^{\dagger}-c_{id_{xy}\downarrow}^{\dagger})
  |0\rangle.\label{eq:jeff-m}
\end{align}
In these states
the spin and the orbital are entangled.

The Coulomb interactions are given by
the multiorbital Hubbard interactions~\cite{Kanamori} of the $t_{2g}$-orbital electrons:
\begin{align}
  &H_{\textrm{int}}=
  \sum_{i}\Bigl\{\sum_{a,b}
  c_{ia\uparrow}^{\dagger}c_{ia\downarrow}^{\dagger}
  [U\delta_{a,b}+J^{\prime}(1-\delta_{a,b})]
  c_{ib\downarrow}c_{ib\uparrow}\notag\\
  &+\sum_{\substack{a,b\\a>b}}\sum_{\sigma,\sigma^{\prime}}
  c_{ia\sigma}^{\dagger}c_{ib\sigma^{\prime}}^{\dagger}
  (U^{\prime}c_{ib\sigma^{\prime}}c_{ia\sigma}
  -J_{\textrm{H}}c_{ib\sigma}c_{ia\sigma^{\prime}})\Bigr\},\label{eq:H_int}
\end{align}
where $U$, $J^{\prime}$, $U^{\prime}$, and $J_{\textrm{H}}$
are the intraorbital Hubbard interaction,
the pair hopping,
the interorbital Hubbard interaction,
and the Hund's coupling, respectively.
In deriving the exchange interactions
of periodically driven $\alpha$-RuCl$_{3}$ (Sec. III A)
we use $H_{\textrm{int}}$ expressed in terms of the irreducible representations
of doubly occupied states~\cite{NA-Jeff,Floquet-NA,Ishihara}:
\begin{align}
  H_{\textrm{int}}=\sum_{i}\sum_{\Gamma,g_{\Gamma}}U_{\Gamma}
  |i;\Gamma,g_{\Gamma}\rangle\langle i;\Gamma,g_{\Gamma}|,\label{eq:H_int-Gam}
\end{align}
where $U_{\Gamma}$'s are given by
\begin{align}
  &U_{A_{1}}=U+2J^{\prime},\label{eq:U_A}\\
  &U_{E}=U-J^{\prime},\\
  &U_{T_{1}}=U^{\prime}-J_{\textrm{H}},\\
  &U_{T_{2}}=U^{\prime}+J_{\textrm{H}},\label{eq:U_T2}
\end{align}
and $|i;\Gamma,g_{\Gamma}\rangle$'s are given by
\begin{align}
  &|i;A_{1}\rangle
  =\frac{1}{\sqrt{3}}
  (c_{id_{yz}\uparrow}^{\dagger}c_{id_{yz}\downarrow}^{\dagger}
  +c_{id_{zx}\uparrow}^{\dagger}c_{id_{zx}\downarrow}^{\dagger}
  +c_{id_{xy}\uparrow}^{\dagger}c_{id_{xy}\downarrow}^{\dagger})|0\rangle,\label{eq:A1}\\
  &|i;E,u\rangle
  =\frac{1}{\sqrt{6}}
  (c_{id_{yz}\uparrow}^{\dagger}c_{id_{yz}\downarrow}^{\dagger}
  +c_{id_{zx}\uparrow}^{\dagger}c_{id_{zx}\downarrow}^{\dagger}\notag\\
  &\ \ \ \ \ \ \ \ \ \ \ \ \ \ \ \ \ \ \
  -2c_{id_{xy}\uparrow}^{\dagger}c_{id_{xy}\downarrow}^{\dagger})|0\rangle,\\
  &|i;E,v\rangle
  =\frac{1}{\sqrt{2}}
  (c_{id_{yz}\uparrow}^{\dagger}c_{id_{yz}\downarrow}^{\dagger}
  -c_{id_{zx}\uparrow}^{\dagger}c_{id_{zx}\downarrow}^{\dagger})|0\rangle,\\
  &|i;T_{1},\alpha_{+}\rangle
  =c_{id_{yz}\uparrow}^{\dagger}c_{id_{zx}\uparrow}^{\dagger}|0\rangle,\\
  &|i;T_{1},\alpha_{-}\rangle
  =c_{id_{yz}\downarrow}^{\dagger}c_{id_{zx}\downarrow}^{\dagger}|0\rangle,\\
  &|i;T_{1},\alpha\rangle
  =\frac{1}{\sqrt{2}}
  (c_{id_{yz}\uparrow}^{\dagger}c_{id_{zx}\downarrow}^{\dagger}
  +c_{id_{yz}\downarrow}^{\dagger}c_{id_{zx}\uparrow}^{\dagger})|0\rangle,\\ 
  &|i;T_{1},\beta_{+}\rangle
  =c_{id_{zx}\uparrow}^{\dagger}c_{id_{xy}\uparrow}^{\dagger}|0\rangle,\\
  &|i;T_{1},\beta_{-}\rangle
  =c_{id_{zx}\downarrow}^{\dagger}c_{id_{xy}\downarrow}^{\dagger}|0\rangle,\\
  &|i;T_{1},\beta\rangle
  =\frac{1}{\sqrt{2}}
  (c_{id_{zx}\uparrow}^{\dagger}c_{id_{xy}\downarrow}^{\dagger}
  +c_{id_{zx}\downarrow}^{\dagger}c_{id_{xy}\uparrow}^{\dagger})|0\rangle,\\
  &|i;T_{1},\gamma_{+}\rangle
  =c_{id_{xy}\uparrow}^{\dagger}c_{id_{yz}\uparrow}^{\dagger}|0\rangle,\\
  &|i;T_{1},\gamma_{-}\rangle
  =c_{id_{xy}\downarrow}^{\dagger}c_{id_{yz}\downarrow}^{\dagger}|0\rangle,\\
  &|i;T_{1},\gamma\rangle
  =\frac{1}{\sqrt{2}}
  (c_{id_{xy}\uparrow}^{\dagger}c_{id_{yz}\downarrow}^{\dagger}
  +c_{id_{xy}\downarrow}^{\dagger}c_{id_{yz}\uparrow}^{\dagger})|0\rangle,\\
  &|i;T_{2},\alpha\rangle
  =\frac{1}{\sqrt{2}}
  (c_{id_{yz}\uparrow}^{\dagger}c_{id_{zx}\downarrow}^{\dagger}
  -c_{id_{yz}\downarrow}^{\dagger}c_{id_{zx}\uparrow}^{\dagger})|0\rangle,\\
  &|i;T_{2},\beta\rangle
  =\frac{1}{\sqrt{2}}
  (c_{id_{zx}\uparrow}^{\dagger}c_{id_{xy}\downarrow}^{\dagger}
  -c_{id_{zx}\downarrow}^{\dagger}c_{id_{xy}\uparrow}^{\dagger})|0\rangle,\\
  &|i;T_{2},\gamma\rangle
  =\frac{1}{\sqrt{2}}
  (c_{id_{xy}\uparrow}^{\dagger}c_{id_{yz}\downarrow}^{\dagger}
  -c_{id_{xy}\downarrow}^{\dagger}c_{id_{yz}\uparrow}^{\dagger})|0\rangle.\label{eq:T2-gam}
\end{align}

\section{Exchange interactions}
In this section
we study the exchange interactions of periodically driven $\alpha$-RuCl$_{3}$.
In Sec. III A
we derive an effective Hamiltonian expressed in terms of the exchange interactions 
in the strong-interaction limit for a high-frequency case of the driving field.
This derivation is performed in the first case of our model,
and the changes in the last four cases are also remarked on.
In Sec. III B
we present the dependences of the exchange interactions
on the dimensionless parameter $|u_{ij}|$ for some non-resonant values of $\omega$.
The reason why we use non-resonant $\omega$'s
is that our theory is valid only for such $\omega$'s
[see the derivation of Eq. (\ref{eq:Heff-t_final})]. 
We also discuss the effects of circularly or linearly polarized light
and the similarities and differences between them.

\subsection{Theory}

We derive the exchange interactions of
our periodically driven Mott insulator with strong SOC.
Since the derivation for circularly polarized light
has been described in our previous paper~\cite{Floquet-NA},
we explain the main points and some changes for linearly polarized light.
Here 
we derive the expression in the first case of our model;
the expressions in the other four cases can be obtained 
from symmetry arguments (see the fifth paragraph of this section).
The derivation consists of three steps. 

First,
we derive an effective Hamiltonian
for the periodically driven Mott insulator with strong SOC.
To do this,
we consider the strong-interaction limit
in which $t_{iajb}$'s in Eq. (\ref{eq:H_KE}) are much smaller than
$U_{\Gamma}$'s in Eq. (\ref{eq:H_int-Gam}).
In this limit
the solution to Schr\"{o}dinger's equation can be approximately expressed 
as $|\Psi\rangle_{t}\approx |\Psi_{0}\rangle_{t}+|\Psi_{1}\rangle_{t}$
with $|\Psi_{0}\rangle_{t}$ and $|\Psi_{1}\rangle_{t}$,
the states without and with, respectively, a doubly occupied site.
Thus,
the solution can be obtained by solving a set of simultaneous equations,  
\begin{align}
  &i\partial_{t}|\Psi_{0}\rangle_{t}
  =\mathcal{P}_{0}H_{\textrm{KE}}|\Psi_{1}\rangle_{t}
  +H_{\textrm{SOC}}|\Psi_{0}\rangle_{t},\label{eq:Psi0}\\
  &i\partial_{t}|\Psi_{1}\rangle_{t}
  =H_{\textrm{KE}}|\Psi_{0}\rangle_{t}
  +(\mathcal{P}_{1}H_{\textrm{KE}}\mathcal{P}_{1}
  +\tilde{H}_{\textrm{int}})|\Psi_{1}\rangle_{t},\label{eq:Psi1}
\end{align}
where
$\mathcal{P}_{0}$ and $\mathcal{P}_{1}$ 
denote the projections onto the subspaces without and with, respectively,
a doubly occupied site, 
and $\tilde{H}_{\textrm{int}}$
is defined as
\begin{align}
  \tilde{H}_{\textrm{int}}=H_{\textrm{int}}+H_{\textrm{SOC}}.
\end{align}
Then, we suppose that 
$\omega$ is much larger than $t_{iajb}$'s.
In this situation
$\mathcal{P}_{1}H_{\textrm{KE}}\mathcal{P}_{1}$ in Eq. (\ref{eq:Psi1})
could be replaced by its time-averaged one $\bar{H}_{\textrm{KE}}$.
As derived in Appendix A, 
$\bar{H}_{\textrm{KE}}$ is given by
\begin{align}
  \bar{H}_{\textrm{KE}}
  =\mathcal{P}_{1}
  \sum_{i,j}\sum_{a,b}\sum_{\sigma}
  t_{iajb}
  \mathcal{J}_{0}[\tilde{u}_{ij}^{(\textrm{p})}]c_{ia\sigma}^{\dagger}c_{jb\sigma}
  \mathcal{P}_{1},\label{eq:barH_KE}
\end{align}
where $\mathcal{J}_{n}(x)$ is the $n$th Bessel function of the first kind,
$\tilde{u}_{ij}^{(\textrm{p})}$'s
for $\bdA(t)=\bdA_{\textrm{circ}}(t)$, $\bdA_{\textrm{linear-}b}(t)$, and $\bdA_{\textrm{linear-}a}(t)$
are given by
\begin{align}
  &\tilde{u}_{ij}^{(\textrm{c})}
  =u_{ij}=\frac{eE_{0}}{\omega}\textrm{sgn}(i-j),\label{eq:uij_circ}\\
  &\tilde{u}_{ij}^{(\textrm{lb})}=
  \begin{cases}
    \frac{1}{2}u_{ij} \ \ (X\ \textrm{or}\ Y\ \textrm{bonds}),\\
    u_{ij} \ \ \ \ (Z\ \textrm{bonds}),\label{eq:uij_linear-b}
  \end{cases}
\end{align}
and
\begin{align}
  \tilde{u}_{ij}^{(\textrm{la})}=
  \begin{cases}
    \frac{\sqrt{3}}{2}u_{ij} \ \ (X\ \textrm{or}\ Y\ \textrm{bonds}),\\
    \ 0 \ \ \ \ \ \ \ (Z\ \textrm{bonds}),\label{eq:uij_linear-a}
  \end{cases}
\end{align} 
respectively,
and $\textrm{sgn}(i-j)$ is $1$ for $i\in A$ or $-1$ for $i\in B$. 
(Note that $i\in A$ or $i\in B$ means
that $i$ is on the $A$ sublattice or on the $B$ sublattice, respectively.) 
By using the replacement
$\mathcal{P}_{1}H_{\textrm{KE}}\mathcal{P}_{1}\rightarrow \bar{H}_{\textrm{KE}}$,
we can write Eq. (\ref{eq:Psi1}) as
\begin{align}
  &(i\partial_{t}-\bar{H}_{\textrm{KE}}-\tilde{H}_{\textrm{int}})|\Psi_{1}\rangle_{t}
  =H_{\textrm{KE}}|\Psi_{0}\rangle_{t}.\label{eq:Psi1-next}
\end{align}
This is equivalent to the following equation:
\begin{align}
  i\partial_{t}[e^{i(\bar{H}_{\textrm{KE}}+\tilde{H}_{\textrm{int}})t}|\Psi_{1}\rangle_{t}]
  =e^{i(\bar{H}_{\textrm{KE}}+\tilde{H}_{\textrm{int}})t}H_{\textrm{KE}}|\Psi_{0}\rangle_{t}.\label{eq:Psi1-rewrite2}
\end{align}
As derived in Appendix B,
the solution to Eq. (\ref{eq:Psi1-rewrite2}) can be expressed as follows:
\begin{align}
  |\Psi_{1}\rangle_{t}=
  \sum_{i,j,a,b,\sigma}\sum_{n=-\infty}^{\infty}
  \frac{t_{iajb}\tilde{\mathcal{J}}_{-n\nu_{ij}}^{(\textrm{p})}(u_{ij})e^{-in\omega t}}
       {n\omega-\bar{H}_{\textrm{KE}}-\tilde{H}_{\textrm{int}}}
       c_{ia\sigma}^{\dagger}c_{jb\sigma}|\Psi_{0}\rangle_{t},\label{eq:Psi1-final}
\end{align}
where $\tilde{\mathcal{J}}_{-n\nu_{ij}}^{(\textrm{p})}(u_{ij})$'s
for $\bdA(t)=\bdA_{\textrm{circ}}(t)$, $\bdA_{\textrm{linear-}b}(t)$, and $\bdA_{\textrm{linear-}a}(t)$
are given by 
\begin{align}
  &\tilde{\mathcal{J}}_{-n\nu_{ij}}^{(\textrm{c})}(u_{ij})=
  \begin{cases}
    \ \mathcal{J}_{-n}(u_{ij})e^{-in\frac{5\pi}{3}} \ \ (\nu_{ij} = X),\\
    \ \mathcal{J}_{-n}(u_{ij})e^{-in\frac{\pi}{3}} \ \ \ (\nu_{ij} = Y),\\
    \ \mathcal{J}_{-n}(u_{ij})e^{-in\pi}\ \ \ \ (\nu_{ij} = Z),
  \end{cases}\label{eq:tilBes_circ}\\
  &\tilde{\mathcal{J}}_{-n\nu_{ij}}^{(\textrm{lb})}(u_{ij})=
  \begin{cases}
    \ \mathcal{J}_{-n}(\frac{u_{ij}}{2})e^{+in\frac{\pi}{2}} \ \ \ (\nu_{ij} = X),\\
    \ \mathcal{J}_{-n}(\frac{u_{ij}}{2})e^{+in\frac{\pi}{2}} \ \ \ (\nu_{ij} = Y),\\
    \ \mathcal{J}_{-n}(u_{ij})e^{-in\frac{\pi}{2}} \ \ \ (\nu_{ij} = Z),
  \end{cases}\label{eq:tilBes_linear-b}
\end{align}
and
\begin{align}
  \tilde{\mathcal{J}}_{-n\nu_{ij}}^{(\textrm{la})}(u_{ij})=
  \begin{cases}
    \ \mathcal{J}_{-n}(\frac{\sqrt{3}}{2}u_{ij})e^{-in\frac{\pi}{2}} \ \ (\nu_{ij} = X),\\
    \ \mathcal{J}_{-n}(\frac{\sqrt{3}}{2}u_{ij})e^{+in\frac{\pi}{2}} \ \ (\nu_{ij} = Y),\\
    \ \delta_{n,0} \ \ \ \ \ \ \ \ \ \ \ \ \ \ \ \ \ \ \ (\nu_{ij} = Z),
  \end{cases}\label{eq:tilBes_linear-a}
\end{align}
respectively.
By substituting Eq. (\ref{eq:Psi1-final}) into Eq. (\ref{eq:Psi0})
and omitting the constant term $H_{\textrm{SOC}}|\Psi_{0}\rangle_{t}$,
we obtain
\begin{align}
  i\partial_{t}|\Psi_{0}\rangle_{t}=H_{\textrm{eff}}(t)|\Psi_{0}\rangle_{t},\label{eq:Psi0-final}
\end{align}
where
\begin{align}
  H_{\textrm{eff}}(t)=&
  \sum_{i,j,i^{\prime},j^{\prime}}
  \sum_{n,m=-\infty}^{\infty}
  \mathcal{P}_{0}T_{i^{\prime}j^{\prime}}
  \tilde{\mathcal{J}}_{m\nu_{i^{\prime}j^{\prime}}}^{(\textrm{p})}(u_{i^{\prime}j^{\prime}})\notag\\
  &\times 
  \frac{e^{i(m-n)\omega t}}
       {n\omega-\bar{H}_{\textrm{KE}}-\tilde{H}_{\textrm{int}}} 
       \tilde{\mathcal{J}}_{-n\nu_{ij}}^{(\textrm{p})}(u_{ij})T_{ij}\mathcal{P}_{0},\label{eq:Heff-t}
\end{align}
and
\begin{align}
  T_{ij}=\sum_{a,b}\sum_{\sigma}t_{iajb}c_{ia\sigma}^{\dagger}c_{jb\sigma}.\label{eq:Tij} 
\end{align}
Furthermore,
since in the denominator of Eq. (\ref{eq:Heff-t})
$H_{\textrm{int}}$ gives the largest contribution of   
$\bar{H}_{\textrm{KE}}$ and $\tilde{H}_{\textrm{int}}(=H_{\textrm{int}}+H_{\textrm{SOC}})$,
we replace $n\omega-\bar{H}_{\textrm{KE}}-\tilde{H}_{\textrm{int}}$ in Eq. (\ref{eq:Heff-t})
by $n\omega-H_{\textrm{int}}$;
this replacement may be sufficient for non-resonant $\omega$'s 
(i.e., the $\omega$'s at which the denominator does not diverge).
As a result,
we have
\begin{align}
  H_{\textrm{eff}}(t)=&
  \sum_{i,j}\sum_{n,m=-\infty}^{\infty} 
  \mathcal{P}_{0}T_{ji}\tilde{\mathcal{J}}_{m\nu_{ji}}^{(\textrm{p})}(u_{ji})\notag\\
  &\times
  \frac{e^{i(m-n)\omega t}}{n\omega-H_{\textrm{int}}} 
  \tilde{\mathcal{J}}_{-n\nu_{ij}}^{(\textrm{p})}(u_{ij})
  T_{ij}\mathcal{P}_{0}.\label{eq:Heff-t_final}
\end{align}

Second,
we derive the leading term of $H_{\textrm{eff}}(t)$.
By expressing $H_{\textrm{eff}}(t)$ as
the Fourier series $H_{\textrm{eff}}(t)=\sum_{l}e^{il\omega t}H_{l}$
and using a high-frequency expansion of the Floquet theory~\cite{review1},
we can write $H_{\textrm{eff}}(t)$ in the form
\begin{align}
  H_{\textrm{eff}}(t)=H_{0}+O\Bigl(\frac{J_{\textrm{ex}}^{2}}{\omega}\Bigr),\label{eq:H-Floquet-next}
\end{align}
where $J_{\textrm{ex}}$ is of the order of the exchange interactions.
If $\omega$ is high enough to satisfy $\omega \gg |J_{\textrm{ex}}|$, 
the leading term of $H_{\textrm{eff}}(t)$ comes from 
the Floquet Hamiltonian $\bar{H}_{\textrm{eff}}(=H_{0})$:
\begin{align}
  \bar{H}_{\textrm{eff}}
  =&\frac{\omega}{2\pi}\int_{0}^{2\pi/\omega}dt H_{\textrm{eff}}(t)\notag\\
  =&\sum_{i,j}\sum_{n=-\infty}^{\infty} 
  \mathcal{P}_{0}T_{ji}
  \frac{\mathcal{J}_{n}[\tilde{u}_{ij}^{(\textrm{p})}]^{2}}{n\omega-H_{\textrm{int}}} 
  T_{ij}\mathcal{P}_{0}.\label{eq:barH}
\end{align}
Furthermore,
by using Eq. (\ref{eq:H_int-Gam})
and expressing the projection operators $\mathcal{P}_{0}$'s in terms of the possible states, 
we can rewrite Eq. (\ref{eq:barH}) as follows:
\begin{align}
  \bar{H}_{\textrm{eff}}
  =&\sum_{i,j}\sum_{n=-\infty}^{\infty}\sum_{\Gamma,g_{\Gamma}} 
  \mathcal{P}_{0}T_{ji}|i;\Gamma,g_{\Gamma}\rangle
  \frac{\mathcal{J}_{n}[\tilde{u}_{ij}^{(\textrm{p})}]^{2}}{n\omega-U_{\Gamma}}\notag\\
  &\times
  \langle i;\Gamma,g_{\Gamma}|T_{ij}\mathcal{P}_{0}\notag\\
  =&\sum_{i,j}\sum_{n=-\infty}^{\infty}\sum_{\Gamma,g_{\Gamma}}\sum_{\textrm{i},\textrm{f}} 
  \langle\textrm{f}|T_{ji}|i;\Gamma,g_{\Gamma}\rangle
  \frac{\mathcal{J}_{n}[\tilde{u}_{ij}^{(\textrm{p})}]^{2}}{n\omega-U_{\Gamma}}\notag\\
  &\times
  \langle i;\Gamma,g_{\Gamma}|T_{ij}|\textrm{i}\rangle
  |\textrm{f}\rangle
  \langle\textrm{i}|,\label{eq:barH-next}
\end{align}
where $|\textrm{i}\rangle$ and $|\textrm{f}\rangle$
are restricted to the $j_{\textrm{eff}}=1/2$ subspace.
Equation (\ref{eq:barH-next}) shows that
a light field affects
the effective Hamiltonian 
through the factor $\mathcal{J}_{n}[\tilde{u}_{ij}^{(\textrm{p})}]^{2}/(n\omega-U_{\Gamma})$
(i.e., the changes due to the Bessel functions
and the energy shifts of intermediate states). 
Note that if we replace
$\sum_{n=-\infty}^{\infty}\mathcal{J}_{n}[\tilde{u}_{ij}^{(\textrm{p})}]^{2}/(n\omega-U_{\Gamma})$
in Eq. (\ref{eq:barH-next})
by $1/(-U_{\Gamma})$,
the resultant equation gives the effective Hamiltonian in the absence of $\bdE(t)$. 

Third,
we express Eq. (\ref{eq:barH-next}) in terms of exchange interactions.
To do this,
we calculate the possible terms
for the $Z$ bonds on the honeycomb lattice;
the other terms can be obtained from symmetry arguments.
$T_{ij}$ for the $Z$ bonds, $T_{ij}^{Z}$, is given by
\begin{align}
  T_{ij}^{Z}
  =&t_{1}\sum_{\sigma}(c_{id_{yz}\sigma}^{\dagger}c_{jd_{yz}\sigma}+c_{id_{zx}\sigma}^{\dagger}c_{jd_{zx}\sigma})\notag\\
  &+t_{2}\sum_{\sigma}(c_{id_{yz}\sigma}^{\dagger}c_{jd_{zx}\sigma}+c_{id_{zx}\sigma}^{\dagger}c_{jd_{yz}\sigma})\notag\\
  &+t_{3}\sum_{\sigma}c_{id_{xy}\sigma}^{\dagger}c_{jd_{xy}\sigma}.\label{eq:Tij^z}
\end{align}
Then,
since Eq. (\ref{eq:barH-next}) is the sum of two-site terms,
we express 
$|\textrm{i}\rangle$ and $|\textrm{f}\rangle$ 
as the products of the $j_{\textrm{eff}}=1/2$ states at two sites:
\begin{align}
  \{|\textrm{i}\rangle,|\textrm{f}\rangle\}
  =\{|+,+\rangle,|+,-\rangle,|-,+\rangle,|-,-\rangle\},\label{eq:i,f}
\end{align}
where
$|+,+\rangle=|+\rangle_{1}|+\rangle_{2}$,
$|+,-\rangle=|+\rangle_{1}|-\rangle_{2}$,
$|-,+\rangle=|-\rangle_{1}|+\rangle_{2}$,
and $|-,-\rangle=|-\rangle_{1}|-\rangle_{2}$; 
$|+\rangle_{i}$ and $|-\rangle_{i}$ have been defined
in Eqs. (\ref{eq:jeff-p}) and (\ref{eq:jeff-m}).
Since $|i;\Gamma,g_{\Gamma}\rangle$'s and $U_{\Gamma}$'s 
are given by
Eqs. (\ref{eq:A1}){--}(\ref{eq:T2-gam}) and Eqs. (\ref{eq:U_A}){--}(\ref{eq:U_T2}),
we can write the finite terms of Eq. (\ref{eq:barH-next})
for the $Z$ bonds in the form 
\begin{align}
  \sum_{\langle i,j\rangle_{Z}}
  \Bigl[J_{Z}\bdS_{i}\cdot\bdS_{j}
    +K_{Z}S_{i}^{z}S_{j}^{z}
    +\Gamma_{Z} (S_{i}^{x}S_{j}^{y}+S_{i}^{y}S_{j}^{x})
  \Bigr],\label{eq:Heff_Z}
\end{align}
where the summation $\sum_{\langle i,j\rangle_{Z}}$ is over all the $Z$ bonds,  
\begin{align}
  J_{Z}=&\sum_{n=-\infty}^{\infty}
  \frac{4\mathcal{J}_{n}(u_{ij}^{Z})^{2}}{27}
  \Bigl[
    \frac{(2t_{1}+t_{3})^{2}}{U+2J^{\prime}-n\omega}
    +\frac{6t_{1}(t_{1}+2t_{3})}{U^{\prime}-J_{\textrm{H}}-n\omega}\notag\\
    &\ \ \ \ +\frac{2[(t_{1}-t_{3})^{2}-3t_{2}^{2}]}{U-J^{\prime}-n\omega}
    +\frac{6t_{2}^{2}}{U^{\prime}+J_{\textrm{H}}-n\omega}
  \Bigr],\label{eq:Jz}\\
  K_{Z}=&\sum_{n=-\infty}^{\infty}
  \frac{4\mathcal{J}_{n}(u_{ij}^{Z})^{2}}{9}
  \Bigl[
    \frac{4t_{2}^{2}}{U-J^{\prime}-n\omega}
    -\frac{(t_{1}-t_{3})^{2}+t_{2}^{2}}{U^{\prime}+J_{\textrm{H}}-n\omega}\notag\\
    &\ \ \ \ -\frac{3t_{2}^{2}-(t_{1}-t_{3})^{2}}{U^{\prime}-J_{\textrm{H}}-n\omega}
  \Bigr],\label{eq:Kz}\\
  \Gamma_{Z}=&\sum_{n=-\infty}^{\infty}
  \frac{8\mathcal{J}_{n}(u_{ij}^{Z})^{2}}{9}
  \Bigl[
    \frac{t_{2}(t_{1}-t_{3})}{U^{\prime}-J_{\textrm{H}}-n\omega}
    -\frac{t_{2}(t_{1}-t_{3})}{U^{\prime}+J_{\textrm{H}}-n\omega}
  \Bigr],\label{eq:Gamz}
\end{align}
and
\begin{align}
  u_{ij}^{Z}
  =
  \begin{cases}
    \ u_{ij} \ \ \textrm{for}\ \bdA(t)=\bdA_{\textrm{circ}}(t),\\
    \ u_{ij} \ \ \textrm{for}\ \bdA(t)=\bdA_{\textrm{linear-}b}(t),\\
    \ 0 \ \ \ \ \textrm{for}\ \bdA(t)=\bdA_{\textrm{linear-}a}(t).
  \end{cases}\label{eq:uz}
\end{align}
The derivation of Eq. (\ref{eq:Heff_Z}) is described in Appendix C.
Equations (\ref{eq:Jz}){--}(\ref{eq:uz}) show that
for $\bdA(t)=\bdA_{\textrm{circ}}(t)$ and $\bdA_{\textrm{linear-}b}(t)$ 
$J_{Z}$, $K_{Z}$, and $\Gamma_{Z}$ can be changed by
varying $\omega$, $E_{0}$, or both,
whereas for $\bdA(t)=\bdA_{\textrm{linear-}a}(t)$
those remain unchanged due to $\mathcal{J}_{n}(0)=\delta_{n,0}$.
This means that
the exchange interactions for the $Z$ bonds
are not affected
if the light is polarized along the $a$-axis.
Then,
replacing $(x,y,z,Z)$ in Eq. (\ref{eq:Heff_Z})
by $(y,z,x,X)$ or $(z,x,y,Y)$, 
we obtain 
the possible terms for the $X$ or the $Y$ bonds, respectively.
As a result, 
$\bar{H}_{\textrm{eff}}$'s for the $X$ and the $Y$ bonds
are given by
\begin{align}
  \sum_{\langle i,j\rangle_{X}}
  \Bigl[J_{X}\bdS_{i}\cdot\bdS_{j}
    +K_{X}S_{i}^{x}S_{j}^{x}
    +\Gamma_{X} (S_{i}^{y}S_{j}^{z}+S_{i}^{z}S_{j}^{y})
  \Bigr],\label{eq:Heff_X}
\end{align}
and
\begin{align}
  \sum_{\langle i,j\rangle_{Y}}
  \Bigl[J_{Y}\bdS_{i}\cdot\bdS_{j}
    +K_{Y}S_{i}^{y}S_{j}^{y}
    +\Gamma_{Y} (S_{i}^{z}S_{j}^{x}+S_{i}^{x}S_{j}^{z})
  \Bigr],\label{eq:Heff_Y}
\end{align}
respectively.
Here
the summations $\sum_{\langle i,j\rangle_{X}}$ and 
$\sum_{\langle i,j\rangle_{Y}}$ are over all the $X$ and the $Y$ bonds, respectively;
$J_{X}$, $J_{Y}$, $K_{X}$, $K_{Y}$,
$\Gamma_{X}$, and $\Gamma_{Y}$
are given by
\begin{align}
  J_{X}&=J_{Y}\notag\\
  &=\sum_{n=-\infty}^{\infty}
  \frac{4\mathcal{J}_{n}(u_{ij}^{X})^{2}}{27}
  \Bigl[
    \frac{(2t_{1}+t_{3})^{2}}{U+2J^{\prime}-n\omega}
    +\frac{6t_{1}(t_{1}+2t_{3})}{U^{\prime}-J_{\textrm{H}}-n\omega}\notag\\
    &\ \ \ \ +\frac{2[(t_{1}-t_{3})^{2}-3t_{2}^{2}]}{U-J^{\prime}-n\omega}
    +\frac{6t_{2}^{2}}{U^{\prime}+J_{\textrm{H}}-n\omega}
  \Bigr],\label{eq:Jx}\\
  K_{X}&=K_{Y}\notag\\
  &=\sum_{n=-\infty}^{\infty}
  \frac{4\mathcal{J}_{n}(u_{ij}^{X})^{2}}{9}
  \Bigl[
    \frac{4t_{2}^{2}}{U-J^{\prime}-n\omega}
    -\frac{(t_{1}-t_{3})^{2}+t_{2}^{2}}{U^{\prime}+J_{\textrm{H}}-n\omega}\notag\\
    &\ \ \ \ -\frac{3t_{2}^{2}-(t_{1}-t_{3})^{2}}{U^{\prime}-J_{\textrm{H}}-n\omega}
  \Bigr],\label{eq:Kx}\\
  \Gamma_{X}&=\Gamma_{Y}\notag\\
  &=\sum_{n=-\infty}^{\infty}
  \frac{8\mathcal{J}_{n}(u_{ij}^{X})^{2}}{9}
  \Bigl[
    \frac{t_{2}(t_{1}-t_{3})}{U^{\prime}-J_{\textrm{H}}-n\omega}
    -\frac{t_{2}(t_{1}-t_{3})}{U^{\prime}+J_{\textrm{H}}-n\omega}
  \Bigr],\label{eq:Gamx}
\end{align}
where
\begin{align}
  u_{ij}^{X}
  =
  \begin{cases}
    \ u_{ij} \ \ \ \ \ \ \textrm{for}\ \bdA(t)=\bdA_{\textrm{circ}}(t),\\
    \ \frac{1}{2}u_{ij} \ \ \ \ \textrm{for}\ \bdA(t)=\bdA_{\textrm{linear-}b}(t),\\
    \ \frac{\sqrt{3}}{2}u_{ij} \ \ \textrm{for}\ \bdA(t)=\bdA_{\textrm{linear-}a}(t).
  \end{cases}\label{eq:ux}
\end{align}
Comparing Eqs. (\ref{eq:Jx}){--}(\ref{eq:ux})
with Eqs. (\ref{eq:Jz}){--}(\ref{eq:uz}),
we see 
linearly polarized light 
can induce the bond anisotropy of the exchange interactions,
i.e.,
the differences between the exchange interactions
for the $Z$ bonds and those for the $X$ or $Y$ bonds, 
even without the bond anisotropy of the hopping integrals.
Such light-induced bond anisotropy does not appear
for circularly polarized light.
The origin of this light-induced bond anisotropy
is the difference in the argument of the Bessel function
[Eqs. (\ref{eq:uz}) and (\ref{eq:ux})].
In Sec. III B
we will analyze the light-induced bond anisotropy quantitatively. 
Then, by combining Eqs. (\ref{eq:Heff_X}) and (\ref{eq:Heff_Y})
with Eq. (\ref{eq:Heff_Z}),
we can express $\bar{H}_{\textrm{eff}}$ as follows:
\begin{align}
  \bar{H}_{\textrm{eff}}
  =\sum_{\langle i,j\rangle}
  \Bigl[J_{\delta}\bdS_{i}\cdot\bdS_{j}
    +K_{\delta}S_{i}^{\gamma}S_{j}^{\gamma}
    +\Gamma_{\delta} (S_{i}^{\alpha}S_{j}^{\beta}+S_{i}^{\beta}S_{j}^{\alpha})
  \Bigr],\label{eq:Heff_1st}
\end{align}
where
\begin{align}
  (\alpha,\beta,\gamma,\delta)
  =
  \begin{cases}
    \ (y,z,x,X) \ \ \ \ (X\ \textrm{bonds}),\\
    \ (z,x,y,Y) \ \ \ \ (Y\ \textrm{bonds}),\\
    \ (x,y,z,Z) \ \ \ \ (Z\ \textrm{bonds}).
  \end{cases}\label{eq:alp-bet-gam}
\end{align}

Before showing the results obtained in numerical calculations,
we comment on the exchange interactions
in the other four cases of our model.
In the second case, in which 
the bond anisotropy of $t_{1}$, $t_{2}$, and $t_{3}$
is considered [Eqs. (\ref{eq:t1-X}){--}(\ref{eq:t3-Y})],
the exchange interactions for the $X$ and $Y$ bonds 
are given by
\begin{align}
  J_{X}&=J_{Y}\notag\\
  &=\sum_{n=-\infty}^{\infty}
  \frac{4\mathcal{J}_{n}(u_{ij}^{X})^{2}}{27}
  \Bigl[
    \frac{(2t_{1}^{\prime}+t_{3}^{\prime})^{2}}{U+2J^{\prime}-n\omega}
    +\frac{6t_{1}^{\prime}(t_{1}^{\prime}+2t_{3}^{\prime})}{U^{\prime}-J_{\textrm{H}}-n\omega}\notag\\
    &\ \ \ \ +\frac{2[(t_{1}^{\prime}-t_{3}^{\prime})^{2}-3(t_{2}^{\prime})^{2}]}{U-J^{\prime}-n\omega}
    +\frac{6(t_{2}^{\prime})^{2}}{U^{\prime}+J_{\textrm{H}}-n\omega}
  \Bigr],\label{eq:Jx-ani}\\
  K_{X}&=K_{Y}\notag\\
  &=\sum_{n=-\infty}^{\infty}
  \frac{4\mathcal{J}_{n}(u_{ij}^{X})^{2}}{9}
  \Bigl[
    \frac{4(t_{2}^{\prime})^{2}}{U-J^{\prime}-n\omega}
    -\frac{(t_{1}^{\prime}-t_{3}^{\prime})^{2}+(t_{2}^{\prime})^{2}}{U^{\prime}+J_{\textrm{H}}-n\omega}\notag\\
    &\ \ \ \ -\frac{3(t_{2}^{\prime})^{2}-(t_{1}^{\prime}-t_{3}^{\prime})^{2}}{U^{\prime}-J_{\textrm{H}}-n\omega}
  \Bigr],\label{eq:Kx-ani}\\
  \Gamma_{X}&=\Gamma_{Y}\notag\\
  &=\sum_{n=-\infty}^{\infty}
  \frac{8\mathcal{J}_{n}(u_{ij}^{X})^{2}}{9}
  \Bigl[
    \frac{t_{2}^{\prime}(t_{1}^{\prime}-t_{3}^{\prime})}{U^{\prime}-J_{\textrm{H}}-n\omega}
    -\frac{t_{2}^{\prime}(t_{1}^{\prime}-t_{3}^{\prime})}{U^{\prime}+J_{\textrm{H}}-n\omega}
  \Bigr],\label{eq:Gamx-ani}
\end{align}
rather than by Eqs. (\ref{eq:Jx}){--}(\ref{eq:Gamx});
those for the $Z$ bonds are given
by Eqs. (\ref{eq:Jz}){--}(\ref{eq:Gamz}).
Thus,
by comparing the results in this case and the first case, 
we can understand how 
the light-induced bond anisotropy of the exchange interactions
is affected by the bond anisotropy of the nearest-neighbor hopping integrals.
Those results are shown in Sec. III B. 
In Sec. IV B
we will analyze the energies of several magnetic states
not only in those cases,
but also in the additional three cases. 
In the latter three cases
we consider the third-neighbor hopping integral $t_{\textrm{3rd}}$,
as well as the nearest-neighbor hopping integrals; 
it leads to the third-neighbor Heisenberg interaction
\begin{align}
  \sum_{\langle\langle\langle i,j\rangle\rangle\rangle}
  J_{\delta}^{\textrm{3rd}}\bdS_{i}\cdot\bdS_{j},\label{eq:H-J3rd}
\end{align}
where the summation $\sum_{\langle\langle\langle i,j\rangle\rangle\rangle}$
is over all the $Z_{3}$, $X_{3}$, and $Y_{3}$ bonds (Fig. \ref{fig1}),
$\delta$ is $Z$, $X$, or $Y$ for the $Z_{3}$, $X_{3}$, or $Y_{3}$ bonds, respectively, 
and $J^{\textrm{3rd}}_{\delta}$ is given (in the second-order perturbation theory) by
\begin{align}
  J_{\delta}^{\textrm{3rd}}=&\sum_{n=-\infty}^{\infty}
  \frac{4\mathcal{J}_{n}(2u_{ij}^{\delta})^{2}}{27}
  \Bigl(
    \frac{t_{\textrm{3rd}}^{2}}{U+2J^{\prime}-n\omega}
    +\frac{2t_{\textrm{3rd}}^{2}}{U-J^{\prime}-n\omega}
    \Bigr).\label{eq:J3rd}
\end{align}
[For example,
$J_{Z}^{\textrm{3rd}}$ is obtained 
by replacing $u_{ij}^{Z}$, $t_{3}$, $t_{1}$, and $t_{2}$ in Eq. (\ref{eq:Jz})
by $2u_{ij}^{Z}$, $t_{\textrm{3rd}}$, $0$, and $0$, respectively.]
In the third case of our model
the effective Hamiltonian consists of the sum of Eqs. (\ref{eq:Heff_1st})
and (\ref{eq:H-J3rd})
with Eqs. (\ref{eq:Jz}){--}(\ref{eq:uz}), (\ref{eq:Jx}){--}(\ref{eq:ux}),
and (\ref{eq:J3rd}).
In the fourth or the fifth case
the effective Hamiltonian is the same as that in the third case
except that the exchange interactions for the $X$ and the $Y$ bonds
are given by Eqs. (\ref{eq:Jx-ani}){--}(\ref{eq:Gamx-ani}). 
Analyses of several magnetic states
in those three cases may be useful to clarify the role of $J_{\delta}^{\textrm{3rd}}$,
which is shown to be important without $\bdE(t)$~\cite{Valenti-PRB},
in determining the magnetic states of periodically driven $\alpha$-RuCl$_{3}$.

\begin{figure*}
  \includegraphics[width=170mm]{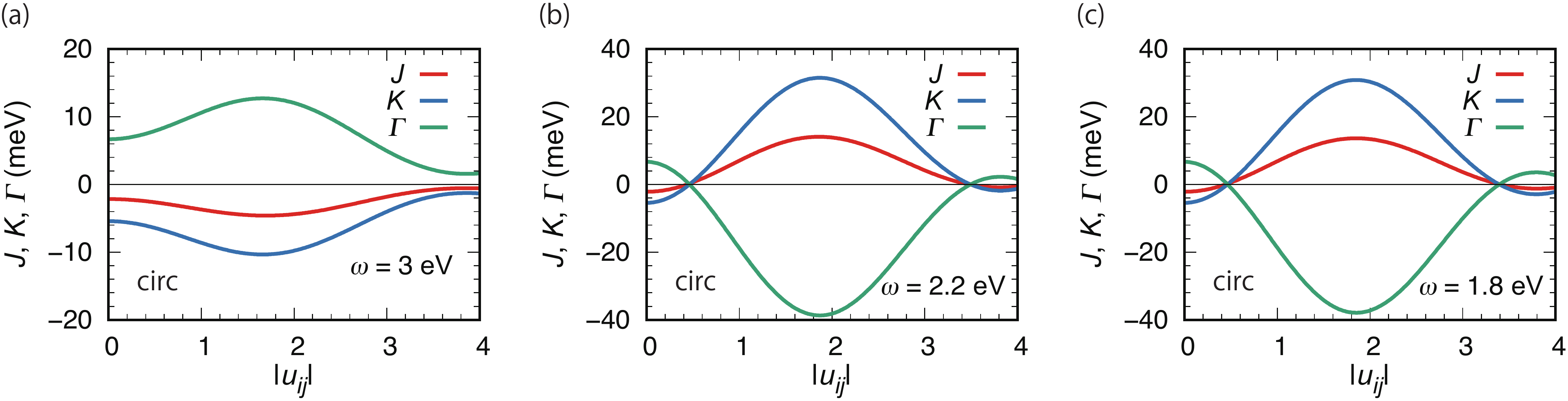}
  \caption{\label{fig2}
    The $|u_{ij}|(=|\frac{eE_{0}}{\omega}|)$ dependences of $J$, $K$, and $\Gamma$
    with $\bdE_{\textrm{circ}}(t)$
    at (a) $\omega=3$, (b) $2.2$, and (c) $1.8$ eV 
    in the first case of our model.
    In this case
    there is no bond anisotropy of the exchange interactions: 
    $J_{Z}=J_{X}=J_{Y}=J$, $K_{Z}=K_{X}=K_{Y}=K$, and $\Gamma_{Z}=\Gamma_{X}=\Gamma_{Y}=\Gamma$. 
  }
\end{figure*}

In our analyses
we do not consider the third-neighbor Kitaev interaction $K_{\delta}^{\textrm{3rd}}$,
although it is also induced by $t_{\textrm{3rd}}$.
This is because the value of $K_{\delta}^{\textrm{3rd}}$ is overestimated
if the third-neighbor hopping integrals other than $t_{\textrm{3rd}}$ are omitted
(the value of $K_{\delta}^{\textrm{3rd}}$
is very small in a more realistic situation~\cite{Valenti-PRB}). 
In contrast to $K_{\delta}^{\textrm{3rd}}$,
the value of $J_{\delta}^{\textrm{3rd}}$ is underestimated.
For more details about the values of $J_{\delta}^{\textrm{3rd}}$ and $K_{\delta}^{\textrm{3rd}}$,
see Appendix D.

\subsection{Results}

\begin{figure*}
  \includegraphics[width=170mm]{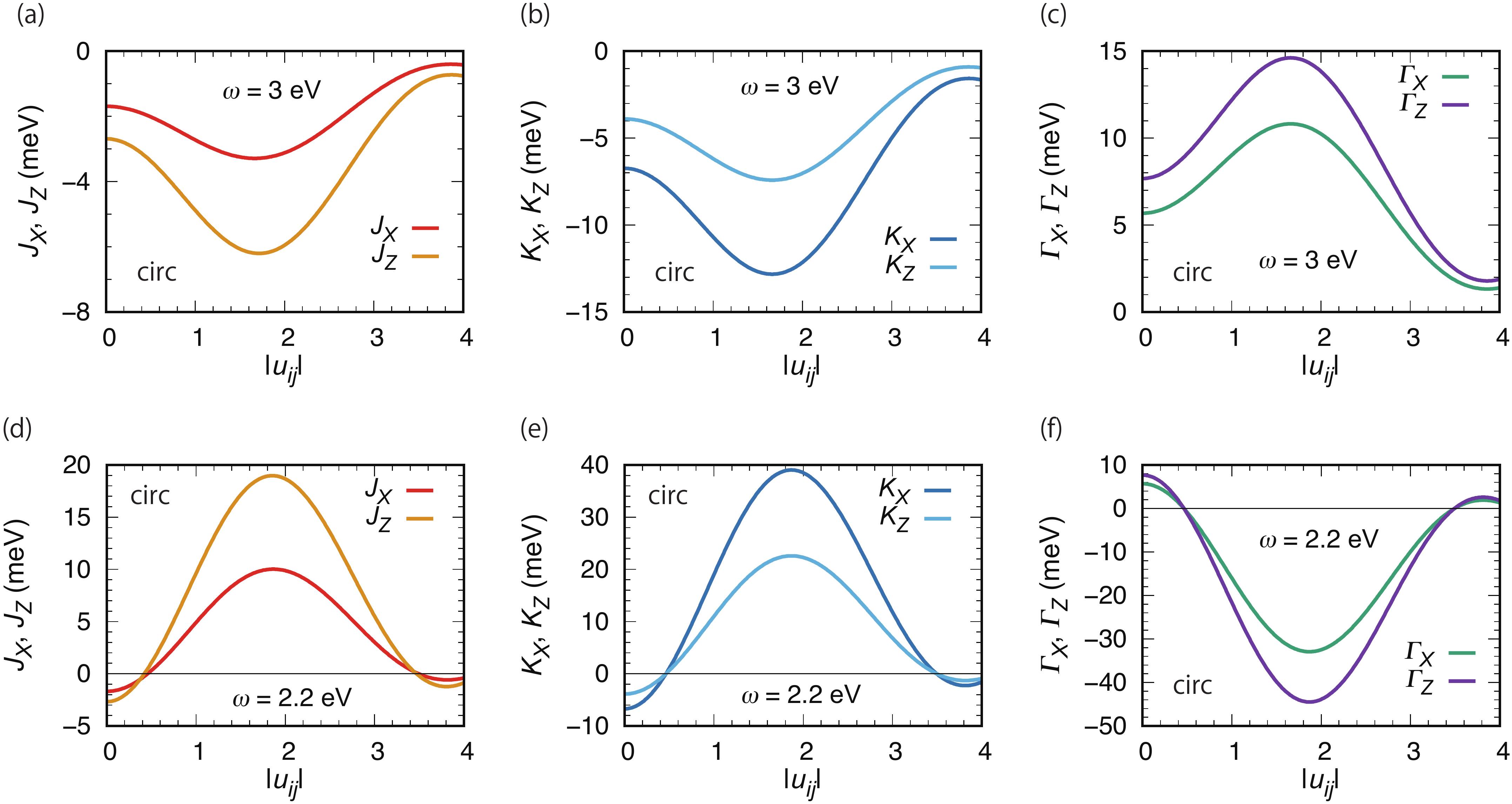}
  \caption{\label{fig3}
    The $|u_{ij}|(=|\frac{eE_{0}}{\omega}|)$ dependences of
    $J_{X}(=J_{Y})$, $J_{Z}$, $K_{X}(=K_{Y})$, $K_{Z}$,
    $\Gamma_{X}(=\Gamma_{Y})$, and $\Gamma_{Z}$ 
    with $\bdE_{\textrm{circ}}(t)$
    in the second case of our model.
    The value of $\omega$ is $3$ eV in (a){--}(c)
    and $2.2$ eV in (d){--}(f). 
    In contrast to the first case with $\bdE_{\textrm{circ}}(t)$ (Fig. \ref{fig2}),
    the exchange interactions become bond-anisotropic due to the bond anisotropy
    of the nearest-neighbor hopping integrals. 
  }
\end{figure*}

We numerically evaluate the exchange interactions
for some non-resonant $\omega$'s in the first two cases of our model.
To do this, 
we replace $\sum_{n=-\infty}^{\infty}$'s in the exchange interactions
by $\sum_{n=-n_{\textrm{max}}}^{n_{\textrm{max}}}$'s and set $n_{\textrm{max}}=500$; 
in the first case the exchange interactions are given 
by Eqs. (\ref{eq:Jz}){--}(\ref{eq:Gamz})
and Eqs. (\ref{eq:Jx}){--}(\ref{eq:Gamx}),
whereas in the second case those are given
by Eqs. (\ref{eq:Jz}){--}(\ref{eq:Gamz})
and Eqs. (\ref{eq:Jx-ani}){--}(\ref{eq:Gamx-ani}).
Furthermore, we set $J^{\prime}=J_{\textrm{H}}$, $U^{\prime}=U-2J_{\textrm{H}}$,
$U=3$ eV, and $J_{\textrm{H}}=0.5$ eV.
We choose the values of the nearest-neighbor hopping integrals as follows:
in the first case $t_{1}=47$ meV, $t_{2}=160$ meV, and $t_{3}=-129$ meV; 
in the second case $t_{1}=51$ meV, $t_{2}=158$ meV, $t_{3}=-154$ meV,
$t_{1}^{\prime}=45$ meV, $t_{2}^{\prime}=162$ meV, and $t_{3}^{\prime}=-103$ meV.
The values in the first case correspond to
the averages~\cite{Floquet-NA} of the values
obtained in the first-principles calculations
(e.g., $t_{1}$ is the average of $t_{1}$, $t_{1a}^{\prime}$, and $t_{1b}^{\prime}$
of Ref. \onlinecite{Valenti-PRB});
and the values in the second case are consistent with 
those obtained in the first-principles calculations~\cite{Valenti-PRB}
except that $t_{1}^{\prime}$ is the average of $t_{1a}^{\prime}$ and $t_{1b}^{\prime}$
(i.e., the tiny difference between them is neglected in our analyses).

First,
we present the $|u_{ij}|$ dependences of the exchange interactions
with circularly polarized light.
Those in the first case at $\omega=3$, $2.2$, and $1.8$ eV
are shown in Figs. \ref{fig2}(a){--}(c).
We see that 
only the magnitudes of the exchange interactions are changed at $\omega=3$ eV,
whereas their magnitudes and signs can be changed at $\omega=2.2$ and $1.8$ eV.
This property remains unchanged
in the second case [Figs. \ref{fig3}(a){--}(f)].
The main difference between the results in these two cases is
that there is no bond anisotropy of the exchange interactions
in the first case (i.e., $J_{Z}=J_{X}=J_{Y}=J$,
$K_{Z}=K_{X}=K_{Y}=K$, and $\Gamma_{Z}=\Gamma_{X}=\Gamma_{Y}=\Gamma$),
whereas it is induced by the bond anisotropy of the
nearest-neighbor hopping integrals
in the second case.
Note that at $|u_{ij}|=0$ in the latter case 
we have $J_{X}/J_{Z}\sim 0.63$, $K_{X}/K_{Z}\sim 1.7$, and
$\Gamma_{X}/\Gamma_{Z}\sim 0.74$,
which are consistent with the values obtained in 
the first-principles calculations~\cite{Valenti-PRB}
(i.e., $J_{X}/J_{Z}\sim 0.64$, $K_{X}/K_{Z}\sim 1.5$, and
$\Gamma_{X}/\Gamma_{Z}\sim 0.74$).

\begin{figure*}
  \includegraphics[width=170mm]{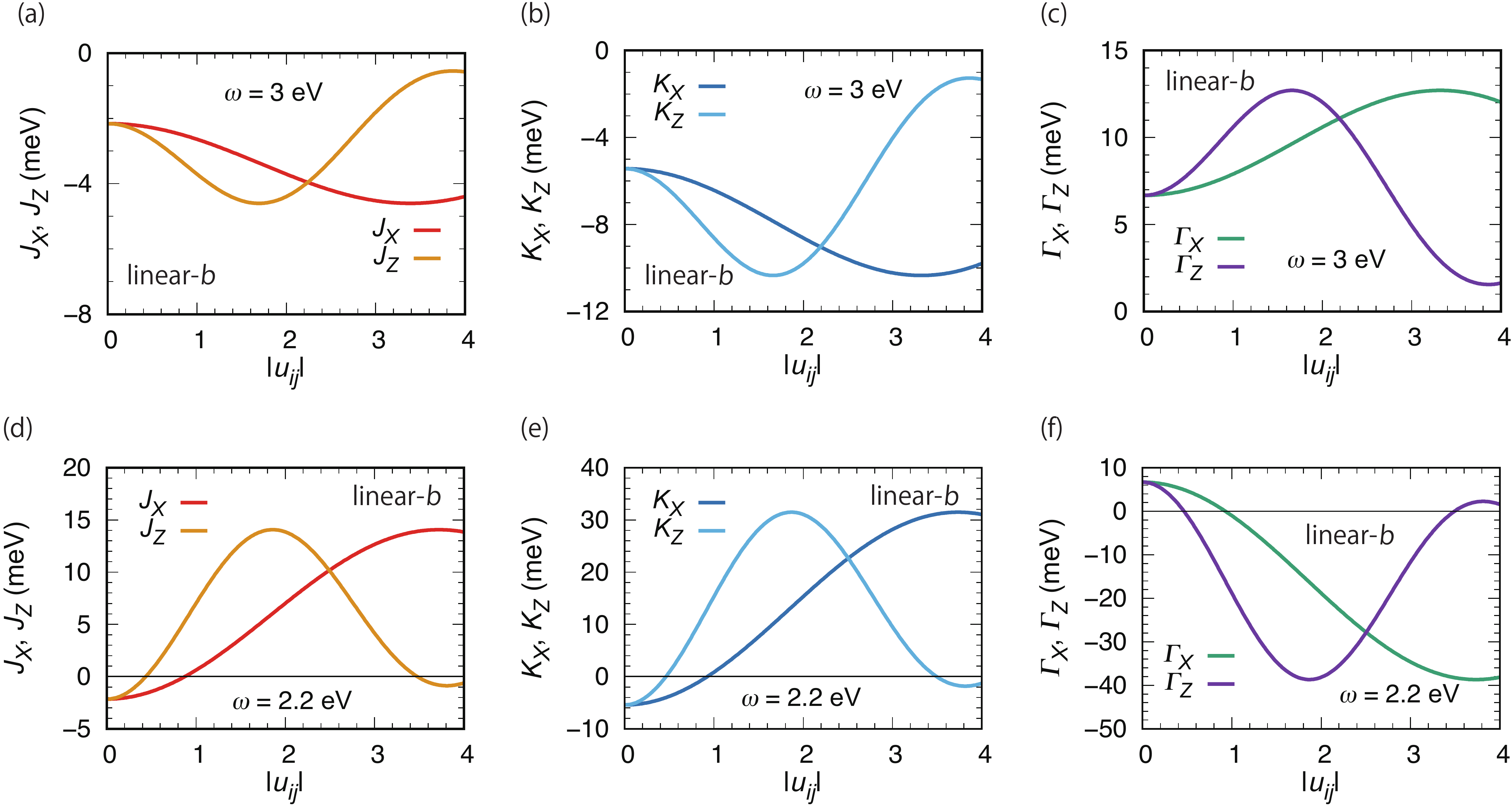}
  \caption{\label{fig4}
    The $|u_{ij}|(=|\frac{eE_{0}}{\omega}|)$ dependences of
    $J_{X}(=J_{Y})$, $J_{Z}$, $K_{X}(=K_{Y})$, $K_{Z}$,
    $\Gamma_{X}(=\Gamma_{Y})$, and $\Gamma_{Z}$ 
    with $\bdE_{\textrm{linear-}b}(t)$
    in the first case of our model.
    The value of $\omega$ is $3$ eV in (a){--}(c)
    and $2.2$ eV in (d){--}(f). 
    In contrast to the first case with $\bdE_{\textrm{circ}}(t)$ (Fig. \ref{fig2}),
    the light field induces the bond anisotropy of the exchange interactions
    even without the bond anisotropy of the hopping integrals. 
  }
\end{figure*}

\begin{figure*}
  \includegraphics[width=170mm]{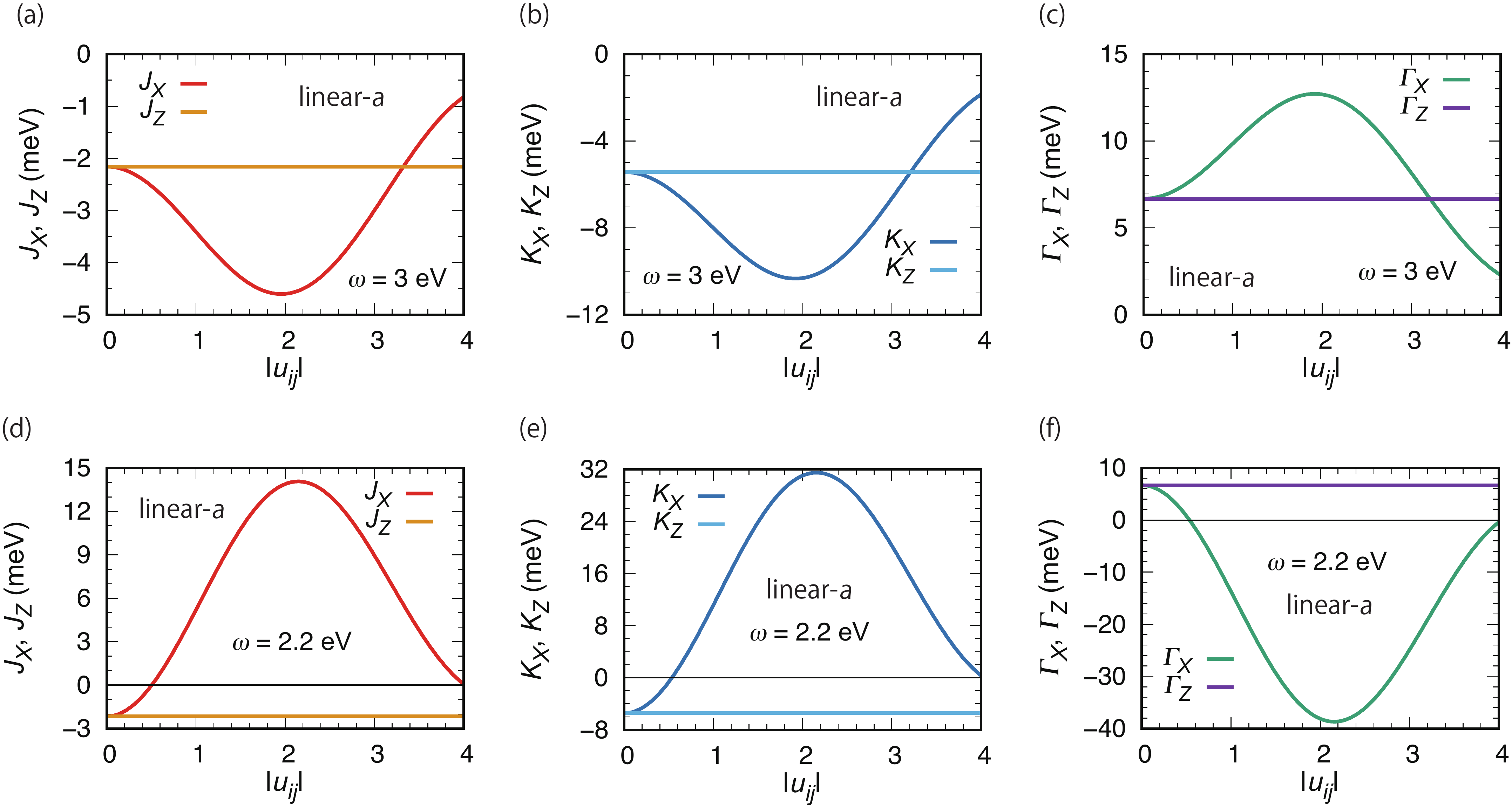}
  \caption{\label{fig5}
    The $|u_{ij}|(=|\frac{eE_{0}}{\omega}|)$ dependences of
    $J_{X}(=J_{Y})$, $J_{Z}$, $K_{X}(=K_{Y})$, $K_{Z}$,
    $\Gamma_{X}(=\Gamma_{Y})$, and $\Gamma_{Z}$ 
    with $\bdE_{\textrm{linear-}a}(t)$
    in the first case of our model.
    The value of $\omega$ is $3$ eV in (a){--}(c)
    and $2.2$ eV in (d){--}(f).
    As well as the first case with $\bdE_{\textrm{linear-}b}(t)$ (Fig. \ref{fig4}),
    the bond anisotropy of the exchange interactions is induced by
    linearly polarized light;
    this contrasts with the case with circularly polarized light (Fig. \ref{fig2}).
    The exchange interactions for the $Z$ bonds are independent of $|u_{ij}|$
    because of Eq. (\ref{eq:uz}).
  }
\end{figure*}

\begin{figure*}
  \includegraphics[width=170mm]{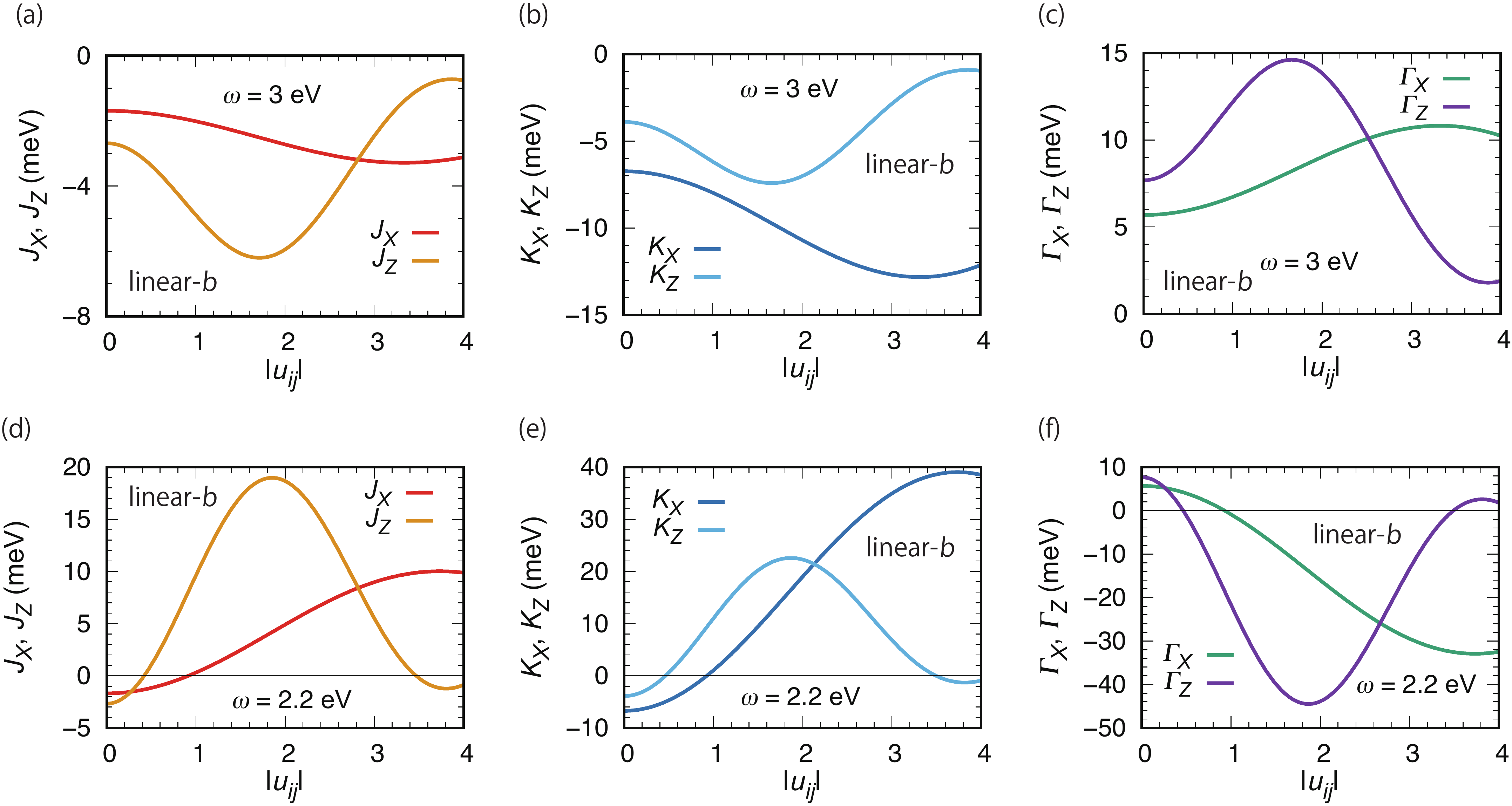}
  \caption{\label{fig6}
    The $|u_{ij}|(=|\frac{eE_{0}}{\omega}|)$ dependences of
    $J_{X}(=J_{Y})$, $J_{Z}$, $K_{X}(=K_{Y})$, $K_{Z}$,
    $\Gamma_{X}(=\Gamma_{Y})$, and $\Gamma_{Z}$ 
    with $\bdE_{\textrm{linear-}b}(t)$
    in the second case of our model.
    The value of $\omega$ is $3$ eV in (a){--}(c)
    and $2.2$ eV in (d){--}(f). 
    The bond-anisotropic $|u_{ij}|$ dependences in this case are similar to
    those obtained in the first case with $\bdE_{\textrm{linear-}b}(t)$ (Fig. \ref{fig4}).
  }
\end{figure*}

\begin{figure*}
  \includegraphics[width=170mm]{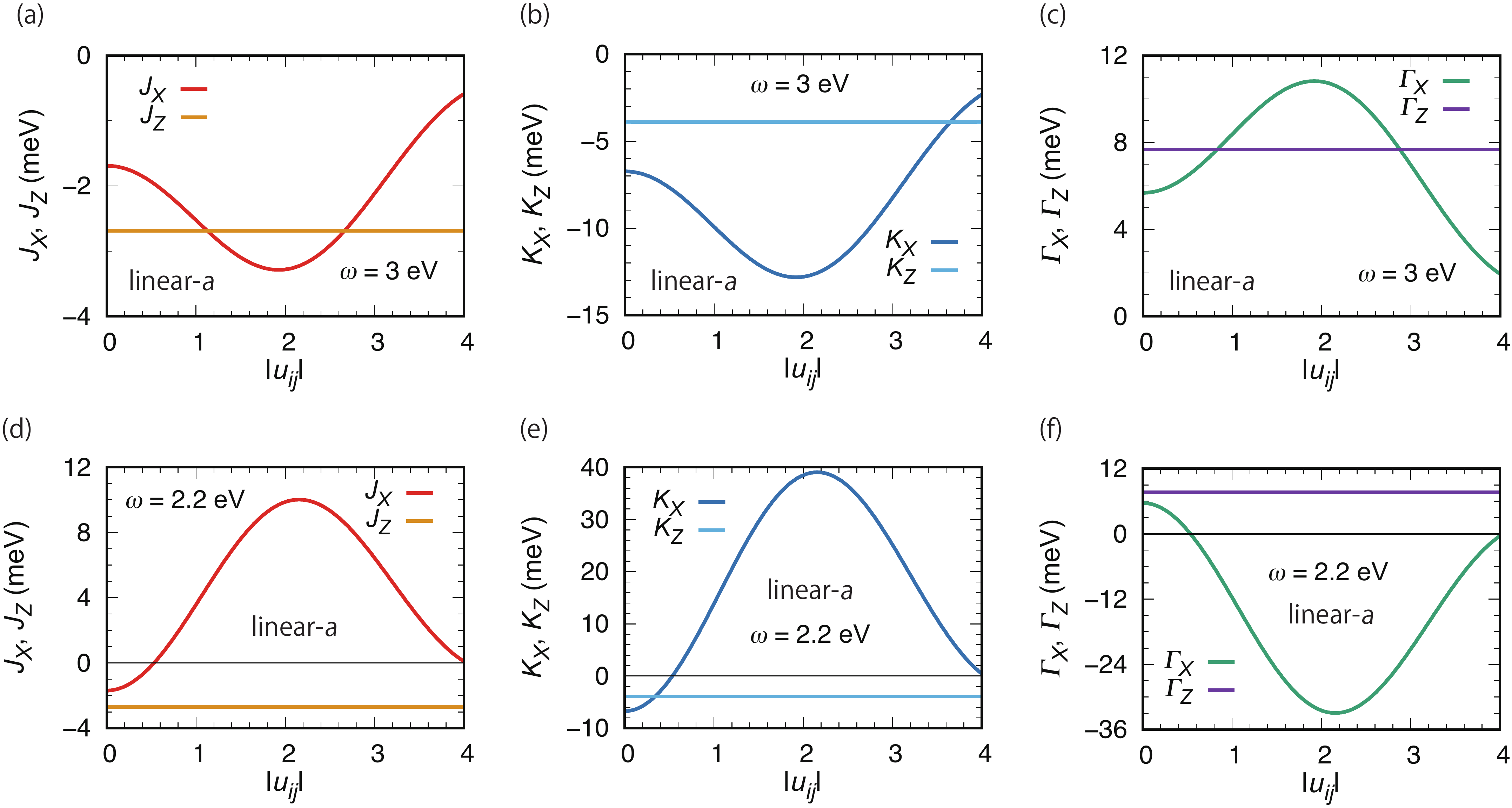}
  \caption{\label{fig7}
    The $|u_{ij}|(=|\frac{eE_{0}}{\omega}|)$ dependences of
    $J_{X}(=J_{Y})$, $J_{Z}$, $K_{X}(=K_{Y})$, $K_{Z}$,
    $\Gamma_{X}(=\Gamma_{Y})$, and $\Gamma_{Z}$ 
    with $\bdE_{\textrm{linear-}a}(t)$
    in the second case of our model.
    The value of $\omega$ is $3$ eV in (a){--}(c)
    and $2.2$ eV in (d){--}(f).
    As well as the case with $\bdE_{\textrm{linear-}b}(t)$,
    the bond-anisotropic $|u_{ij}|$ dependences in this case are similar to
    those obtained in the first case with $\bdE_{\textrm{linear-}a}(t)$ (Fig. \ref{fig5}).    
  }
\end{figure*}

Before showing the results with linearly polarized light,
we explain the mechanism of
the magnitude or sign changes in the exchange interactions.
Here we focus on the $|u_{ij}|$ dependences of $J$
in the first case at $\omega=3$ and $2.2$ eV.
This is enough in understanding the magnitude or sign changes shown above 
because of the following three facts: 
$J$, $K$ and $\Gamma$ have the similar $|u_{ij}|$ dependences;
the results at $\omega=1.8$ eV are essentially the same as those at $\omega=2.2$ eV; 
and the $|u_{ij}|$ dependences of the exchange interactions in the second case
are similar to those in the first case.
As we explain below,
the magnitude changes and the difference in the sign changes
can be understood by considering
the leading terms of Eq. (\ref{eq:Jz}), the $n=0$ and the $n=1$ terms.
At $\omega=3$ eV in the first case
we can express the leading terms of $J(=J_{Z})$ as follows:
\begin{align}
  J&\approx
  (J_{1}+J_{2}+J_{3})\mathcal{J}_{0}(u_{ij})^{2}
  +(c_{1}^{\prime}J_{1}-c_{2}^{\prime}J_{2})\mathcal{J}_{1}(u_{ij})^{2}\notag\\
  &\approx
  (J_{2}+J_{3})\mathcal{J}_{0}(u_{ij})^{2}
  -c_{2}^{\prime}J_{2}\mathcal{J}_{1}(u_{ij})^{2},\label{eq:J-w3}
\end{align}
where
$J_{1}=\frac{4(2t_{1}+t_{3})^{2}}{27(U+2J_{\textrm{H}})}$,
$J_{2}=\frac{8(t_{1}-t_{3})^{2}}{27(U-J_{\textrm{H}})}$,
$J_{3}=\frac{8t_{1}(t_{1}+2t_{3})}{9(U-3J_{\textrm{H}})}$,
$c_{1}^{\prime}=\frac{U+2J_{\textrm{H}}}{\delta\omega_{1}^{\prime}}$,
$c_{2}^{\prime}=\frac{U-J_{\textrm{H}}}{\delta\omega_{2}^{\prime}}$,
and $\omega=U-J_{\textrm{H}}+\delta\omega_{2}^{\prime}=U+2J_{\textrm{H}}-\delta\omega_{1}^{\prime}$
(i.e., $\delta\omega_{2}^{\prime}=0.5$ eV and $\delta\omega_{1}^{\prime}=1$ eV at $\omega=3$ eV).
In deriving the second line of Eq. (\ref{eq:J-w3})
we have used $J_{1}\ll J_{2}, |J_{3}|$, which is satisfied in $\alpha$-RuCl$_{3}$.
Because of this property,
the $|u_{ij}|$ dependence of $J$ is similar to those of $K$ and $\Gamma$,
as described in Ref. \onlinecite{Floquet-NA}. 
Since $J_{2}>0$, $J_{3}<0$, and $J_{2}+J_{3}<0$ are also satisfied,
Eq. (\ref{eq:J-w3}) shows that
$J$ is always negative, i.e., its sign remains unchanged,
although its magnitude is changed due to the Bessel functions.
Then, at $\omega=2.2$ eV in the first case 
the leading terms of Eq. (\ref{eq:Jz}) are given by
\begin{align}
  J&\approx
  (J_{2}+J_{3})\mathcal{J}_{0}(u_{ij})^{2}
  +(c_{2}J_{2}-c_{3}J_{3})\mathcal{J}_{1}(u_{ij})^{2},\label{eq:J-w22}
\end{align}
where $c_{2}=\frac{U-J_{\textrm{H}}}{\delta\omega_{2}}$, 
$c_{3}=\frac{U-3J_{\textrm{H}}}{\delta\omega_{3}}$,
and $\omega=U-3J_{\textrm{H}}+\delta\omega_{3}=U-J_{\textrm{H}}-\delta\omega_{2}$
(i.e., $\delta\omega_{3}=0.7$ eV and $\delta\omega_{2}=0.3$ eV at $\omega=2.2$ eV).
[We have used $J_{1}\ll J_{2}, |J_{3}|$ again in the derivation of Eq. (\ref{eq:J-w22}).] 
In contrast to Eq. (\ref{eq:J-w3}),
the term including $\mathcal{J}_{1}(u_{ij})^{2}$ in Eq. (\ref{eq:J-w22}) 
gives the positive-sign contribution.
Thus, we can see from Eq. (\ref{eq:J-w22}) that
the competition between 
the negative-sign term including $\mathcal{J}_{0}(u_{ij})^{2}$ 
and the positive-sign term including $\mathcal{J}_{1}(u_{ij})^{2}$
is the origin of the sign changes at $|u_{ij}|\sim 0.4$, $3.5$ in Fig. \ref{fig2}(b);
and that the magnitude changes come from the Bessel functions.
Since the leading terms at $\omega=1.8$ eV is also written in the form of Eq. (\ref{eq:J-w22}),
we can similarly understand the magnitude and sign changes at $\omega=1.8$ eV.
Note that a similar argument is applicable to
Eqs. (\ref{eq:Kz}), (\ref{eq:Gamz}), (\ref{eq:Jx}){--}(\ref{eq:Gamx}),
and (\ref{eq:Jx-ani}){--}(\ref{eq:Gamx-ani}).

\begin{figure}
  \includegraphics[width=86mm]{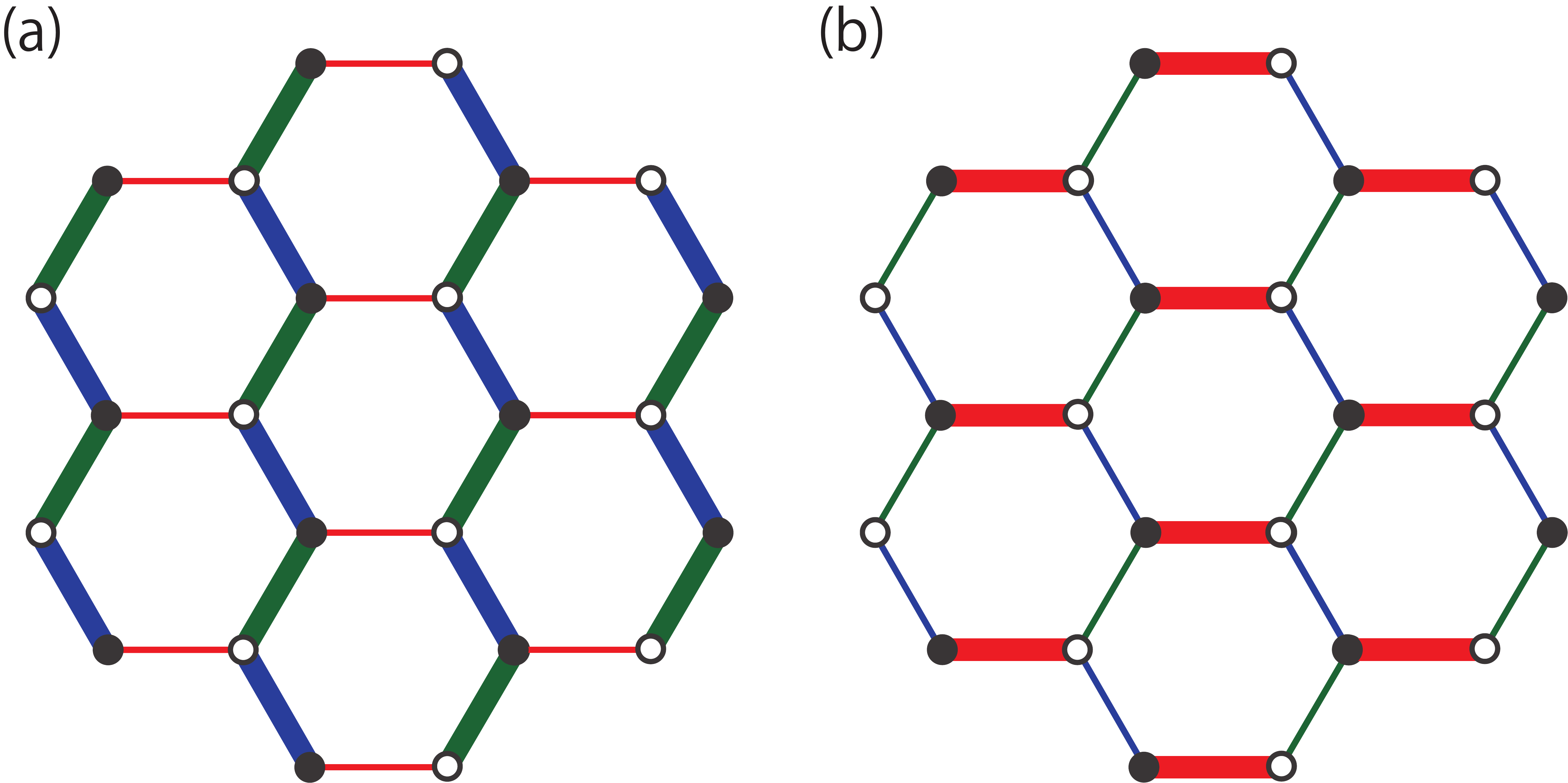}
  \caption{\label{fig8}
    Weakly coupled (a) zigzag and (b) step spin chains
    with $\bdE_{\textrm{linear-}b}$ and $\bdE_{\textrm{linear-}a}$, respectively.
    The definitions of blue, green, and red lines
    and black and white circles are the same as those in Fig. \ref{fig1}.
    Thicker bonds represent the bonds with the exchange interactions
    which are larger in magnitude.
  }
\end{figure}

We turn to the $|u_{ij}|$ dependences of the exchange interactions
with linearly polarized light.
The $|u_{ij}|$ dependences with $\bdE_{\textrm{linear-}b}(t)$
are shown in Figs. \ref{fig4} and \ref{fig6},
and those with $\bdE_{\textrm{linear-}a}(t)$
are shown in Figs. \ref{fig5} and \ref{fig7}.
(Note that the results at $\omega=1.8$ eV are not shown
because they are qualitatively the same as those at $\omega=2.2$ eV.)
Some properties are similar to
those with circularly polarized light (Figs. \ref{fig2} and \ref{fig3}):
the sign changes in the exchange interactions are absent at $\omega=3$ eV
and present at $\omega=2.2$ eV;
and the results obtained in the first case remain qualitatively unchanged
even in the second case.
The magnitude and sign changes in the exchange interactions can be understood
in a way similar to those with circularly polarized light.
We also see from Figs. \ref{fig4}{--}\ref{fig7} that 
linearly polarized light can change 
the ratios $J_{X}/J_{Z}$, $K_{X}/K_{Z}$, and $\Gamma_{X}/\Gamma_{Z}$
even without the bond anisotropy of the hopping integrals.
[As explained below Eq. (\ref{eq:ux}),
this property results from
the difference between $u_{ij}^{Z}$ and $u_{ij}^{X}$.]
Because of this property,
those ratios can have values which cannot be realized in non-driven $\alpha$-RuCl$_{3}$:
although 
$|J_{X}|<|J_{Z}|$, $|K_{X}|>|K_{Z}|$, and $\Gamma_{X}<\Gamma_{Z}$
hold in non-driven $\alpha$-RuCl$_{3}$,
$|J_{X}|>|J_{Z}|$, $|K_{X}|<|K_{Z}|$, and $\Gamma_{X}>\Gamma_{Z}$ are possible
in $\alpha$-RuCl$_{3}$ driven by a field of linearly polarized light.  
In addition,
it is possible to change
the signs of the exchange interactions
only for the $Z$ bonds or only for the $X$ and $Y$ bonds;
for example, at $|u_{ij}|\sim 0.5$ in Figs. \ref{fig4}(d){--}\ref{fig4}(f)
we can change the signs for the $Z$ bonds without changing those for the $X$ and $Y$ bonds.
Then,
the honeycomb-network spin system 
could be transformed either into
weakly coupled zigzag spin chains
in the case of $\bdE_{\textrm{linear-}b}(t)$
for $|u_{ij}|\approx 0.4{-}0.42$ at $\omega=2.2$ eV
[e.g., see Figs. \ref{fig4}(d){--}\ref{fig4}(f) and \ref{fig6}(d){--}\ref{fig6}(f)]
or into weakly coupled step spin chains
in the case of $\bdE_{\textrm{linear-}a}(t)$
for $|u_{ij}|\approx 0.48{-}0.5$ at $\omega=2.2$ eV
[e.g., see Figs. \ref{fig5}(d){--}\ref{fig5}(f) and \ref{fig7}(d){--}\ref{fig7}(f)].
In the weakly coupled zigzag spin chains [Fig. \ref{fig8}(a)]
$J_{X}=J_{Y}$, $K_{X}=K_{Y}$, and $\Gamma_{X}=\Gamma_{Y}$ are dominant
and $J_{Z}$, $K_{Z}$, and $\Gamma_{Z}$ give
the weak coupling between zigzag chains;
in the weakly coupled step spin chains [Fig. \ref{fig8}(b)]
$J_{Z}$, $K_{Z}$, and $\Gamma_{Z}$ are dominant
and $J_{X}=J_{Y}$, $K_{X}=K_{Y}$, and $\Gamma_{X}=\Gamma_{Y}$
give the weak coupling between step chains.
Note that
in the first case of our model
with $\bdE_{\textrm{linear-}b}(t)$ for $|u_{ij}|\approx 0.42$ at $\omega=2.2$ eV
$J_{X}/J_{Z}\sim 9.8$, $K_{X}/K_{Z}\sim 4.6$, and $\Gamma_{X}/\Gamma_{Z}\sim 4.6$;
and that
in the first case with $\bdE_{\textrm{linear-}a}(t)$ for $|u_{ij}|\approx 0.48$ at $\omega=2.2$ eV
$J_{X}/J_{Z}\sim 0.1$, $K_{X}/K_{Z}\sim 0.2$, and $\Gamma_{X}/\Gamma_{Z}\sim 0.2$.

\section{Magnetic states}

In this section
we study several magnetic states in periodically driven $\alpha$-RuCl$_{3}$.
In Sec. IV A
we evaluate the expectation value of our Floquet Hamiltonian within the MFA
and express it in a quadratic form.  
Then, we explain
how to obtain the energies and spin configurations of magnetic states.  
We also remark on the magnetic states considered in our analyses.
In Sec. IV B
we present the $|u_{ij}|$ dependences of the energies of the magnetic states
for some non-resonant $\omega$'s 
and discuss the effects of one of the light fields
and the differences due to the polarization of light.
(The reason why we use non-resonant $\omega$'s has been explained in Sec. III.)

\subsection{Theory}

Applying the MFA to our Floquet Hamiltonian, 
we derive an expression of its expectation value in a quadratic form. 
Since the MFA for Mott insulators with strong SOC has been explained,
for example, in Ref. \onlinecite{NA-Jeff},
we explain the main points here.
By using the MFA,
we can write the expectation value of Eq. (\ref{eq:Heff_1st}) as
\begin{align}
  \langle \bar{H}_{\textrm{eff}}\rangle
  =\sum_{\langle i,j\rangle}\sum_{\mu,\nu=x,y,z}
  M_{\mu\nu}^{ij}\langle S_{i}^{\mu}\rangle\langle S_{j}^{\nu}\rangle,\label{eq:H-MFA-start}
\end{align}
where $M_{\mu\nu}^{ij}$ is given in the first or second case of our model by
\begin{align}
  M_{\mu\nu}^{ij}=
    \begin{cases}
    \ J_{\delta}+K_{\delta} \ \ (\mu=\nu=\gamma),\\
    \ J_{\delta} \ \ \ \ \ \ \ \ \ (\mu=\nu=\alpha\ \textrm{or}\ \beta),\\
    \ \Gamma_{\delta} \ \ \ \ \ \ \ \ \ (\mu=\alpha, \nu=\beta),\\
    \ \Gamma_{\delta} \ \ \ \ \ \ \ \ \ (\mu=\beta, \nu=\alpha),\\
    \ 0 \ \ \ \ \ \ \ \ \ \ \ (\textrm{otherwise}).
  \end{cases}
\end{align}
[Note that $\delta$, $\gamma$, $\alpha$, and $\beta$
have been defined in Eq. (\ref{eq:alp-bet-gam}).] 
In the third, fourth, or fifth case of our model
the contribution from Eq. (\ref{eq:H-J3rd}) is added to Eq. (\ref{eq:H-MFA-start}).
In the MFA
the expectation value of spin operators should satisfy the hard-spin constraints:
\begin{align}
  |\langle\bdS_{i}\rangle|^{2}=S^{2},\
  |\langle\bdS_{j}\rangle|^{2}=S^{2},\label{eq:local-constraints} 
\end{align}
where $i$ and $j$ belong to
the $A$ and the $B$ sublattices (Fig. \ref{fig1}), respectively,
and $S$ is $1/2$ in the case of $\alpha$-RuCl$_{3}$.
Since
$\langle S_{i}^{\mu}\rangle$ and $\langle S_{j}^{\nu}\rangle$ are expressed as 
\begin{align}
  \langle S_{i}^{\mu}\rangle
  &=\sqrt{\frac{2}{N}}\sum_{\bdq}\langle S_{\bdq A}^{\mu}\rangle e^{i\bdq\cdot\bdR_{i}},\label{eq:Sq_A}\\
  \langle S_{j}^{\nu}\rangle
  &=\sqrt{\frac{2}{N}}\sum_{\bdq}\langle S_{\bdq B}^{\nu}\rangle e^{i\bdq\cdot\bdR_{j}},\label{eq:Sq_B}
\end{align}
where $N$ is the total number of sites,
we can express Eq. (\ref{eq:H-MFA-start}) in the following quadratic form:
\begin{align}
  \langle \bar{H}_{\textrm{eff}}\rangle
  =&\sum_{\bdq}\sum_{\mu,\nu=x,y,z}\sum_{l,l^{\prime}=A,B}
  \langle S_{\bdq l}^{\mu}\rangle^{\ast}
  [M_{ll^{\prime}}(\bdq)]_{\mu\nu}
  \langle S_{\bdq l^{\prime}}^{\nu}\rangle\notag\\
  =&\sum_{\bdq}\sum_{\mu,\nu=x,y,z}
  \langle S_{-\bdq A}^{\mu}\rangle[M(\bdq)]_{\mu\nu}\langle S_{\bdq B}^{\nu}\rangle\notag\\
  &+\sum_{\bdq}\sum_{\mu,\nu=x,y,z}
  \langle S_{-\bdq B}^{\mu}\rangle[M(\bdq)^{\dagger}]_{\mu\nu}\langle S_{\bdq A}^{\nu}\rangle,
  \label{eq:H-MFA-2nd}
\end{align}
where the $\mu\times\nu$ matrix $M(\bdq)$ is given by the matrix 
\begin{align}
 \left(
  \begin{array}{@{\,}ccc@{\,}}
    J(\bdq)+K_{x}(\bdq) & \Gamma_{z}(\bdq) & \Gamma_{y}(\bdq)\\[3pt]
    \Gamma_{z}(\bdq) & J(\bdq)+K_{y}(\bdq) & \Gamma_{x}(\bdq)\\[3pt]
    \Gamma_{y}(\bdq) & \Gamma_{x}(\bdq) & J(\bdq)+K_{z}(\bdq)
  \end{array}
  \right),\label{eq:Mq}
\end{align}
and $J(\bdq)$, $K_{\mu}(\bdq)$'s, and $\Gamma_{\mu}(\bdq)$'s are defined as
\begin{align}
  &J(\bdq)=
  \frac{J_{X}}{2}e^{-i\frac{q_{x}}{2}+i\frac{\sqrt{3}}{2}q_{y}}
  +\frac{J_{Y}}{2}e^{-i\frac{q_{x}}{2}-i\frac{\sqrt{3}}{2}q_{y}}
  +\frac{J_{Z}}{2}e^{iq_{x}}\notag\\
  &\ \
  +\frac{J_{X}^{\textrm{3rd}}}{2}e^{iq_{x}-i\sqrt{3}q_{y}}
  +\frac{J_{Y}^{\textrm{3rd}}}{2}e^{iq_{x}+i\sqrt{3}q_{y}}
  +\frac{J_{Z}^{\textrm{3rd}}}{2}e^{-2iq_{x}},\label{eq:Jq}\\
  &K_{x}(\bdq)=\frac{K_{X}}{2}e^{-i\frac{q_{x}}{2}+i\frac{\sqrt{3}}{2}q_{y}},\label{eq:Kxq}\\
  &K_{y}(\bdq)=\frac{K_{Y}}{2}e^{-i\frac{q_{x}}{2}-i\frac{\sqrt{3}}{2}q_{y}},\label{eq:Kyq}\\
  &K_{z}(\bdq)=\frac{K_{Z}}{2}e^{iq_{x}},\label{eq:Kzq}\\
  &\Gamma_{x}(\bdq)=\frac{\Gamma_{X}}{2}e^{-i\frac{q_{x}}{2}+i\frac{\sqrt{3}}{2}q_{y}},\label{eq:GamXq}\\
  &\Gamma_{y}(\bdq)=\frac{\Gamma_{Y}}{2}e^{-i\frac{q_{x}}{2}-i\frac{\sqrt{3}}{2}q_{y}},\label{eq:GamYq}\\
  &\Gamma_{z}(\bdq)=\frac{\Gamma_{Z}}{2}e^{iq_{x}}.\label{eq:GamZq}
\end{align}
The details of the derivation of Eq. (\ref{eq:H-MFA-2nd}) are described in Appendix E.
Furthermore,
we make some remarks about momentum in the case of the honeycomb lattice
in Appendix F. 
Note that the MFA can reproduce
the phase diagram obtained
in the Luttinger-Tizsa method~\cite{Luttinger-Tizsa1,Luttinger-Tizsa2}
[i.e., Fig. 2(a) of Ref. \onlinecite{Rau-PRL}
except the region surrounded by the dashed white line~\cite{remark-Rau}].

\begin{figure*}
  \includegraphics[width=170mm]{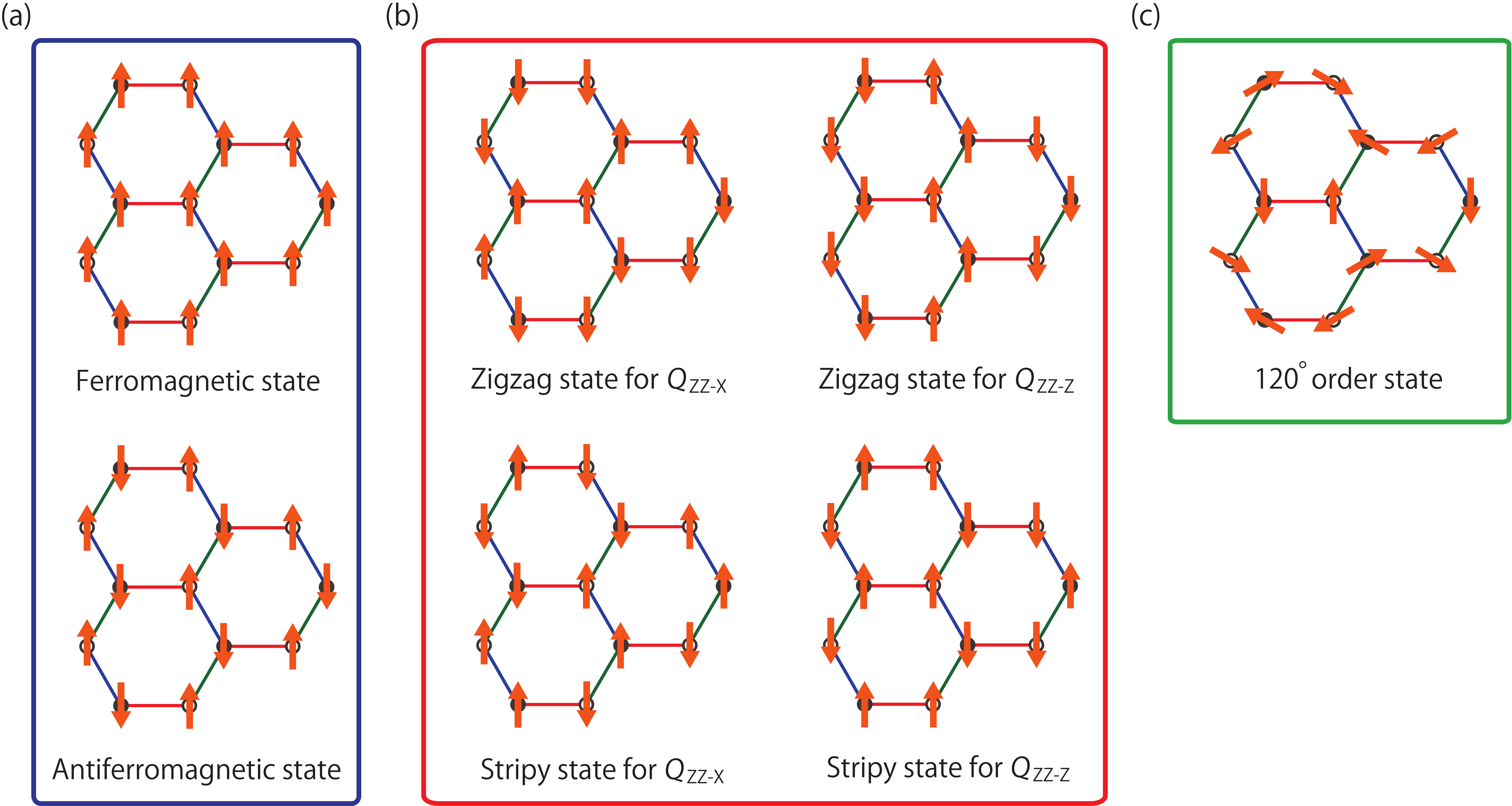}
  \caption{\label{fig9}
    Spin configurations of the magnetic states considered in our analyses:
    (a) the magnetic states with $\bdQ=\bdzero$,
    the ferromagnetic state and the antiferromagnetic state,
    (b) the magnetic states
    with $\bdQ=\bdQ_{\textrm{ZZ-}X}(=\bdK_{01}/2)$
    or $\bdQ=\bdQ_{\textrm{ZZ-}Z}(=\bdK_{10}/2)$, 
    the zigzag states and the stripy states,
    and (c) the magnetic state with $\bdQ=\bdQ_{120}$, the $120^{\circ}$ order state.
    Specific directions of spins, which are represented by the arrows,
    are chosen 
    because the relative angles between the neighboring spins are essential
    for understanding the differences in the spin configurations. 
  }
\end{figure*}

By using Eqs. (\ref{eq:local-constraints}){--}(\ref{eq:GamZq}),
we can obtain the energies and spin configurations of magnetic states. 
Since Eq. (\ref{eq:H-MFA-2nd}) is quadratic in spin variables,
we obtain six eigenvalues and the corresponding eigenvectors for each $\bdq$
by diagonalizing the matrix
\begin{align}
 \left(
  \begin{array}{@{\,}cc@{\,}}
    0 & M(\bdq)\\[3pt]
    M(\bdq)^{\dagger} & 0
  \end{array}
  \right),\label{eq:Matrix-diag}
\end{align}
where $M(\bdq)$ has been defined in Eq. (\ref{eq:Mq})
and $0$ represents the zero matrix
(i.e., the $\mu\times\nu$ matrix of which the components are all zero).
At a certain $\bdq$, for example $\bdq=\bdQ$, 
the eigenvalue which is the smallest of the six ones
gives the energy of a magnetic state characterized by the ordering vector $\bdQ$.
Then,
its spin configuration can be obtained by combining  
the corresponding eigenvectors
and Eqs. (\ref{eq:local-constraints}){--}(\ref{eq:Sq_B});
the obtained spin configuration is valid
only if it is consistent with Eq. (\ref{eq:local-constraints}). 

In the analyses of Sec. IV B
we consider several magnetic states,
which can be classified into three groups.
The first group consists of the magnetic states with $\bdQ=\bdzero$,
which include a ferromagnetic state and
an antiferromagnetic state [Fig. \ref{fig9}(a)].
Note that
the $\bdQ$ of the antiferromagnetic state 
becomes $\bdzero$ in the presence of a sublattice structure
(because of it, the spins on a sublattice
are all ferromagnetic, i.e., parallel).
The second group consists of the magnetic states with $\bdQ=\bdK/2$,
where $\bdK$ is the reciprocal lattice vector [Eq. (\ref{eq:K})];
they include the zigzag states and the stripy states [Fig. \ref{fig9}(b)].
If we see a unit consisting of
one site on the $A$ or $B$ sublattice and its three neighbors,
in the zigzag states three spins are ferromagnetic (parallel)
and the other is antiferromagnetic (antiparallel);
in the stripy states
two spins are ferromagnetic and the others are antiferromagnetic [Fig. \ref{fig9}(b)].
In our analyses
we consider two kinds of $\bdQ=\bdK/2$,
i.e., one is $\bdQ_{\textrm{ZZ-}X}=\bdK_{01}/2$ and
the other is $\bdQ_{\textrm{ZZ-}Z}=\bdK_{10}/2$
(for the details of $\bdK_{01}$ and $\bdK_{10}$ see Appendix F). 
In the zigzag state 
with $\bdQ=\bdQ_{\textrm{ZZ-}X}$ or $\bdQ=\bdQ_{\textrm{ZZ-}Z}$
the spins on the $X$ or the $Z$ bonds, respectively, are antiferromagnetic;
in the stripy state with $\bdQ=\bdQ_{\textrm{ZZ-}X}$ or $\bdQ=\bdQ_{\textrm{ZZ-}Z}$
those are ferromagnetic.
This difference between
the zigzag (or stripy) states with $\bdQ=\bdQ_{\textrm{ZZ-}X}$ and $\bdQ_{\textrm{ZZ-}Z}$
is partly due to the momentum dependences of $K_{\mu}(\bdq)$'s.
Namely, since 
Eqs. (\ref{eq:Kxq}){--}(\ref{eq:Kzq}) 
show
\begin{align}
  &K_{x}(\bdQ_{\textrm{ZZ-}X})=\frac{K_{X}}{2}e^{i\pi}e^{-i\pi/3}
  =\frac{-K_{X}}{2}e^{-i\pi/3},\label{eq:Kx-QZZx}\\
  &K_{y}(\bdQ_{\textrm{ZZ-}X})=\frac{K_{Y}}{2}e^{-i\pi/3},\label{eq:Ky-QZZx}\\
  &K_{z}(\bdQ_{\textrm{ZZ-}X})=\frac{K_{Z}}{2}e^{-i\pi/3},\label{eq:Kz-QZZx}
\end{align}
and
\begin{align}
  &K_{x}(\bdQ_{\textrm{ZZ-}Z})=\frac{K_{X}}{2}e^{-i\pi/3},\\
  &K_{y}(\bdQ_{\textrm{ZZ-}Z})=\frac{K_{Y}}{2}e^{-i\pi/3},\\
  &K_{z}(\bdQ_{\textrm{ZZ-}Z})=\frac{K_{Z}}{2}e^{i\pi}e^{-i\pi/3}
  =\frac{-K_{Z}}{2}e^{-i\pi/3},\label{eq:Kz-QZZz}
\end{align}
the effective Kitaev interaction of the $X$ bonds for $\bdQ=\bdQ_{\textrm{ZZ-}X}$
or of the $Z$ bonds for $\bdQ=\bdQ_{\textrm{ZZ-}Z}$
has the opposite sign 
and, as a result,
the spins on the $X$ bonds for $\bdQ=\bdQ_{\textrm{ZZ-}X}$
or the $Z$ bonds for $\bdQ=\bdQ_{\textrm{ZZ-}Z}$
are aligned in the opposite direction
compared with the spins on the other bonds. 
The third group consists of the magnetic state with $\bdQ=\bdQ_{120}$,
the $120^{\circ}$ order state [Fig. \ref{fig9}(c)].
The magnetic states explained above are realized
in Mott insulators with strong SOC on the honeycomb lattice~\cite{Rau-PRL,PD1}. 

\subsection{Results}

\begin{figure*}
  \includegraphics[width=170mm]{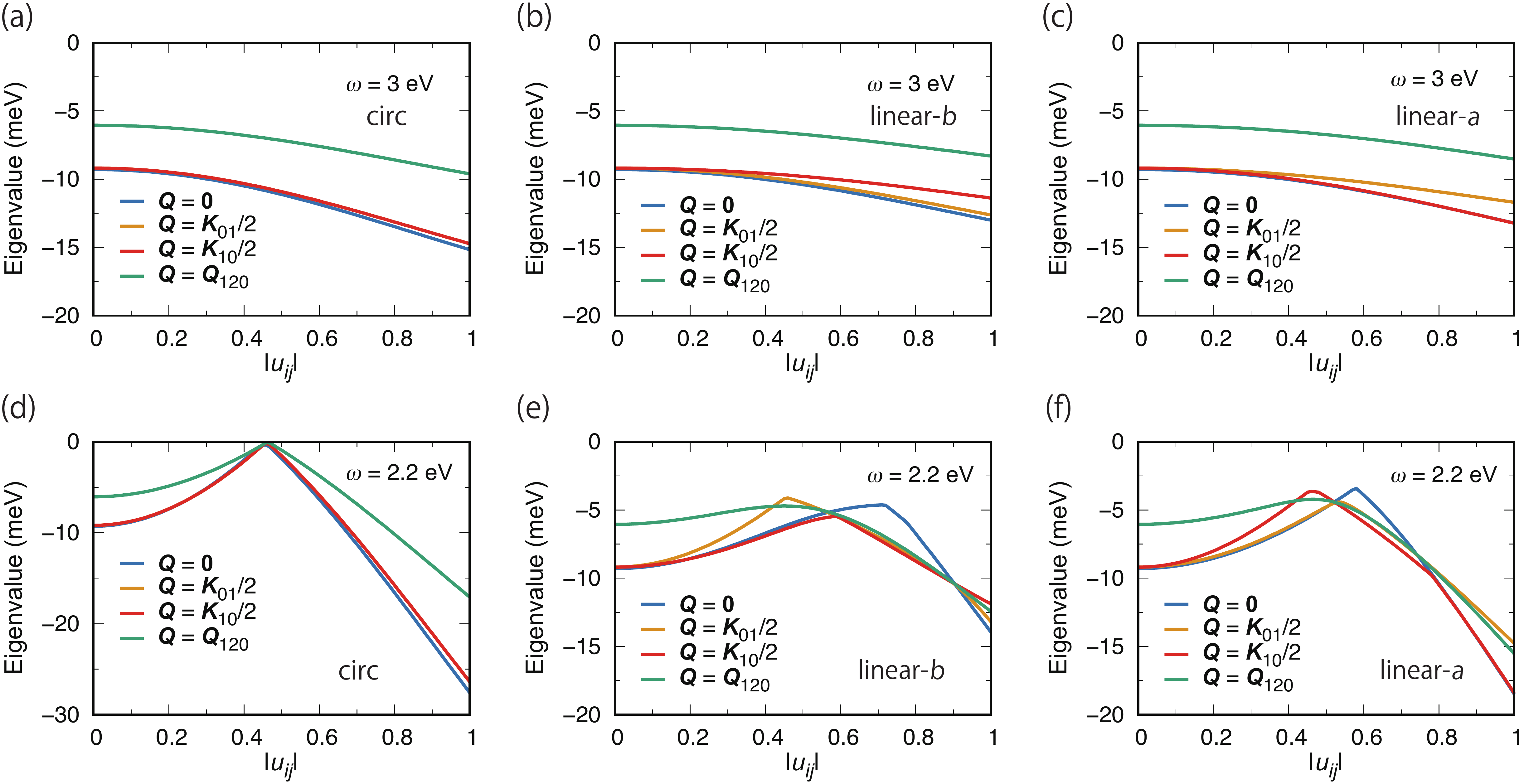}
  \caption{\label{fig10}
    The $|u_{ij}|$ dependences of the energies of the magnetic states
    for $\bdQ=\bdzero$, $\bdQ_{\textrm{ZZ-}X}(=\bdK_{01}/2)$,
    $\bdQ_{\textrm{ZZ-}Z}(=\bdK_{10}/2)$, and $\bdQ_{120}$
    within the MFA
    in the first case of our model with $\bdE_{\textrm{circ}}(t)$ [(a) and (d)],
    $\bdE_{\textrm{linear-}b}(t)$ [(b) and (e)],
    and $\bdE_{\textrm{linear-}a}(t)$ [(c) and (f)].
    The value of $\omega$ is $3$ eV in (a){--}(c)
    and $2.2$ eV in (d){--}(f).
    In this case the bond-averaged nearest-neighbor hopping integrals are considered. 
  }
\end{figure*}

\begin{figure*}
  \includegraphics[width=170mm]{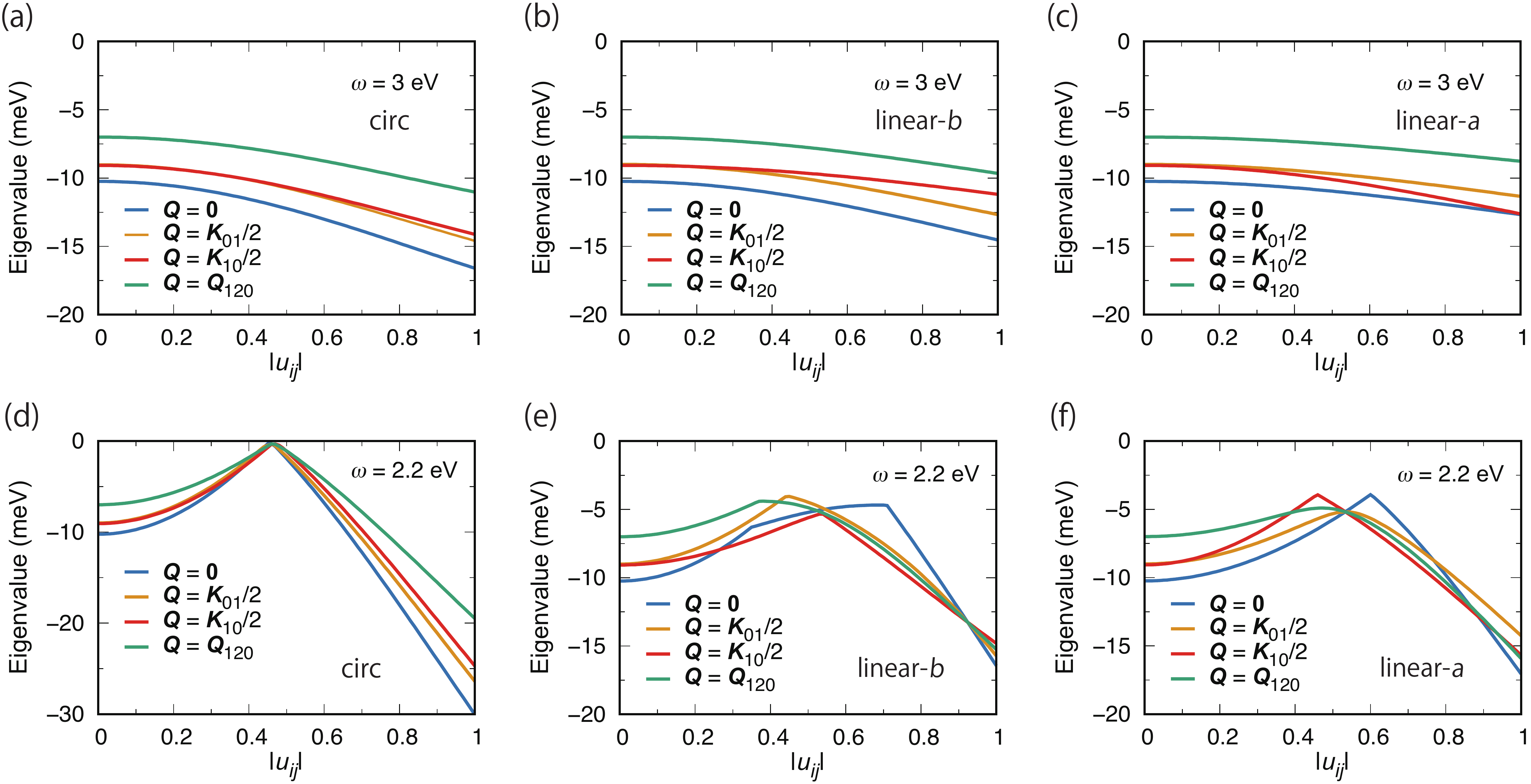}
  \caption{\label{fig11}
    The $|u_{ij}|$ dependences of the energies of the magnetic states
    for $\bdQ=\bdzero$, $\bdQ_{\textrm{ZZ-}X}(=\bdK_{01}/2)$,
    $\bdQ_{\textrm{ZZ-}Z}(=\bdK_{10}/2)$, and $\bdQ_{120}$
    within the MFA
    in the second case of our model with $\bdE_{\textrm{circ}}(t)$ [(a) and (d)],
    $\bdE_{\textrm{linear-}b}(t)$ [(b) and (e)],
    and $\bdE_{\textrm{linear-}a}(t)$ [(c) and (f)].
    The value of $\omega$ is $3$ eV in (a){--}(c)
    and $2.2$ eV in (d){--}(f).
    In this case
    the bond-anisotropic nearest-neighbor hopping integrals are considered.
    In contrast to the first case (Fig. \ref{fig10}),
    the degeneracy of the magnetic states
    for $\bdQ=\bdK_{01}/2$ and $\bdK_{10}/2$
    is lifted even for $\bdE_{\textrm{circ}}(t)$
    due to the bond anisotropy of the nearest-neighbor hopping integrals. 
  }
\end{figure*}

\begin{figure*}
  \includegraphics[width=170mm]{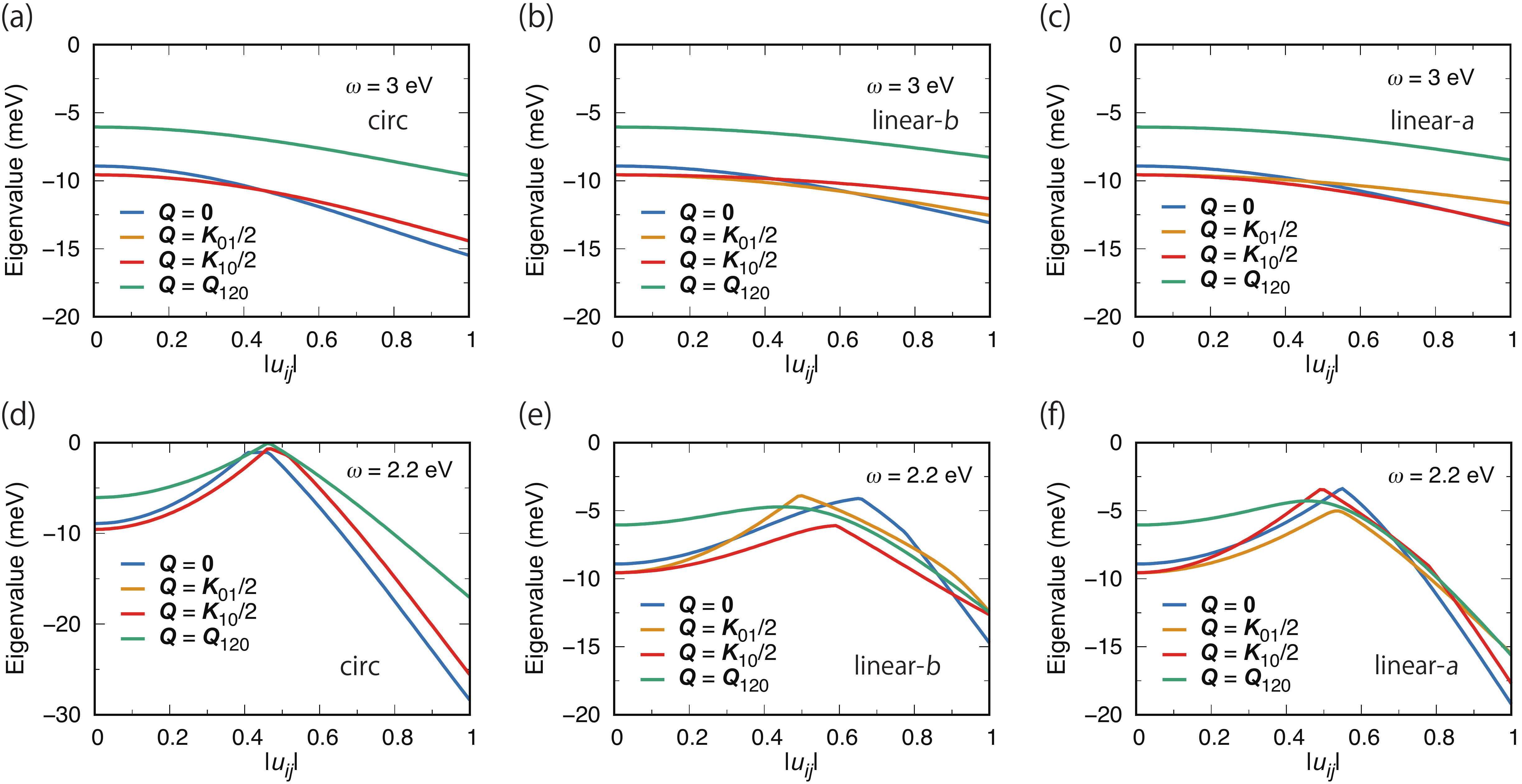}
  \caption{\label{fig12}
    The $|u_{ij}|$ dependences of the energies of the magnetic states
    for $\bdQ=\bdzero$, $\bdQ_{\textrm{ZZ-}X}(=\bdK_{01}/2)$,
    $\bdQ_{\textrm{ZZ-}Z}(=\bdK_{10}/2)$, and $\bdQ_{120}$
    within the MFA
    in the third case of our model with $\bdE_{\textrm{circ}}(t)$ [(a) and (d)],
    $\bdE_{\textrm{linear-}b}(t)$ [(b) and (e)],
    and $\bdE_{\textrm{linear-}a}(t)$ [(c) and (f)].
    The value of $\omega$ is $3$ eV in (a){--}(c)
    and $2.2$ eV in (d){--}(f).
    In this case the bond-averaged nearest-neighbor hopping integrals
    and the third-neighbor one are considered.
    In contrast to the first case (Fig. \ref{fig10}),
    the magnetic state for $\bdQ=\bdK_{01}/2$ or $\bdK_{10}/2$
    has the lower energy than that for $\bdQ=\bdzero$
    owing to finite $J^{\textrm{3rd}}_{\delta}$
    induced by the third-neighbor hopping integral. 
  }
\end{figure*}

\begin{figure}
  \includegraphics[width=86mm]{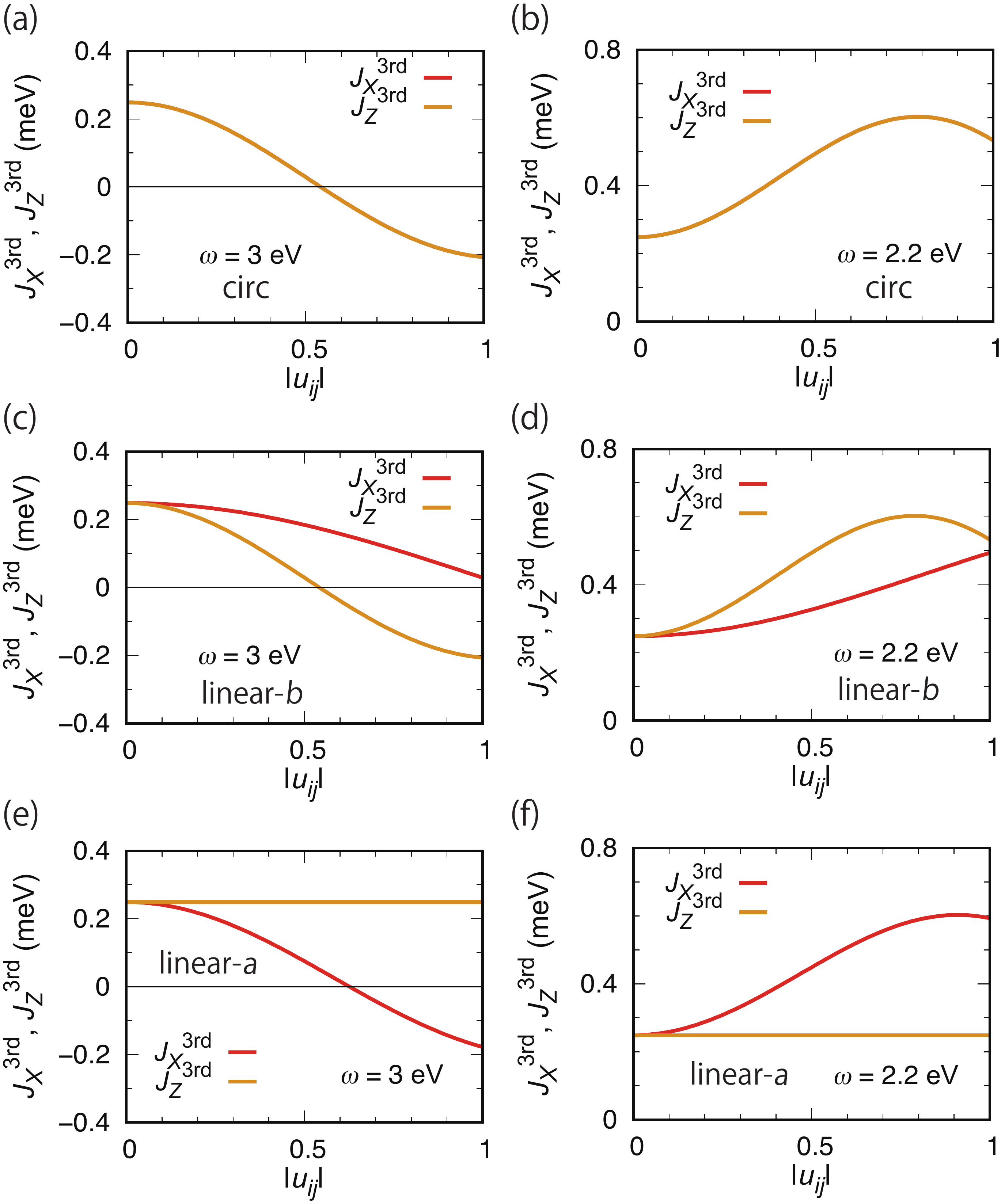}
  \caption{\label{fig13}
    The $|u_{ij}|$ dependences of $J_{\delta}^{\textrm{3rd}}$
    in the third or fourth case of our model,
    in which $t_{\textrm{3rd}}=-40$ meV, 
    with $\bdE_{\textrm{circ}}(t)$ [(a) and (b)],
    $\bdE_{\textrm{linear-}b}(t)$ [(c) and (d)],
    and $\bdE_{\textrm{linear-}a}(t)$ [(e) and (f)].
    The value of $\omega$ is $3$ eV in (a){--}(c)
    and $2.2$ eV in (d){--}(f).   
    In (a) and (b)
    $J_{X}^{\textrm{3rd}}=J_{Y}^{\textrm{3rd}}=J_{Z}^{\textrm{3rd}}=J_{\textrm{3rd}}$.
    The similar $|u_{ij}|$ dependences are obtained in the fifth case. 
  }
\end{figure}

\begin{figure*}
  \includegraphics[width=170mm]{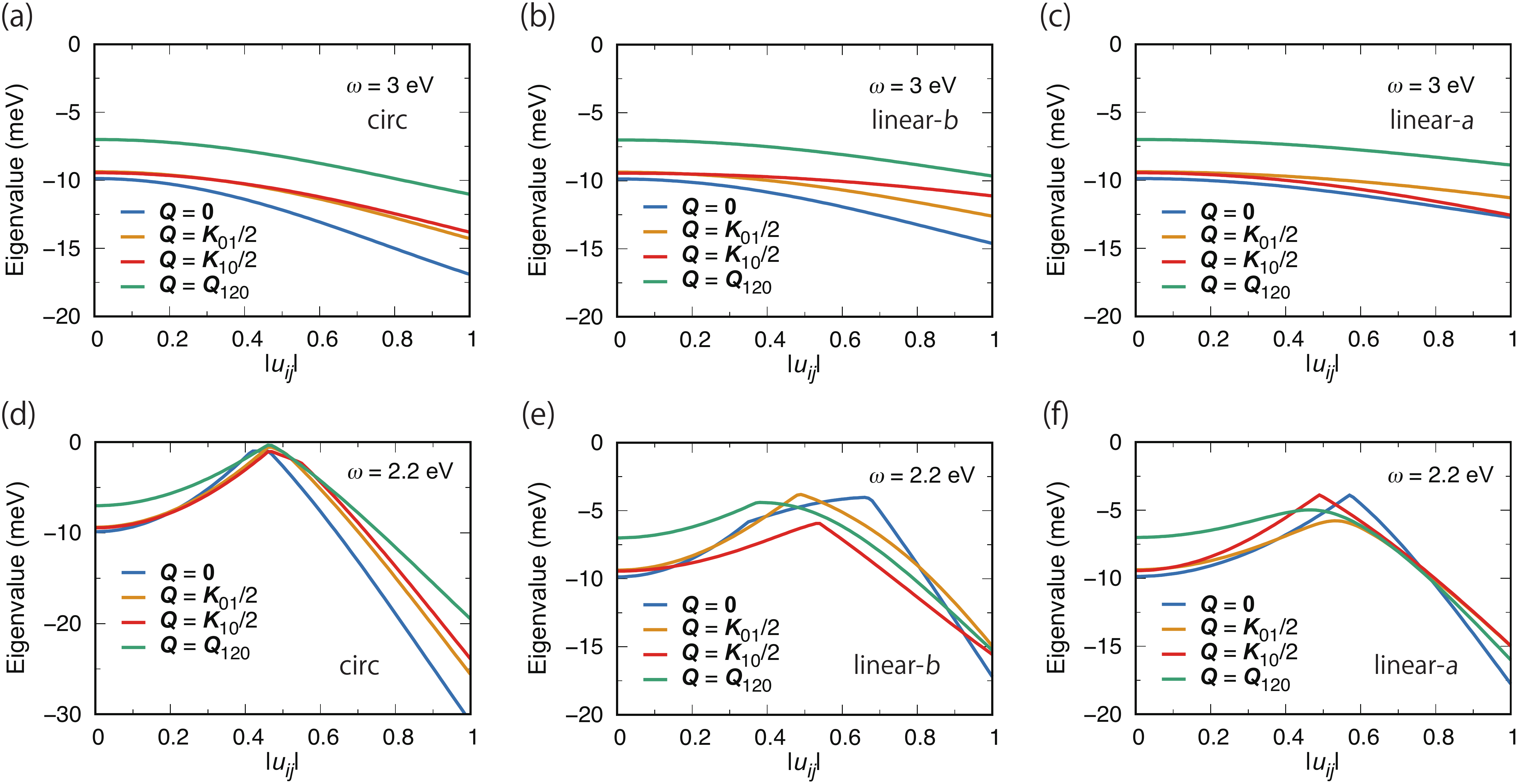}
  \caption{\label{fig14}
    The $|u_{ij}|$ dependences of the energies of the magnetic states
    for $\bdQ=\bdzero$, $\bdQ_{\textrm{ZZ-}X}(=\bdK_{01}/2)$,
    $\bdQ_{\textrm{ZZ-}Z}(=\bdK_{10}/2)$, and $\bdQ_{120}$
    within the MFA
    in the fourth case of our model
    with $\bdE_{\textrm{circ}}(t)$ [(a) and (d)],
    $\bdE_{\textrm{linear-}b}(t)$ [(b) and (e)],
    and $\bdE_{\textrm{linear-}a}(t)$ [(c) and (f)].  
    The value of $\omega$ is $3$ eV in (a){--}(c)
    and $2.2$ eV in (d){--}(f).
    In this case the bond-anisotropic nearest-neighbor hopping integrals
    and the third-neighbor one are considered.
  }
\end{figure*}

\begin{figure*}
  \includegraphics[width=170mm]{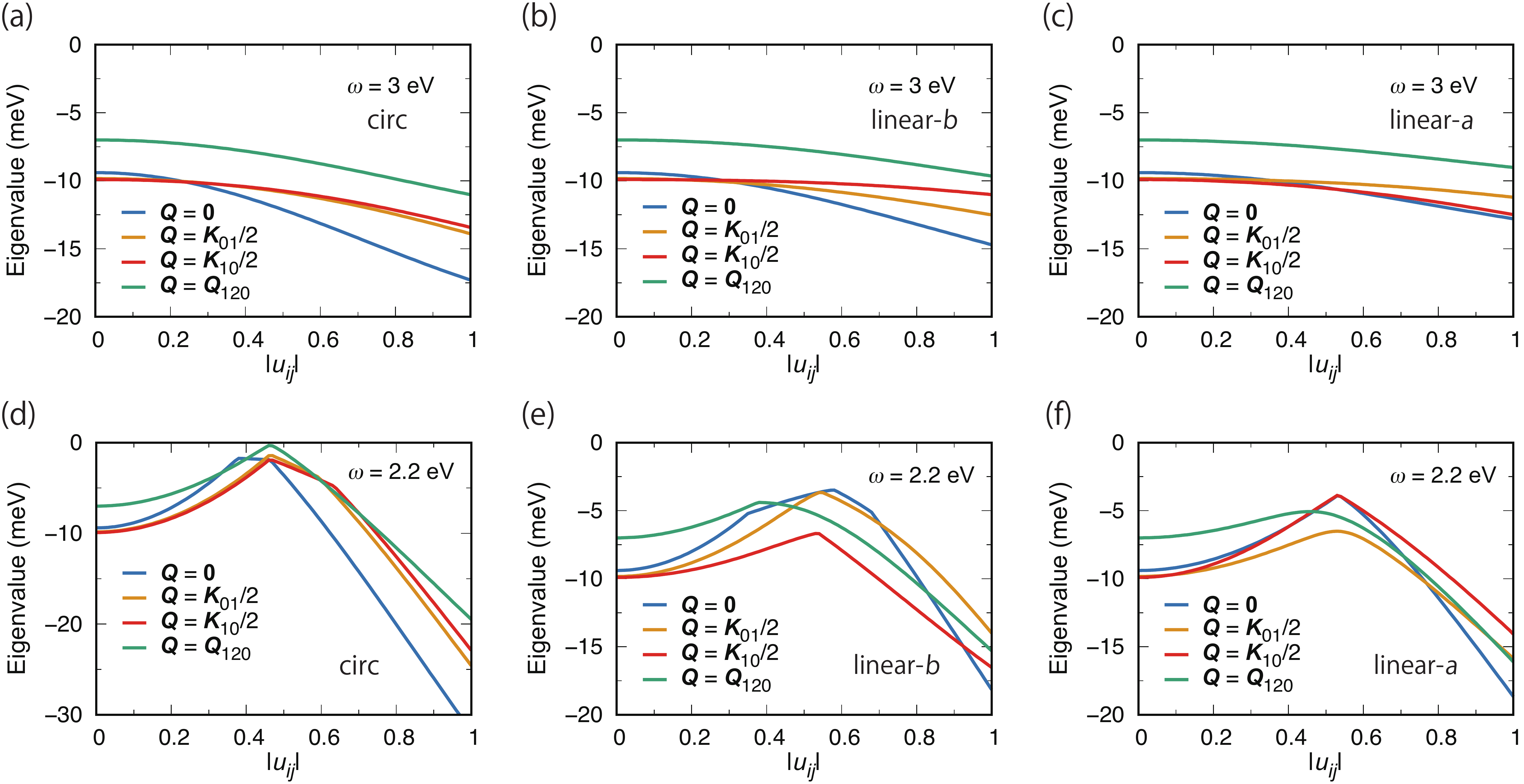}
  \caption{\label{fig15}
    The $|u_{ij}|$ dependences of the energies of the magnetic states
    for $\bdQ=\bdzero$, $\bdQ_{\textrm{ZZ-}X}(=\bdK_{01}/2)$,
    $\bdQ_{\textrm{ZZ-}Z}(=\bdK_{10}/2)$, and $\bdQ_{120}$
    within the MFA
    in the fifth case of our model
    with $\bdE_{\textrm{circ}}(t)$ [(a) and (d)],
    $\bdE_{\textrm{linear-}b}(t)$ [(b) and (e)],
    and $\bdE_{\textrm{linear-}a}(t)$ [(c) and (f)].  
    The value of $\omega$ is $3$ eV in (a){--}(c)
    and $2.2$ eV in (d){--}(f).
    In this case the bond-anisotropic nearest-neighbor hopping integrals
    and the third-neighbor one are considered;
    the difference between the parameters in the fourth and fifth cases is
    in the values of $t_{\textrm{3rd}}$.
  }
\end{figure*}

To study how the light fields affect the magnetic states,
we numerically calculate their energies within the MFA
in the five cases of our model at $\omega=3$ and $2.2$ eV.
(The results at $\omega=1.8$ eV, which are not shown,
are qualitatively the same as those at $\omega=2.2$ eV.)
As explained in Sec. IV A,
the energy of a magnetic state characterized by $\bdQ$
corresponds to the lowest eigenvalue
obtained by diagonalizing Eq. (\ref{eq:Matrix-diag}) at $\bdq=\bdQ$.
Furthermore,
the $\bdQ$'s considered in this study
are $\bdzero$, $\bdQ_{\textrm{ZZ-}X}(=\bdK_{01}/2)$,
$\bdQ_{\textrm{ZZ-}Z}(=\bdK_{10}/2)$, and $\bdQ_{120}$.
In the numerical calculations
we impose the periodic boundary condition
and set $N_{1}=N_{2}=120$ in Eq. (\ref{eq:q}). 
The parameters of our model except $t_{\textrm{3rd}}$ are chosen 
in the way described in Sec. III B.
We set 
$t_{\textrm{3rd}}=-40$ meV in the third or fourth case
and $t_{\textrm{3rd}}=-60$ meV in the fifth case. 
Note that the former value  
corresponds to the average of 
the intraorbital hopping integral of the $d_{xy}$ orbital for the $Z_{3}$ bonds,  
that of the $d_{yz}$ orbital for the $X_{3}$ bonds,
and that of the $d_{xz}$ orbital for the $Y_{3}$ bonds
which are obtained in the first-principles calculations~\cite{Valenti-PRB};
and that
the latter value is also considered
to clarify the effects of $J_{\delta}^{\textrm{3rd}}$ in detail.

Before discussing the properties for $|u_{ij}|\neq 0$,
we comment on the properties at $u_{ij}=0$ in the five cases of our model.
First,
the ferromagnetic state, 
a magnetic state with $\bdQ=\bdzero$, 
has the lowest energy
in the first, second, and fourth cases,
whereas
the zigzag state for $\bdQ=\bdQ_{\textrm{ZZ-}Z}$, 
a magnetic state with $\bdQ=\bdK_{10}/2$,  
is the lowest-energy state
in the third and fifth cases;
the $120^{\circ}$ order state has the highest energy in all the cases.
This result indicates that 
the stability of the zigzag or ferromagnetic state
is sensitive to the value of $t_{\textrm{3rd}}$ 
and the degree of the bond anisotropy of the hopping integrals.
Since the main effect of $t_{\textrm{3rd}}$ is to induce $J_{\delta}^{\textrm{3rd}}$,
our result is consistent with
the result obtained in a minimal model of $\alpha$-RuCl$_{3}$~\cite{Valenti-PRB}.
Furthermore, 
the competition between these magnetic states agrees with
the experimental result~\cite{al-RuCl3-FM-compete}.
The sensitivity to $J_{\delta}^{\textrm{3rd}}$ can be understood by 
estimating its energy in the MFA:
since the expectation value of Eq. (\ref{eq:H-J3rd}) can be written in the MFA as
$\sum_{\langle\langle\langle i,j\rangle\rangle\rangle}J_{\delta}^{\textrm{3rd}}
\langle\bdS_{i}\rangle\cdot\langle\bdS_{j}\rangle$,
the corresponding energies per spin surrounded by three third neighbors
for the ferromagnetic, zigzag, and $120^{\circ}$ order states
are $3J_{\textrm{3rd}}S^{2}$, $-3J_{\textrm{3rd}}S^{2}$, and $0$, respectively, 
where $J_{X}^{\textrm{3rd}}=J_{Y}^{\textrm{3rd}}=J_{Z}^{\textrm{3rd}}=J_{\textrm{3rd}}$.
Then,
the zigzag states for $\bdQ=\bdQ_{\textrm{ZZ-}Z}$ and $\bdQ_{\textrm{ZZ-}X}$, 
ones of the magnetic states with $\bdQ=\bdK_{10}/2$ and $\bdK_{01}/2$,
are degenerate in the first and third cases; 
in the second, fourth, and fifth cases 
this degeneracy is lifted and the former state is of lower energy.
This lifting is due to the bond anisotropy of the hopping integrals; 
in addition,
the lower energy of the zigzag state for $\bdQ=\bdQ_{\textrm{ZZ-}Z}$
is mainly due to $|J_{X}+K_{X}|>|J_{Z}+K_{Z}|$,
which makes the ferromagnetic spin alignment of the $X$ bonds
more stable than that of the $Z$ bonds. 
(As explained in Sec. IV A,
two spins on a $X$ or $Z$ bond are antiferromagnetic
in the zigzag state for $\bdQ=\bdQ_{\textrm{ZZ-}X}$ or $\bdQ_{\textrm{ZZ-}Z}$, respectively.)
Note that
the energy difference between these zigzag states
is about $0.1$ meV per spin. 

We now present the $|u_{ij}|$ dependences of the energies of the magnetic states
in the first case of our model
with $\bdE_{\textrm{circ}}(t)$, $\bdE_{\textrm{linear-}b}(t)$, or $\bdE_{\textrm{linear-}a}(t)$.
These dependences at $\omega=3$ and $2.2$ eV are shown
in Fig. \ref{fig10}.
The results with $\bdE_{\textrm{circ}}(t)$ [Figs. \ref{fig10}(a) and \ref{fig10}(d)]
show that
the energies of the magnetic states with $\bdQ=\bdzero$, $\bdQ_{\textrm{ZZ-}X}$,
and $\bdQ_{\textrm{ZZ-}Z}$ are close even in the range of $0<|u_{ij}|\leq 1$
and that the energy of the $120^{\circ}$ order state is much higher than them
except near
the $|u_{ij}|$'s at which the exchange interactions are
very small in magnitude [see Figs. \ref{fig2}(b) and \ref{fig10}(d)].
The similar properties hold
even in the results with $\bdE_{\textrm{linear-}b}(t)$ or $\bdE_{\textrm{linear-}a}(t)$
at $\omega=3$ eV [Fig. \ref{fig10}(b) or \ref{fig10}(c)].

There are several properties characteristic of the linearly polarized light fields.
One is the lifting of the degeneracy of the magnetic states
with $\bdQ=\bdQ_{\textrm{ZZ-}X}$ and $\bdQ_{\textrm{ZZ-}Z}$. 
For example,
in the case with $\bdE_{\textrm{linear-}b}(t)$ at $\omega=3$ eV
the magnetic state with $\bdQ=\bdQ_{\textrm{ZZ-}X}$ is of lower energy
than that with $\bdQ=\bdQ_{\textrm{ZZ-}Z}$ [Fig. \ref{fig10}(b)],
whereas in the case with $\bdE_{\textrm{linear-}a}(t)$ at $\omega=3$ eV 
the latter is of lower energy [Fig. \ref{fig10}(c)]. 
This lifting is due to the light-induced bond anisotropy of the exchange interactions,
one of the characteristics of linearly polarized light.
The property that   
the magnetic state with $\bdQ=\bdQ_{\textrm{ZZ-}X}$ is of lower energy
than that with $\bdQ=\bdQ_{\textrm{ZZ-}Z}$
in the case with $\bdE_{\textrm{linear-}b}(t)$ at $\omega=3$ eV 
comes from the facts that 
the exchange interactions for the $Z$ bonds are larger in magnitude
than those for the $X$ or $Y$ bonds 
and that $J_{Z}$ and $K_{Z}$ are ferromagnetic
[see Figs. \ref{fig4}(a){--}(c) in the range of $0\leq |u_{ij}|\leq 1$]; 
the property
in the case with $\bdE_{\textrm{linear-}a}(t)$ at $\omega=3$ eV
can be similarly understood.
The other characteristic properties are
the changes in the competing magnetic states.
From Fig. \ref{fig10}(e)
we see 
the competing magnetic states 
in the range of $0<|u_{ij}|\leq 0.4$ are 
the magnetic states with $\bdQ=\bdzero$ and $\bdQ_{\textrm{ZZ-}Z}$,  
whereas those in the range of $0.6\leq |u_{ij}|\leq 0.8$
become the magnetic states with $\bdQ=\bdQ_{\textrm{ZZ-}Z}$,
$\bdQ_{\textrm{ZZ-}X}$, and $\bdQ_{120}$.
In addition,
from Fig. \ref{fig10}(f)
we see a similar change in the competing magnetic states.
These results are related to
the light-induced bond anisotropy of the exchange interactions 
because in the range of $0.6\leq |u_{ij}|\leq 0.8$
with $\bdE_{\textrm{linear-}b}(t)$ or $\bdE_{\textrm{linear-}a}(t)$
the signs of the exchange interactions only
for the $Z$ bonds or for the $X$ and $Y$ bonds, respectively,
are changed 
and their magnitudes become larger than those for the other bonds
[see Figs. \ref{fig4}(d){--}(f) for $\bdE_{\textrm{linear-}b}(t)$
and Figs. \ref{fig5}(d){--}(f) for $\bdE_{\textrm{linear-}a}(t)$].

We turn to the results in the second case of our model.
Figure \ref{fig11} shows the $|u_{ij}|$ dependences of the energies of the magnetic states
in this case with $\bdE_{\textrm{circ}}(t)$,
$\bdE_{\textrm{linear-}b}(t)$, or $\bdE_{\textrm{linear-}a}(t)$.
These results are qualitatively the same
as those in the first case
except that the degeneracy of the magnetic states
with $\bdQ=\bdQ_{\textrm{ZZ-}X}$ and $\bdQ_{\textrm{ZZ-}Z}$ is lifted even for $\bdE_{\textrm{circ}}(t)$.
(As described above,
this lifting results from the bond anisotropy of the hopping integrals,
which is absent in the first case and present in the second case.)
Combining this result with the results at $u_{ij}=0$,
we find that
the main effects of the bond anisotropy of the hopping integrals
are
to lift the degeneracy of the magnetic states
with $\bdQ=\bdQ_{\textrm{ZZ-}X}$ and $\bdQ_{\textrm{ZZ-}Z}$
and 
to decrease the energy of the ferromagnetic state
compared with those of the zigzag states at $u_{ij}=0$.

As well as the results in the second cases,
the results in the third case (Fig. \ref{fig12}) are similar to those in the first case
except for two differences. 
The two differences are
that at $u_{ij}=0$ the energies of the zigzag states
are lower than that of the ferromagnetic state
and that the magnetic state with $\bdq=\bdzero$
becomes of lower energy than those of the magnetic states with
$\bdQ=\bdQ_{\textrm{ZZ-}X}$ and $\bdQ_{\textrm{ZZ-}Z}$ above a certain value of $|u_{ij}|$
[e.g., see the values at $|u_{ij}|=0.46$ in Fig. \ref{fig12}(a)].
These differences result from the effects of $J_{\delta}^{\textrm{3rd}}$
because the latter difference can be understood
from the $|u_{ij}|$ dependences of $J_{\delta}^{\textrm{3rd}}$ (Fig. \ref{fig13});
for example, in the case with $\bdE_{\textrm{circ}}(t)$ at $\omega=3$ eV
the blue and red lines of Fig. \ref{fig12}(a) cross
at the value of $|u_{ij}|$ at which $J_{\textrm{3rd}}$ is small in magnitude [Fig. \ref{fig13}(a)]. 
Then, as described above,
the former difference can be understood from
the difference in the energies due to $J_{\delta}^{\textrm{3rd}}$.

The similar effects of $J_{\delta}^{\textrm{3rd}}$ appear
in the fourth and fifth cases (Figs. \ref{fig14} and \ref{fig15}).
However,
in the presence of the bond anisotropy of the nearest-neighbor hopping integrals 
$t_{\textrm{3rd}}=-40$ meV is not sufficient for making
the energies of the zigzag states lower at $u_{ij}=0$
than that of the ferromagnetic state (Fig. \ref{fig14}),
although the zigzag states are of lower energy
at $u_{ij}=0$ for $t_{\textrm{3rd}}=-60$ meV (Fig. \ref{fig15}).
Nevertheless,
we believe this does not contradict the experimental result
that the zigzag state is stabilized in $\alpha$-RuCl$_{3}$ 
because $J_{\delta}^{\textrm{3rd}}$ is underestimated 
in our simplified treatment (see Sec. III A and Appendix D);
a more accurate calculation is beyond the scope of this paper.

The most important thing seen
from Figs. \ref{fig11}, \ref{fig12}, \ref{fig14}, and \ref{fig15} is that
the characteristic properties found in the first case 
remain qualitatively unchanged even in the other four cases.
Thus, we believe
the lifting of the magnetic states with $\bdQ=\bdQ_{\textrm{ZZ-}X}$ and $\bdQ_{\textrm{ZZ-}Z}$
and the change in the competing magnetic states
are characteristic of linearly polarized light.
Although the former can be realized by using
the bond-anisotropic nearest-neighbor hopping integrals,
the latter is a unique effect of linearly polarized light. 

\section{Discussion}

First, we discuss the validity of our model.
Our model has
the three nearest-neighbor hopping integrals, including their bond anisotropy,
and the third-neighbor one. 
For the Mott insulating state 
the effective Hamiltonian consists of 
the nearest-neighbor Heisenberg, Kitaev,
and off-diagonal symmetric exchange interactions
and the third-neighbor Heisenberg interaction
(i.e., $J_{\delta}$, $K_{\delta}$, $\Gamma_{\delta}$, and $J_{\delta}^{\textrm{3rd}}$).
We believe 
our model is sufficient for describing the magnetic properties
of $\alpha$-RuCl$_{3}$ 
because its first-principles calculations~\cite{Valenti-PRB} showed that
the leading hopping integrals are $t_{2}$ and $t_{3}$,
which are at least an order of magnitude larger than the others
and suggested that
its minimal spin model consists of the bond-averaged
$J_{\delta}$, $K_{\delta}$, $\Gamma_{\delta}(=-K_{\delta})$, and $J_{\delta}^{\textrm{3rd}}$.
In addition, 
the signs of the exchange interactions of our model
are consistent with those estimated
by fitting a magnetization measurement~\cite{Exp-FM-Kitaev}.
Note that according to this measurement,
another off-diagonal symmetric exchange interaction $\Gamma^{\prime}$,
which does not appear in our model, is smaller than 
these four exchange interactions.
The similar results are obtained
in other studies~\cite{al-RuCl3-FM-compete,Valenti-PRB}.
This is reasonable because
$\Gamma^{\prime}$ is proportional to
the other nearest-neighbor interorbital hopping integral~\cite{Rau-PRL},
which is small in the case of $\alpha$-RuCl$_{3}$~\cite{Valenti-PRB}.

We should note that although the effect of SOC
on the coefficients of the exchange interactions
is necessary for discussing their values quantitatively~\cite{Valenti-PRB},
the following arguments indicate that
its effect may be not large in the case of $\alpha$-RuCl$_{3}$.
In general, 
$H_{\textrm{SOC}}$ affects the coefficients of the exchange interactions of Mott insulators, 
as we can see from Eq. (\ref{eq:Heff-t}).   
Since the $LS$-type SOC induces the onsite interorbital excitations,
it can connect $|i;\Gamma,g_{\Gamma}\rangle$
with $|i;\Gamma,g_{\Gamma}^{\prime}\rangle$ for $g_{\Gamma}^{\prime}\neq g_{\Gamma}$;
such off-diagonal terms result in the degeneracy lifting of
the states $|i;\Gamma,g_{\Gamma}\rangle$'s for given $\Gamma$.
Thus, 
the main effect of $H_{\textrm{SOC}}$ on the coefficients of the exchange interactions is
to change the energies of the intermediate states
in the second-order perturbation processes
considered to derive the exchange interactions; 
the energy for a certain $\Gamma$, including this effect of $H_{\textrm{SOC}}$, 
could be expressed as $E_{\Gamma}\pm c_{\Gamma}\lambda$,
where $\lambda$ is the coupling constant of $H_{\textrm{SOC}}$
and $c_{\Gamma}=O(1)$ or $O(0.1)$.
Accordingly,
the modulations of the exchange interactions which come from
the intermediate states for $\Gamma$ 
are roughly given by
$E_{\Gamma}/(E_{\Gamma}\pm c_{\Gamma}\lambda)\sim 1\mp c_{\Gamma}(\lambda/E_{\Gamma})$.
Since $\lambda=O(0.1 \textrm{eV})$ in $\alpha$-RuCl$_{3}$,
we have $(\lambda/E_{\Gamma})=O(0.1)$, and thus
the corrections due to SOC are small for $\alpha$-RuCl$_{3}$. 
From the above arguments, we conclude that
our treatment,
in which the effect of SOC on the coefficients of the exchange interactions is neglected, 
is sufficient for discussing the exchange interactions of $\alpha$-RuCl$_{3}$ qualitatively. 

Next, we make some remarks about heating effects.
Since the periodic driving field causes the system to heat up,
it eventually approaches an infinite-temperature state~\cite{Heat-InfT1}.
However, there are intermediate times $t\lesssim \tau$ at which
the periodically driven system can be approximately described
by the Floquet Hamiltonian~\cite{Heat-Floq1,Heat-Floq2}.
Since $\tau$ is roughly given by $\tau\approx T\exp(\omega/J_{\textrm{ex}})$~\cite{Heat-Floq2}
and our parameters satisfy
$\omega=O(1\textrm{eV})$ and
$J_{\textrm{ex}}=\textrm{max}(J_{\delta},K_{\delta},\Gamma_{\delta},J_{\delta}^{\textrm{3rd}})=O(10\textrm{meV})$,
the intermediate times of our system may be sufficiently large.
Note that because of these values of $\omega$ and $J_{\textrm{ex}}$,
the correction to the Floquet Hamiltonian,
the second term of Eq. (\ref{eq:H-Floquet-next}), is negligible.
Then,
since our $\omega$ is non-resonant,
the heating effect due to the doublon creation induced by the driving field
is also negligible~\cite{Floquet-MultiMott}.
(Since this heating effect is non-negligible for resonant or nearly resonant $\omega$,
we have shown the results for some non-resonant $\omega$'s
in Secs. III B and IV B.)
It should be noted that
the effects of the doublon creation due to the driving field
can be described by
$\bar{H}_{\textrm{KE}}$ in Eq. (\ref{eq:Heff-t}), 
and that they become negligible
if $\omega$ is non-resonant
in the sense that
the denominator of Eq. (\ref{eq:Heff-t})
does not diverge~\cite{Floquet-MultiMott}.
(If $\omega$ is resonant or nearly resonant,
$\bar{H}_{\textrm{KE}}$ induces a non-negligible imaginary part,
resulting in the heating effects~\cite{Floquet-MultiMott}.)
Thus, we believe our results based on the Floquet Hamiltonian are meaningful
as the properties of periodically driven $\alpha$-RuCl$_{3}$.

We also remark on 
a property induced by the field of circularly polarized light.
It has been shown for a single-orbital Hubbard model
driven by circularly polarized light~\cite{Chirality1,Chirality2} 
that
when the light frequency is comparable with the Hubbard interaction,
the effective Hamiltonian of the Mott insulator could acquire 
a spin scalar chirality, which is non-negligible
compared with the antiferromagnetic Heisenberg interaction.
Note that the spin scalar chirality terms come from
the fourth-order perturbation processes in which the kinetic terms are treated as perturbation,
whereas the exchange interactions come from the second-order ones. 
Although a similar mechanism might work in more complicated models,
our rough estimate shown below indicates that
such contributions may be negligible in our cases.
Since the dominant Bessel functions appearing in the exchange interactions
and the spin scalar chirality terms are 
$\mathcal{J}_{0}(u_{ij})$ and $\mathcal{J}_{1}(u_{ij})$, 
a ratio of $J_{\textrm{chi}}$, one of the spin scalar chirality terms,
to $J_{\textrm{exch}}$, one of the exchange interactions,
may be roughly given by
$|J_{\textrm{chi}}/J_{\textrm{exch}}|\sim t^{2}\mathcal{J}_{1}(u_{ij})^{2}/(U_{\textrm{int}}|U_{\textrm{int}}-\omega|)$,
where
$t$ is of the order of the hopping integrals
and $U_{\textrm{int}}$ is of the order of the onsite Coulomb interactions;
in some cases of our analyses
our parameters correspond to  
$t=O(0.1\textrm{eV})$, $U_{\textrm{int}}=O(1\textrm{eV})$,
and $|U_{\textrm{int}}-\omega|=O(0.1\textrm{eV})$.
(Precisely speaking,
our $|U_{\Gamma}-n\omega|$'s for any $\Gamma$ and allowed $n$
are larger than the hopping integrals.) 
In addition,
since $\mathcal{J}_{1}(u_{ij})^{2}=O(0.1)$,  
we have $|J_{\textrm{chi}}/J_{\textrm{ex}}|\sim 10^{-2}$.
Although there is another contribution to the spin scalar chirality terms,
$J_{\textrm{chi}}^{\prime}$,
it may be smaller than the above contribution
because of an additional factor $|U_{\textrm{int}}-\omega|/\omega=O(10^{-1})$
[i.e., $|J_{\textrm{chi}}^{\prime}/J_{\textrm{chi}}|\sim
  |U_{\textrm{int}}-\omega|/\omega=O(10^{-1})$]. 
Thus, we believe the spin scalar chirality terms are negligible
and the results shown in Sec. IV B remain qualitatively unchanged.

We now address an experimental observation of our results.
Controlling the exchange interactions via a light field, in principle,
is achievable by performing pump-probe measurements.
However,
a strong light field is necessary 
because the $\omega$'s considered in our study are high.
For example,
to realize the periodically driven $\alpha$-RuCl$_{3}$
at $\omega=2.2$ eV for $|u_{ij}|\sim 0.4$ or $0.6$,
the amplitude of a light field, $E_{0}$,
should be $E_{0}\sim 26$ or $39$ MVcm$^{-1}$, respectively 
[for the relation between $u_{ij}$ and $E_{0}$ see Eq. (\ref{eq:uij_circ})];
in the cases at $\omega=1.8$ eV for $|u_{ij}|\sim 0.4$ or $0.6$,
we have $E_{0}\sim 21$ or $32$ MVcm$^{-1}$, respectively. 
In these estimates we have used $a_{\textrm{NN}}\sim 3.4\times 10^{-8}$cm,
which corresponds to the average of
the experimentally observed lengths of two nearest-neighbor bonds~\cite{al-RuCl3-lattice}.
Then, the properties of the magnetic states could be observed
by using, for example, neutron scattering measurements.
Specifically,
the changes in the competing magnetic states due to linearly polarized light
could be detected as the evolution of short-range correlations characterized by
the corresponding ordering vectors.
Since
a light field of the order of $10$ MVcm$^{-1}$ can be realized experimentally~\cite{Iwai-review},
we hope
our main results,
i.e., 
the changes in the exchange interactions, their bond anisotropy,
and the competing magnetic states via linearly polarized light,  
will be observed by experiments.

Finally, we comment on several directions for further relevant research.
First,
our theory can be extended to
the cases of 
$\alpha$-RuCl$_{3}$ with both a light field and an external magnetic field
and of other periodically driven Mott insulators on the honeycomb lattice (e.g., some Ir oxides). 
Our theory may be also useful for formulating a theory
in the case on another lattice with strong SOC. 
The studies in these cases are contained in the possible research directions.
Another research direction is to study a possibility of Kitaev spin liquids
in a more elaborate method than the MFA. 
Such recent studies include Ref. \onlinecite{Floquet-StrongSOC}.
In addition, our results may be useful
for realizing a gapped spin liqud, a toric code phase~\cite{Kitaev,PRR-Fujimoto},
because it could be stabilized
in the presence of strong bond anisotropy of the exchange interactions.
One of the possible situations might be $\alpha$-RuCl$_{3}$ with
both an inplane magnetic field $H_{ab}$ and a linearly polarized light field
because
a gapless spin liquid could be stabilized 
for $\alpha$-RuCl$_{3}$
in the range of $7.5 \textrm{T} < H_{ab} < 16 \textrm{T}$~\cite{Hind-SL}
and a linearly polarized light field could induce the strong bond anisotropy.
Then,
our results about the exchange interactions could be used to
control magnetization dynamics and spintronics phenomena of
Mott insulators with strong SOC
because the key quantities to describe them
are the exchange interactions~\cite{review-opt,review-spintronics}.
Thus, the extensions to dynamical or transport properties are
ones of the important future research directions.
Another important future research direction is an extension to the case for small $\omega$,
at which the heating effects are no longer negligible. 

\section{Conclusion}

We have studied the magnetic properties of $\alpha$-RuCl$_{3}$ driven
by circularly or linearly polarized light.
We showed that,
as well as the magnitudes and signs of the exchange interactions,
their bond anisotropy can be changed by tuning
the amplitude and frequency of one of the fields of linearly polarized light.
This is one of the characteristics of linearly polarized light
because the bond anisotropy is not induced by circularly polarized light.
Since the light-induced bond anisotropy can be used to
change the ratios of $J$, $K$, and $\Gamma$ for the $Z$ bonds
to those for the $X$ or $Y$ bonds, 
the honeycomb-network spin system could be transformed into
weakly coupled zigzag or step chains
for $\bdE_{\textrm{linear-}b}(t)$ or $\bdE_{\textrm{linear-}a}(t)$,
respectively.
We also showed that 
the $|u_{ij}|$ dependences of the exchange interactions obtained in the first case
remain qualitatively unchanged except for the degeneracy lifting
of the exchange interactions for the $Z$ bonds and for the $X$ and $Y$ bonds. 
Then, we showed that
the competing magnetic states
can be changed only for linearly polarized light.
Such a situation could be realized by using 
the strong field of the order of $10$MVcm$^{-1}$
in pump-probe measurements. 
We also showed that
the bond anisotropy of the nearest-neighbor hopping integrals
and the third-neighbor hopping integral do not change qualitatively
the results obtained in the first case of our model 
except for the stability of the zigzag states at $|u_{ij}|=0$
and the degeneracy lifting of the magnetic states with 
$\bdQ=\bdQ_{\textrm{ZZ-}X}$ and $\bdQ_{\textrm{ZZ-}Z}$.
We believe
this paper is useful for further research
of $\alpha$-RuCl$_{3}$ and the relevant materials such as the honeycomb iridates
and provides an important step towards a comprehensive understanding of
magnetic properties of periodically driven Mott insulators with strong SOC.

\begin{acknowledgments}
  This work was supported by
  JST CREST Grant No. JPMJCR1901, 
  JSPS KAKENHI Grants No. JP19K14664 and No. JP16K05459,
  and MEXT Q-LEAP Grant No. JP-MXS0118067426.
\end{acknowledgments}

\appendix
\section{Derivation of Eq. (\ref{eq:barH_KE})}

We calculate $\bar{H}_{\textrm{KE}}$'s for $\bdA(t)=\bdA_{\textrm{circ}}(t)$,
$\bdA_{\textrm{linear-}b}(t)$, and $\bdA_{\textrm{linear-}a}(t)$.
$\bar{H}_{\textrm{KE}}$ is given by
\begin{align}
  \bar{H}_{\textrm{KE}}
  =&\mathcal{P}_{1}
  \frac{\omega}{2\pi}\int_{0}^{2\pi/\omega}dt H_{\textrm{KE}}
  \mathcal{P}_{1}\notag\\
  =&\mathcal{P}_{1}
  \frac{\omega}{2\pi}\int_{0}^{2\pi/\omega}dt
  \sum_{i,j}\sum_{a,b}\sum_{\sigma}
  t_{iajb}e^{-ie(\bdR_{i}-\bdR_{j})\cdot\bdA(t)}\notag\\
  &\times c_{ia\sigma}^{\dagger}c_{jb\sigma}
  \mathcal{P}_{1}.\label{eq:barH_KE-start}
\end{align}
To perform the time integral,
we rewrite the Peierls phase factor 
using Eqs. (\ref{eq:A_c}){--}(\ref{eq:A_la}). 
For $\bdA(t)=\bdA_{\textrm{circ}}(t)$,
we write it as follows~\cite{Floquet-NA}:
\begin{align}
  e^{-ie(\bdR_{i}-\bdR_{j})\cdot\bdA(t)}=
  \begin{cases}
    \ e^{iu_{ij}\sin(\omega t+\frac{5\pi}{3})} \ \ (X\ \textrm{bonds}),\\
    \ e^{iu_{ij}\sin(\omega t+\frac{\pi}{3})} \ \ \ (Y\ \textrm{bonds}),\\
    \ e^{iu_{ij}\sin(\omega t+\pi)} \ \ \ \ (Z\ \textrm{bonds}),
  \end{cases}\label{eq:Peierls_circ}
\end{align}
where $u_{ij}$ is defined in Eq. (\ref{eq:uij_circ}).
Similarly,
we obtain
\begin{align}
  e^{-ie(\bdR_{i}-\bdR_{j})\cdot\bdA(t)}=
  \begin{cases}
    \ e^{i\frac{1}{2}u_{ij}\sin(\omega t-\frac{\pi}{2})} \ \ (X\ \textrm{or}\ Y\ \textrm{bonds}),\\
    \ e^{iu_{ij}\sin(\omega t+\frac{\pi}{2})} \ \ \ \ (Z\ \textrm{bonds})
  \end{cases}\label{eq:Peierls_linear-b}
\end{align}
for $\bdA(t)=\bdA_{\textrm{linear-}b}(t)$, and
\begin{align}
  e^{-ie(\bdR_{i}-\bdR_{j})\cdot\bdA(t)}=
  \begin{cases}
    \ e^{i\frac{\sqrt{3}}{2}u_{ij}\sin(\omega t+\frac{\pi}{2})} \ \ (X\ \textrm{bonds}),\\
    \ e^{i\frac{\sqrt{3}}{2}u_{ij}\sin(\omega t-\frac{\pi}{2})} \ \ (Y\ \textrm{bonds}),\\
    \ 1 \ \ \ \ \ \ \ \ \ \ \ \ \ \ \ \ \ \ \ \ \ (Z\ \textrm{bonds})
  \end{cases}\label{eq:Peierls_linear-a}
\end{align}
for $\bdA(t)=\bdA_{\textrm{linear-}a}(t)$.
By combining Eqs. (\ref{eq:Peierls_circ}){--}(\ref{eq:Peierls_linear-a})
with Eq. (\ref{eq:barH_KE-start}) and using the relations
\begin{align}
  e^{ix\sin\theta}
  =&\sum_{n=-\infty}^{\infty}\mathcal{J}_{n}(x)e^{in\theta}\label{eq:Bess-def}
\end{align}
and
\begin{align}
  \frac{\omega}{2\pi}\int_{0}^{2\pi/\omega}dt e^{in\omega t}=\delta_{n,0},
\end{align}
we obtain Eq. (\ref{eq:barH_KE}).

\section{Derivation of Eq. (\ref{eq:Psi1-final})}

We solve Eq. (\ref{eq:Psi1-rewrite2}). 
By integrating both sides 
and choosing for the initial state the state
in which the lower-limit contributions of the integrals cancel each other out,
we have
\begin{align}
  ie^{i(\bar{H}_{\textrm{KE}}+\tilde{H}_{\textrm{int}})t}|\Psi_{1}\rangle_{t}
  =&\int^{t} dt^{\prime}e^{i(\bar{H}_{\textrm{KE}}+\tilde{H}_{\textrm{int}})t^{\prime}}H_{\textrm{KE}}
  |\Psi_{0}\rangle_{t^{\prime}}.\label{eq:Psi1-rewrite3}
\end{align}
Furthermore,
since the time variation of $|\Psi_{0}\rangle_{t^{\prime}}$ is slow,
we could write Eq. (\ref{eq:Psi1-rewrite3}) as
\begin{align}
  ie^{i(\bar{H}_{\textrm{KE}}+\tilde{H}_{\textrm{int}})t}|\Psi_{1}\rangle_{t}
  \approx &\int^{t} dt^{\prime}e^{i(\bar{H}_{\textrm{KE}}+\tilde{H}_{\textrm{int}})t^{\prime}}H_{\textrm{KE}}
  |\Psi_{0}\rangle_{t}.\label{eq:Psi1-rewrite4}
\end{align}
By combining Eq. (\ref{eq:Psi1-rewrite4})
with Eq. (\ref{eq:H_KE}) and Eqs. (\ref{eq:Peierls_circ}){--}(\ref{eq:Bess-def})
and performing the time integral,
we obtain Eq. (\ref{eq:Psi1-final}).

\section{Derivation of Eq. (\ref{eq:Heff_Z})}

We calculate the possible terms of Eq. (\ref{eq:barH-next}) for the $Z$ bonds.
The calculations consist of two steps.

First, we calculate $\langle i;\Gamma,g_{\Gamma}|T_{ij}|\textrm{i}\rangle$ for $T_{ij}=T_{ij}^{Z}$. 
Since
\begin{align}
|\textrm{i}\rangle=\{|+\rangle_{1}|+\rangle_{2}, |+\rangle_{1}|-\rangle_{2},
|-\rangle_{1}|+\rangle_{2}, |-\rangle_{1}|-\rangle_{2}\},
\end{align}
we calculate the finite terms of $\langle i;\Gamma,g_{\Gamma}|T_{12}^{Z}|\textrm{i}\rangle$;
the contributions from $\langle i;\Gamma,g_{\Gamma}|T_{21}^{Z}|\textrm{i}\rangle$
to $\bar{H}_{\textrm{eff}}$
can be taken into account by multiplying
those from $\langle i;\Gamma,g_{\Gamma}|T_{12}^{Z}|\textrm{i}\rangle$
by two.
Using Eqs. (\ref{eq:jeff-p}), (\ref{eq:jeff-m}), and (\ref{eq:Tij^z}),
we have
\begin{align}
  T_{12}^{Z}|+,+\rangle
  =&\frac{1}{3}\Bigl[(t_{3}-t_{1})c_{1d_{yz}\downarrow}^{\dagger}c_{1d_{xy}\uparrow}^{\dagger}
    -i(t_{1}-t_{3})c_{1d_{zx}\downarrow}^{\dagger}c_{1d_{xy}\uparrow}^{\dagger}\notag\\
    &+2t_{2}c_{1d_{yz}\downarrow}^{\dagger}c_{1d_{zx}\downarrow}^{\dagger}
    -it_{2}c_{1d_{yz}\downarrow}^{\dagger}c_{1d_{xy}\uparrow}^{\dagger}\notag\\
    &-t_{2}c_{1d_{zx}\downarrow}^{\dagger}c_{1d_{xy}\uparrow}^{\dagger}
    \Bigr]|0\rangle,\label{eq:T12_++}\\
  T_{12}^{Z}|-,-\rangle
  =&\frac{1}{3}\Bigl[(t_{1}-t_{3})c_{1d_{yz}\uparrow}^{\dagger}c_{1d_{xy}\downarrow}^{\dagger}
    -i(t_{1}-t_{3})c_{1d_{zx}\uparrow}^{\dagger}c_{1d_{xy}\downarrow}^{\dagger}\notag\\
    &+2t_{2}c_{1d_{yz}\uparrow}^{\dagger}c_{1d_{zx}\uparrow}^{\dagger}
    -it_{2}c_{1d_{yz}\uparrow}^{\dagger}c_{1d_{xy}\downarrow}^{\dagger}\notag\\    
    &+t_{2}c_{1d_{zx}\uparrow}^{\dagger}c_{1d_{xy}\downarrow}^{\dagger}
    \Bigr]|0\rangle,\label{eq:T12_--}\\
  T_{12}^{Z}|+,-\rangle
  =&\frac{1}{3}\Bigl[-(t_{1}-it_{2})c_{1d_{yz}\uparrow}^{\dagger}c_{1d_{yz}\downarrow}^{\dagger}
    -(t_{1}+it_{2})c_{1d_{zx}\uparrow}^{\dagger}c_{1d_{zx}\downarrow}^{\dagger}\notag\\
    &-(it_{1}+t_{2})c_{1d_{yz}\uparrow}^{\dagger}c_{1d_{zx}\downarrow}^{\dagger}
    +(it_{1}-t_{2})c_{1d_{zx}\uparrow}^{\dagger}c_{1d_{yz}\downarrow}^{\dagger}\notag\\
    &-(t_{1}-it_{2})c_{1d_{yz}\uparrow}^{\dagger}c_{1d_{xy}\uparrow}^{\dagger}
    +(it_{1}-t_{2})c_{1d_{zx}\uparrow}^{\dagger}c_{1d_{xy}\uparrow}^{\dagger}\notag\\
    &-t_{3}c_{1d_{yz}\downarrow}^{\dagger}c_{1d_{xy}\downarrow}^{\dagger}
    -it_{3}c_{1d_{zx\downarrow}}^{\dagger}c_{1d_{xy}\downarrow}^{\dagger}\notag\\
    &-t_{3}c_{1d_{xy}\uparrow}^{\dagger}c_{1d_{xy}\downarrow}^{\dagger}
    \Bigr]|0\rangle,\label{eq:T12_+-}\\
  T_{12}^{Z}|-,+\rangle
  =&\frac{1}{3}\Bigl[(t_{1}+it_{2})c_{1d_{yz}\uparrow}^{\dagger}c_{1d_{yz}\downarrow}^{\dagger}
    +(t_{1}-it_{2})c_{1d_{zx}\uparrow}^{\dagger}c_{1d_{zx}\downarrow}^{\dagger}\notag\\
    &+(it_{1}+t_{2})c_{1d_{yz}\uparrow}^{\dagger}c_{1d_{zx}\downarrow}^{\dagger}
    -(it_{1}-t_{2})c_{1d_{zx}\uparrow}^{\dagger}c_{1d_{yz}\downarrow}^{\dagger}\notag\\
    &+(t_{1}+it_{2})c_{1d_{yz}\downarrow}^{\dagger}c_{1d_{xy}\downarrow}^{\dagger}
    +(it_{1}+t_{2})c_{1d_{zx}\downarrow}^{\dagger}c_{1d_{xy}\downarrow}^{\dagger}\notag\\
    &+t_{3}c_{1d_{yz}\uparrow}^{\dagger}c_{1d_{xy}\uparrow}^{\dagger}
    -it_{3}c_{1d_{zx\uparrow}}^{\dagger}c_{1d_{xy}\uparrow}^{\dagger}\notag\\
    &+t_{3}c_{1d_{xy}\uparrow}^{\dagger}c_{1d_{xy}\downarrow}^{\dagger}
    \Bigr]|0\rangle.\label{eq:T12_-+}
\end{align}
By using Eqs. (\ref{eq:T12_++}){--}(\ref{eq:T12_-+})
and Eqs. (\ref{eq:A1}){--}(\ref{eq:T2-gam}), 
we can calculate
$\langle i;\Gamma,g_{\Gamma}|T_{12}^{Z}|\textrm{i}\rangle$'s;
as a result,
the finite terms are given by 
\begin{align}
  \langle i;A_{1}|T_{12}^{Z}|+,-\rangle
  &=-\langle i;A_{1}|T_{12}^{Z}|-,+\rangle\notag\\
  &=-\frac{1}{3\sqrt{3}}(2t_{1}+t_{3}),\label{eq:MatEle-first}\\
  \langle i;E,u|T_{12}^{Z}|+,-\rangle
  &=-\langle i;E,u|T_{12}^{Z}|-,+\rangle\notag\\
  &=-\frac{2}{3\sqrt{6}}(t_{1}-t_{3}),\\
  \langle i;E,v|T_{12}^{Z}|+,-\rangle
  &=\langle i;E,v|T_{12}^{Z}|-,+\rangle
  =\frac{2}{3\sqrt{2}}it_{2},\\
  \langle i;T_{1},\alpha_{+}|T_{12}^{Z}|-,-\rangle
  &=\langle i;T_{1},\alpha_{-}|T_{12}^{Z}|+,+\rangle
  =\frac{2}{3}t_{2},\\
  \langle i;T_{1},\alpha|T_{12}^{Z}|+,-\rangle
  &=-\langle i;T_{1},\alpha|T_{12}^{Z}|-,+\rangle=-\frac{2}{3\sqrt{2}}it_{1},\\
  \langle i;T_{2},\alpha|T_{12}^{Z}|+,-\rangle
  &=-\langle i;T_{2},\alpha|T_{12}^{Z}|-,+\rangle=-\frac{2}{3\sqrt{2}}t_{2},\\
  \langle i;T_{1},\beta_{+}|T_{12}^{Z}|+,-\rangle
  &=\frac{1}{3}(it_{1}-t_{2}),
\end{align}
\begin{align}
  \langle i;T_{1},\beta_{+}|T_{12}^{Z}|-,+\rangle
  &=\langle i;T_{1},\beta_{-}|T_{12}^{Z}|+,-\rangle=-\frac{i}{3}t_{3},\\
  \langle i;T_{1},\beta_{-}|T_{12}^{Z}|-,+\rangle
  &=\frac{1}{3}(it_{1}+t_{2}),\\
  \langle i;T_{1},\beta|T_{12}^{Z}|+,+\rangle
  &=-\frac{1}{3\sqrt{2}}[i(t_{1}-t_{3})+t_{2}],\\
  \langle i;T_{1},\beta|T_{12}^{Z}|-,-\rangle
  &=-\frac{1}{3\sqrt{2}}[i(t_{1}-t_{3})-t_{2}],\\
  \langle i;T_{2},\beta|T_{12}^{Z}|+,+\rangle
  &=\frac{1}{3\sqrt{2}}[i(t_{1}-t_{3})+t_{2}],\\
  \langle i;T_{2},\beta|T_{12}^{Z}|-,-\rangle
  &=-\frac{1}{3\sqrt{2}}[i(t_{1}-t_{3})-t_{2}],\\
  \langle i;T_{1},\gamma_{+}|T_{12}^{Z}|+,-\rangle
  &=\frac{1}{3}(t_{1}-it_{2}),\\
  \langle i;T_{1},\gamma_{+}|T_{12}^{Z}|-,+\rangle
  &=-\langle i;T_{1},\gamma_{-}|T_{12}^{Z}|+,-\rangle\notag\\
  &=-\frac{1}{3}t_{3},\\
  \langle i;T_{1},\gamma_{-}|T_{12}^{Z}|-,+\rangle
  &=-\frac{1}{3}(t_{1}+it_{2}),\\
  \langle i;T_{1},\gamma|T_{12}^{Z}|+,+\rangle
  &=\frac{1}{3\sqrt{2}}[(t_{1}-t_{3})+it_{2}],\\
  \langle i;T_{1},\gamma|T_{12}^{Z}|-,-\rangle
  &=-\frac{1}{3\sqrt{2}}[(t_{1}-t_{3})-it_{2}],\\
  \langle i;T_{2},\gamma|T_{12}^{Z}|+,+\rangle
  &=\frac{1}{3\sqrt{2}}[(t_{1}-t_{3})+it_{2}],\\
  \langle i;T_{2},\gamma|T_{12}^{Z}|-,-\rangle
  &=\frac{1}{3\sqrt{2}}[(t_{1}-t_{3})-it_{2}],\label{eq:MatEle-last}
\end{align}
and the others are zero.

Then, we express Eq. (\ref{eq:barH-next}) for the $Z$ bonds
in terms of the exchange interactions. 
By combining Eqs. (\ref{eq:MatEle-first}){--}(\ref{eq:MatEle-last})
and Eqs. (\ref{eq:U_A}){--}(\ref{eq:U_T2})
with Eq. (\ref{eq:barH-next}),
we can write $\bar{H}_{\textrm{eff}}$ for the $Z$ bonds, $\bar{H}_{\textrm{eff}}^{Z}$, as follows: 
\begin{align}
\bar{H}_{\textrm{eff}}^{Z}=H_{A_{1}}+H_{E}+H_{T_{1}}+H_{T_{2}},\label{eq:Heff-rewrite}
\end{align}
where
\begin{align}
  H_{A_{1}}
  &=\sum_{\langle i,j\rangle_{Z}}\sum_{n=-\infty}^{\infty}
  \frac{4(2t_{1}+t_{3})^{2}\mathcal{J}_{n}(u_{ij}^{Z})^{2}}{27(U+2J^{\prime}-n\omega)}
  \bdS_{i}\cdot\bdS_{j},\label{eq:H_A1}\\
  H_{E}
  &=\sum_{\langle i,j\rangle_{Z}}\sum_{n=-\infty}^{\infty}
  \frac{8\mathcal{J}_{n}(u_{ij}^{Z})^{2}}{27(U-J^{\prime}-n\omega)}\notag\\
  &\times \{[(t_{1}-t_{3})^{2}-3t_{2}^{2}]\bdS_{i}\cdot\bdS_{j}
  +6t_{2}^{2}S_{i}^{z}S_{j}^{z}\},\label{eq:H_E}\\
  H_{T_{1}}
  &=\sum_{\langle i,j\rangle_{Z}}\sum_{n=-\infty}^{\infty}
  \frac{4\mathcal{J}_{n}(u_{ij}^{Z})^{2}}{9(U^{\prime}-J_{\textrm{H}}-n\omega)}\notag\\
  &\times 
  \{-[(t_{1}-t_{3})^{2}+t_{2}^{2}-2t_{1}^{2}-4t_{1}t_{3}]\bdS_{i}\cdot\bdS_{j}\notag\\
  &+[(t_{1}-t_{3})^{2}+t_{2}^{2}](S_{i}^{y}S_{j}^{y}+S_{i}^{x}S_{j}^{x})\notag\\
  &+2t_{2}(t_{1}-t_{3})(S_{i}^{x}S_{j}^{y}+S_{i}^{y}S_{j}^{x})\notag\\
  &+2[(t_{1}^{2}+t_{2}^{2}+t_{3}^{2})-2t_{2}^{2}-2t_{1}t_{3}]S_{i}^{z}S_{j}^{z}\},\label{eq:H_T1}
  \end{align}
\begin{align}
  H_{T_{2}}
  &=\sum_{\langle i,j\rangle_{Z}}\sum_{n=-\infty}^{\infty}
  \frac{4\mathcal{J}_{n}(u_{ij}^{Z})^{2}}{9(U^{\prime}+J_{\textrm{H}}-n\omega)}\notag\\
  &\times 
  \{[(t_{1}-t_{3})^{2}+t_{2}^{2}](S_{i}^{x}S_{j}^{x}+S_{i}^{y}S_{j}^{y})\notag\\
  &-2t_{2}(t_{1}-t_{3})(S_{i}^{x}S_{j}^{y}+S_{i}^{y}S_{j}^{x})\notag\\
  &-[(t_{1}-t_{3})^{2}-t_{2}^{2}]\bdS_{i}\cdot\bdS_{j}\}.\label{eq:H_T2}
\end{align}
In deriving these equations
we have used the relations of operators in the $j_{\textrm{eff}}=1/2$ subspace
[e.g., $|-,+\rangle\langle -,+|=(\frac{1}{2}-S_{1}^{z})(\frac{1}{2}+S_{2}^{z})$
and $|-,+\rangle\langle +,-|=S_{1}^{-}S_{2}^{+}$]
and omitted the constant terms.
By using the identity $S_{i}^{x}S_{j}^{x}+S_{i}^{y}S_{j}^{y}=\bdS_{i}\cdot\bdS_{j}-S_{i}^{z}S_{j}^{z}$,
we can rewrite Eqs. (\ref{eq:H_T1}) and (\ref{eq:H_T2}).
A combination of the resultant equations and  
Eqs. (\ref{eq:Heff-rewrite}){--}(\ref{eq:H_E})
gives Eq. (\ref{eq:Heff_Z}).

\section{Estimates of the values of $J_{\delta}^{\textrm{3rd}}$ and $K_{\delta}^{\textrm{3rd}}$}

We estimate $J_{\delta}^{\textrm{3rd}}$ and $K_{\delta}^{\textrm{3rd}}$ at $E_{0}=0$ in two cases.
In the following estimation
we calculate their values for the $Z_{3}$ bonds (Fig. \ref{fig1}) 
because the bond anisotropy of the third-neighbor hopping integrals
is weak~\cite{Valenti-PRB};
the values of the third-neighbor hopping integrals used below
are consistent with those of Ref. \onlinecite{Valenti-PRB}. 
Then, 
we set $J^{\prime}=J_{\textrm{H}}$, $U^{\prime}=U-2J_{\textrm{H}}$,
$U=3$ eV, and $J_{\textrm{H}}=0.5$ eV.

If we consider only $t_{\textrm{3rd}}$,
the intraorbital hopping integral of the $d_{xy}$ orbital,
among the third-neighbor hopping integrals,  
$J_{Z}^{\textrm{3rd}}$ and $K_{Z}^{\textrm{3rd}}$ at $E_{0}=0$ are given by
\begin{align}
  J_{Z}^{\textrm{3rd}}=
  \frac{4t_{\textrm{3rd}}^{2}}{27(U+2J_{\textrm{H}})}
  +\frac{8t_{\textrm{3rd}}^{2}}{27(U-J_{\textrm{H}})},
  \label{eq:J3rd-case1}
\end{align}
and
\begin{align}
  K_{Z}^{\textrm{3rd}}=
  \frac{4}{9}t_{\textrm{3rd}}^{2}
  \Bigl(
  \frac{1}{U-3J_{\textrm{H}}}
  -\frac{1}{U-J_{\textrm{H}}}
  \Bigr),
  \label{eq:K3rd-case1}
\end{align}
respectively. 
If we set $t_{\textrm{3rd}}\sim -40$ meV,
we have
$J_{Z}^{\textrm{3rd}}\sim 0.25$ meV and
$K_{Z}^{\textrm{3rd}}\sim 0.19$ meV.

Next,
we consider not only $t_{\textrm{3rd}}$, 
but also two additional terms of
third-neighbor hopping integrals, $t_{\textrm{3rd}}^{\prime}$ and $t_{\textrm{3rd}}^{\prime\prime}$.
Here 
$t_{\textrm{3rd}}^{\prime}$ represents 
the intraorbital hopping integral
of the $d_{yz}$ or $d_{zx}$ orbital 
and $t_{\textrm{3rd}}^{\prime\prime}$ represents 
the interorbital hopping integral between these orbitals. 
Then,
we can write $J_{Z}^{\textrm{3rd}}$ and $K_{Z}^{\textrm{3rd}}$ at $E_{0}=0$ as follows:
\begin{align}
  J_{Z}^{\textrm{3rd}}=&
  \frac{4(2t_{\textrm{3rd}}^{\prime}+t_{\textrm{3rd}})^{2}}{27(U+2J_{\textrm{H}})}
  +\frac{8(t_{\textrm{3rd}}^{\prime}-t_{\textrm{3rd}})^{2}}{27(U-J_{\textrm{H}})}\notag\\
  &+\frac{8t_{\textrm{3rd}}^{\prime}(t_{\textrm{3rd}}^{\prime}+2t_{\textrm{3rd}})}{9(U-3J_{\textrm{H}})},
  \label{eq:J3rd-case2}\\ 
  K_{Z}^{\textrm{3rd}}=&
  \frac{4}{9}
  \Bigl[(t_{\textrm{3rd}}^{\prime}-t_{\textrm{3rd}})^{2}-3(t_{\textrm{3rd}}^{\prime\prime})^{2}\Bigr]\notag\\
  &\times 
  \Bigl(
  \frac{1}{U-3J_{\textrm{H}}}
  -\frac{1}{U-J_{\textrm{H}}}
  \Bigr).
  \label{eq:K3rd-case2}
\end{align}
Setting $t_{\textrm{3rd}}\sim -40$ meV,
$t_{\textrm{3rd}}^{\prime}\sim -8$ meV,
and $t_{\textrm{3rd}}^{\prime\prime}\sim -7$ meV,
we have
$J_{Z}^{\textrm{3rd}}\sim 0.65$ meV and 
$K_{Z}^{\textrm{3rd}}\sim 0.1$ meV.

Comparing the estimated values in the above two cases,
we see
$J_{Z}^{\textrm{3rd}}$ and $K_{Z}^{\textrm{3rd}}$ are underestimated and overestimated,
respectively,
if we consider only $t_{\textrm{3rd}}$ among the third-neighbor hopping integrals.
Since $K_{\delta}^{\textrm{3rd}}$ is much smaller than $J_{\delta}^{\textrm{3rd}}$
even in a more realistic situation~\cite{Valenti-PRB},
we consider only $J_{\delta}^{\textrm{3rd}}$ and
neglect $K_{\delta}^{\textrm{3rd}}$ in our analyses.

\section{Derivation of Eq. (\ref{eq:H-MFA-2nd})}

We rewrite Eq. (\ref{eq:H-MFA-start}) using Eqs. (\ref{eq:Sq_A}) and (\ref{eq:Sq_B}).
First,
we can rewrite the terms of $J_{\delta}$'s as 
\begin{align}
  &\sum_{\langle i,j\rangle}J_{\delta}
  \langle \bdS_{i}\rangle\cdot\langle \bdS_{j}\rangle\notag\\
  =&\frac{1}{2}\sum_{i=1}^{N/2}\sum_{j=1}^{z_{\textrm{NN}}}J_{\delta}
  \langle \bdS_{i}\rangle\cdot\langle \bdS_{j}\rangle
  +\frac{1}{2}\sum_{j=1}^{N/2}\sum_{i=1}^{z_{\textrm{NN}}}J_{\delta}
  \langle \bdS_{i}\rangle\cdot\langle \bdS_{j}\rangle\notag\\
  =&\sum_{\bdq}\sum_{\mu=x,y,z}\Bigl[
    J_{1}(\bdq)\langle S_{-\bdq A}^{\mu}\rangle\langle S_{\bdq B}^{\mu}\rangle
    +J_{1}(\bdq)^{\ast}\langle S_{-\bdq B}^{\mu}\rangle\langle S_{\bdq A}^{\mu}\rangle\Bigr],
  \label{eq:J_App}
\end{align}
where
\begin{align}
  J_{1}(\bdq)
  =&\sum_{j=1}^{z_{\textrm{NN}}}\frac{J_{\delta}}{2}e^{-i\bdq\cdot(\bdR_{i}-\bdR_{j})}\notag\\
  =&\frac{J_{X}}{2}e^{-i\frac{q_{x}}{2}+i\frac{\sqrt{3}}{2}q_{y}}
  +\frac{J_{Y}}{2}e^{-i\frac{q_{x}}{2}-i\frac{\sqrt{3}}{2}q_{y}}
  +\frac{J_{Z}}{2}e^{iq_{x}},\label{eq:J1q_App}
\end{align}
and $z_{\textrm{NN}}$ denotes the number of nearest-neighbor sites
at a certain cite on the honeycomb lattice. 
In Eq. (\ref{eq:J1q_App})
the first, second, and third terms  
correspond to the contributions from
the $X$, $Y$, and $Z$ bonds (Fig. \ref{fig1}), respectively.
Similarly,
we can express the other terms as follows:
\begin{align}
  &\sum_{\langle i,j\rangle}K_{\delta}
  \langle S_{i}^{\gamma}\rangle\langle S_{j}^{\gamma}\rangle\notag\\
  =&\sum_{\bdq}\Bigl[
    K_{x}(\bdq)\langle S_{-\bdq A}^{x}\rangle\langle S_{\bdq B}^{x}\rangle
    +K_{y}(\bdq)\langle S_{-\bdq A}^{y}\rangle\langle S_{\bdq B}^{y}\rangle\notag\\
    &\ \ \ \
    +K_{z}(\bdq)\langle S_{-\bdq A}^{z}\rangle\langle S_{\bdq B}^{z}\rangle
    \Bigr]\notag\\
  &+\sum_{\bdq}\Bigl[
    K_{x}(\bdq)^{\ast}\langle S_{-\bdq B}^{x}\rangle\langle S_{\bdq A}^{x}\rangle
    +K_{y}(\bdq)^{\ast}\langle S_{-\bdq B}^{y}\rangle\langle S_{\bdq A}^{y}\rangle\notag\\
    &\ \ \ \
    +K_{z}(\bdq)^{\ast}\langle S_{-\bdq B}^{z}\rangle\langle S_{\bdq A}^{z}\rangle\Bigr],
  \label{eq:K_App}
\end{align}
\begin{align}
  &\sum_{\langle i,j\rangle}\Gamma_{\delta}
    (\langle S_{i}^{\alpha}\rangle\langle S_{j}^{\beta}\rangle
    +\langle S_{i}^{\beta}\rangle\langle S_{j}^{\alpha}\rangle)\notag\\
  =&\sum_{\bdq}\Bigl[
    \Gamma_{x}(\bdq)
    (\langle S_{-\bdq A}^{y}\rangle\langle S_{\bdq B}^{z}\rangle
    +\langle S_{-\bdq A}^{z}\rangle\langle S_{\bdq B}^{y}\rangle)\notag\\
    &\ \ \ \
    +\Gamma_{y}(\bdq)
    (\langle S_{-\bdq A}^{z}\rangle\langle S_{\bdq B}^{x}\rangle
    +\langle S_{-\bdq A}^{x}\rangle\langle S_{\bdq B}^{z}\rangle)\notag\\
    &\ \ \ \
    +\Gamma_{z}(\bdq)
    (\langle S_{-\bdq A}^{x}\rangle\langle S_{\bdq B}^{y}\rangle
    +\langle S_{-\bdq A}^{y}\rangle\langle S_{\bdq B}^{x}\rangle)\Bigr]\notag\\
  &+\sum_{\bdq}\Bigl[
    \Gamma_{x}(\bdq)^{\ast}
    (\langle S_{-\bdq B}^{y}\rangle\langle S_{\bdq A}^{z}\rangle
    +\langle S_{-\bdq B}^{z}\rangle\langle S_{\bdq A}^{y}\rangle)\notag\\
    &\ \ \ \
    +\Gamma_{y}(\bdq)^{\ast}
    (\langle S_{-\bdq B}^{z}\rangle\langle S_{\bdq A}^{x}\rangle
    +\langle S_{-\bdq B}^{x}\rangle\langle S_{\bdq A}^{z}\rangle)\notag\\
    &\ \ \ \
    +\Gamma_{z}(\bdq)^{\ast}
    (\langle S_{-\bdq B}^{x}\rangle\langle S_{\bdq A}^{y}\rangle
    +\langle S_{-\bdq B}^{y}\rangle\langle S_{\bdq A}^{x}\rangle)\Bigr],
  \label{eq:Gam_App}\\
  &\sum_{\langle\langle\langle i,j\rangle\rangle\rangle}J_{\delta}^{\textrm{3rd}}
  \langle \bdS_{i}\rangle\cdot\langle \bdS_{j}\rangle\notag\\
  =&\sum_{\bdq}\sum_{\mu=x,y,z}\Bigl[
    J_{3}(\bdq)\langle S_{-\bdq A}^{\mu}\rangle\langle S_{\bdq B}^{\mu}\rangle
    +J_{3}(\bdq)^{\ast}\langle S_{-\bdq B}^{\mu}\rangle\langle S_{\bdq A}^{\mu}\rangle\Bigr],
  \label{eq:J-3rd_App}
\end{align}
where $K_{\mu}(\bdq)$'s and $\Gamma_{\mu}(\bdq)$'s have been defined
in Eq. (\ref{eq:Kxq}){--}(\ref{eq:GamZq}),
and $J_{3}(\bdq)$ is given by
\begin{align}
  J_{3}(\bdq)
  =&\frac{J_{X}^{\textrm{3rd}}}{2}e^{iq_{x}-i\sqrt{3}q_{y}}
  +\frac{J_{Y}^{\textrm{3rd}}}{2}e^{iq_{x}+i\sqrt{3}q_{y}}
  +\frac{J_{Z}^{\textrm{3rd}}}{2}e^{-2iq_{x}}.\label{eq:J3q_App}
\end{align}
By combining Eqs. (\ref{eq:J_App}){--}(\ref{eq:J3q_App})
and setting $J(\bdq)=J_{1}(\bdq)+J_{3}(\bdq)$,
we obtain Eq. (\ref{eq:H-MFA-2nd}).

\section{Some remarks on momentum}

We remark on momentum in the case of the honeycomb lattice.
The remarks are about its expression and the reciprocal lattice vector
with the periodic boundary condition.
In the case of the honeycomb lattice (Fig. \ref{fig1})
a set of primitive vectors, $\bda_{1}$ and $\bda_{2}$, can be written as
\begin{align}
  &\bda_{1}
  ={}^{t}\Bigl(\frac{\sqrt{3}}{2}a_{\textrm{2nd}}\ \ \frac{1}{2}a_{\textrm{2nd}}\Bigr)
  ={}^{t}\Bigl(\frac{3}{2}\ \ \frac{\sqrt{3}}{2}\Bigr),\label{eq:a1}\\
  &\bda_{2}
  ={}^{t}\Bigl(0\ \ a_{\textrm{2nd}}\Bigr)
  ={}^{t}\Bigl(0\ \ \sqrt{3}\Bigr),\label{eq:a2}
\end{align}
and thus the primitive vectors for the reciprocal lattice,
$\bdb_{1}$ and $\bdb_{2}$, are given by
\begin{align}
  &\bdb_{1}
  ={}^{t}\Bigl(\frac{4\pi}{3}\ \ 0\Bigr),\label{eq:b1}\\
  &\bdb_{2}
  ={}^{t}\Bigl(-\frac{2\pi}{3}\ \ \frac{2\pi}{\sqrt{3}}\Bigr).\label{eq:b2}
\end{align}
(As described in the caption of Fig. \ref{fig1},
we can represent the honeycomb lattice as
a triangular Bravais lattice with a two-sublattice structure~\cite{Ashcro-Merm}.) 
By imposing the periodic boundary condition~\cite{Ashcro-Merm},
we can express momentum $\bdq$ as 
\begin{align}
  \bdq=\frac{l_{1}}{N_{1}}\bdb_{1}+\frac{l_{2}}{N_{2}}\bdb_{2},\label{eq:q}
\end{align}
where the integers $l_{1}$ and $l_{2}$ satisfy
$0\leq l_{1}<N_{1}$ and $0\leq l_{2}<N_{2}$ with $N_{1}N_{2}=N$.
As described in Sec. IV B,
we set $N_{1}=N_{2}=120$ in our analyses. 
The values of $l_{1}$ and $l_{2}$
for the magnetic states considered in Sec. IV B are given as follows:
for $\bdq=\bdQ_{\textrm{ZZ-}X}$ $l_{1}=N_{1}/4$ and $l_{2}=N_{1}/2$;
for $\bdq=\bdQ_{\textrm{ZZ-}Z}$ $l_{1}=N_{1}/2$ and $l_{2}=0$;
and for $\bdq=\bdQ_{120}$ $l_{1}=l_{2}=(2N_{1}/3)$. 
Then, 
the reciprocal lattice vector $\bdK$ can be written as
\begin{align}
  \bdK
  ={}^{t}(K_{x}\ \ K_{y})
  =m_{1}\bdb_{1}+m_{2}\bdb_{2},\label{eq:K}
\end{align}
where
\begin{align}
  K_{x}=\frac{2\pi}{3}(2m_{1}-m_{2}),\
  K_{y}=\frac{2\pi}{\sqrt{3}}m_{2},
\end{align}
and $m_{1}$ and $m_{2}$ are integers. 
From these equations
we see that 
$\bdK$ for $(m_{1},m_{2})=(0,1)$, $\bdK_{01}$, equals $2\bdQ_{\textrm{ZZ-}X}$,
and that $\bdK$ for $(m_{1},m_{2})=(1,0)$, $\bdK_{10}$, equals $2\bdQ_{\textrm{ZZ-}Z}$.
Namely, we have
\begin{align}
  \bdQ_{\textrm{ZZ-}X}=\frac{\bdK_{01}}{2},\\
  \bdQ_{\textrm{ZZ-}Z}=\frac{\bdK_{10}}{2}.
\end{align}  
In contrast to $\bdQ_{\textrm{ZZ-}X}$ and $\bdQ_{\textrm{ZZ-}Z}$,
$\bdQ_{120}$ is not equal to $\bdK/2$ for any allowed $m_{1}$ and $m_{2}$.

\end{document}